\documentclass[fleqn,usenatbib]{mnras}

\usepackage{newtxtext,newtxmath}

\usepackage[T1]{fontenc}

\DeclareRobustCommand{\VAN}[3]{#2}
\let\VANthebibliography\thebibliography
\def\thebibliography{\DeclareRobustCommand{\VAN}[3]{##3}\VANthebibliography}

\usepackage{graphicx}	
\usepackage{amsmath}	
\usepackage{pdflscape}
\usepackage{color}
\newcommand{\hii}{\mbox{H\,{\scshape ii}}\,}

\title[PAHs in Seyfert and star-forming galaxies]{Polycyclic Aromatic Hydrocarbons in Seyfert and star-forming galaxies}

\author[I. Garc\'ia-Bernete et al.]
{\parbox{\textwidth}{I. Garc\'ia-Bernete$^{1}$\thanks{E-mail: igbernete@gmail.com}, D. Rigopoulou$^{1}$, A. Alonso-Herrero$^{2}$, M. Pereira-Santaella$^{3}$, P.F. Roche,$^{1}$ and B. Kerkeni$^{1, 4, 5}$\\
}\vspace{0.4cm}\\
\parbox{\textwidth}{$^1$Department of Physics, University of Oxford, Oxford OX1 3RH, UK \\
$^2$Centro de Astrobiolog\'ia, CSIC-INTA, ESAC Campus, E-28692, Villanueva de la Ca\~nada, Madrid, Spain\\
$^3$Centro de Astrobiolog\'ia, CSIC-INTA, Carretera de Torrej\'on a Ajalvir, E-28880 Torrej\'on de Ardoz, Madrid, Spain\\
$^4$D\'epartement de Physique, LPMC, Facult\'e des Sciences de Tunis, Universit\'e de Tunis el Manar, Tunis 2092, Tunisia\\
$^5$ISAMM, Universit\'e de la Manouba, La Manouba 2010 Tunisia\\
}
}

\date{Accepted 2021 October 22. Received 2021 October 20; in original form 2020 November 23}

\pubyear{2021}

\begin{document}
\label{firstpage}
\pagerange{\pageref{firstpage}--\pageref{lastpage}}
\maketitle

\begin{abstract}
Polycyclic Aromatic Hydrocarbons (PAHs) are carbon-based molecules resulting from the union of aromatic rings and related species, which are likely responsible for strong infrared emission features. In this work, using a sample of 50 Seyfert galaxies (D$_{\rm L}<$100~Mpc) we compare the circumnuclear (inner kpc) PAH emission of AGN to that of a control sample of star-forming galaxies (22 luminous infrared galaxies and 30 \hii galaxies), and investigate the differences between central and extended PAH emission. Using Spitzer/InfraRed Spectrograph spectral data of Seyfert and star-forming galaxies
and newly developed PAH diagnostic model grids, derived from theoretical spectra, we compare the predicted and observed PAH ratios. We find that star-forming galaxies and AGN-dominated systems are located in different regions of the PAH diagnostic diagrams. This suggests that not only are the size and charge of the PAH molecules different, but also the nature and hardness of the radiation field that excite them. We find tentative evidence that PAH ratios in AGN-dominated systems are consistent with emission from larger PAH molecules (N$_{\rm c}>$300-400) as well as neutral species. By subtracting the spectrum of the central source from the total, we compare the PAH emission in the central vs extended region of a small sample of AGN.  In contrast to the findings for the central regions of AGN-dominated systems, the PAH ratios measured in the extended regions of both type 1 and type 2 Seyfert galaxies can be explained assuming similar PAH molecular size distribution and ionized fractions of molecules to those seen in central regions of star-forming galaxies (100$<$N$_{\rm c}$ $<$300). 
\end{abstract}




\section{Introduction}

Previous infrared (IR) observations have demonstrated that Polycyclic Aromatic Hydrocarbons (PAH) features are ubiquitous in a variety of astrophysical objects and environments. Indeed, the variation between PAH features indicate different physical conditions (see e.g. \citealt{Li20} for a review). Recently, \citet{Rigopoulou20} presented newly developed PAH diagnostic grids derived from theoretically computed PAH spectra (see also \citealt{Maragkoudakis20}). These grids enable disentangling the properties of the PAH molecules such as the size of the molecule, its ionization fraction and, the hardness of the radiation field.

PAH molecules absorb a significant fraction of ultraviolet (UV)$/$optical photons resulting in their excitation. The excited molecules produce IR features via vibrational relaxation (e.g. \citealt{Draine07}). PAH molecules are likely responsible for the family of strong infrared emission features (3.3, 6.2, 7.7, 8.6, 11.3 and 12.7~$\mu$m; e.g. \citealt{Tielens08}) observed in galactic and extragalactic sources. PAHs are abundant ($\sim$10$^{-7}$ relative to H; e.g. \citealt{Allamandola89}) and they account for up to 20\% of the total IR emission of metal rich galaxies \citep{Smith07a}. Interestingly, these molecules are considered the building blocks of the larger interstellar dust grains (e.g. \citealt{Tielens13}).

The 3.3, 8.6 and 11.3~$\mu$m bands are attributed to the stretching modes of C--H bonds, whereas 6.2 and 7.7~$\mu$m bands are due to C--C bond (e.g. \citealt{Allamandola89}). Previous studies used the 6.2 and 7.7~$\mu$m features as a diagnostic of PAH molecular size (e.g. \citealt{Jourdain90,Draine01,Sales10,Draine20,Rigopoulou20}). Recently, the 11.3/3.3 PAH ratio has also been proposed as a PAH molecule size indicator \citep{Maragkoudakis20}. Moreover, \citet{Rigopoulou20} found that the dependency of the 11.3/3.3 PAH ratio with the molecular size is not a linear one, and that the tight relation with the number of carbons for small molecules (N$_{\rm c}<$100) progressively flattens for larger molecules. These authors also found a tight correlation with the intensity of the radiation field
with the 11.3/3.3 PAH ratio decreasing with increasing energy. 

It is well established that ionization dramatically enhances the emission in the 6--9~$\mu$m region, suggesting that these IR bands result from highly vibrationally excited cations, while the 3.3, 8.6 and 11.3~$\mu$m features originate mostly from neutral PAH molecules (e.g. \citealt{Allamandola89,Draine01,Draine20}).  Therefore, the ratios between the 11.3~$\mu$m feature and those related to the charged PAHs (i.e. 6.2 or 7.7~$\mu$m features) have been proposed as a good indicator of the PAH ionization fraction (e.g. \citealt{Draine01, Galliano08,Draine20, Rigopoulou20}).

PAH features are often used to measure the star formation rate (SFR) of galaxies (see e.g. \citealt{Rigopoulou99,Peeters04,Wu05}) but also in AGN (see e.g. \citealt{Diamond12,Esquej14}). In addition, low ionization potential (IP) mid-infrared (MIR) emission lines such as [Ne\,II]$\lambda$12.81~$\mu$m have been used as an SFR indicator in AGN (e.g. \citealt{Spinoglio02,Ho07,Pereira10}). However, this emission line could be significantly contaminated by the AGN \citep{Bernete17}. These SFR indicators are mainly important in the vicinity of AGN-dominated systems where classical SF indicators normally fail. In a study by \citet{Jensen17} the authors suggest that PAH molecules can also be excited by AGN photons. If this is indeed the case, then the use of PAH emission as an SFR tracer needs to be re-evaluated.  

In the present work, we employ the newly computed PAH grids presented in \citet{Rigopoulou20} to investigate the properties of the PAH emitting molecules in a sample of AGN (our main sample) and a {\it control sample} of star-forming galaxies with no AGN detections. Our AGN sample includes Seyfert galaxies with a wide range of properties (see Section \ref{sample_selection}) enabling us to obtain statistically significant results. The control sample of star-forming galaxies has a similar redshift distribution. Our aim is to study the effect of the nuclear activity on the properties of the PAH molecules located in the central/extended regions of AGN and also in star-forming galaxies.

The paper is organized as follows. Section \ref{sample_selection} and \ref{observations} describe the sample selection and the observations used in this study, respectively. The MIR spectral modelling and spectral decomposition are presented in Section \ref{PAHFIT} and \ref{fagn}. The main results of the PAH emission are presented in Section \ref{results}. In Section \ref{pahdiagram}, we compare the PAH band ratios of AGN and star-forming galaxies with the theoretical models. In Section \ref{pahband}, we compare the average PAH spectra of the different source groups. Finally, in Section \ref{conclusions}, we summarize the main conclusions of this work. Throughout this paper, we assumed a cosmology with H$_0$=70 km~s$^{-1}$~Mpc$^{-1}$, $\Omega_m$=0.3, and $\Omega_{\Lambda}$=0.7, and a velocity-field corrected using the \citet{Mould00} model, which includes the influence of the Virgo cluster, the Great Attractor, and the Shapley supercluster.

\begin{table*}
\centering
\begin{tabular}{lccccccc}
\hline
Name &	R.A.&	Dec.&D$_{L}$ &Spatial&Inclination& log N$_{\rm H}^{\rm X-ray}$&Spectral\\
 	&	(J2000)&	(J2000)&(Mpc) &scale&(deg) & (cm$^{-2}$)& type\\
 	&				& &(pc arcsec$^{-1}$)& && &\\
 (1)&(2)&(3)&(4)&(5)&(6)&(7)&(8)\\	
\hline
Mrk\,352        &   00h59m53.3s&    31d49h37s&62.5&303   &34&20.0       &1.0*                \\
NGC\,931        &   02h28m14.5s&    31d18m42s&69.5&337      &81&21.09   &1.5*                   \\
NGC\,1097       &   02h46m19.0s&    -30d16m28s&17.2&83    &55&20.36$^a$ &LINERb/1.0*    \\	
NGC\,1275       &   03h19m48.2s&    +41d30m42s&73.9&358    &58&21.68    &1.5 /2.0*/LINER$^h$      \\
NGC\,1365		&	03h33m36.4s&	-36d08m24s&22.4&108 &46&22.21       &1.8		    \\
ESO\,548-G081   &   03h42m03.7s&    -21d14m40s&60.7&294 &64&20.0        &1.0*           \\
NGC\,1566       &   04h20m00.7s&    -54d56m17s&20.2&98     &38&20.0     &1.5	       \\
NGC\,1667       &   04h48m37.2s&    -06d19m12s&63.8&310 &38&24.43$^b$   &2.0            \\	
MCG-01-13-025   &   04h51m41.5s&    -03d48m34s&66.9&324 &50&20.0        &1.2*           \\
4U0517$+$17     &   05h10m45.5s& +16d29m55.7s&75.5&366  &45&21.08       &1.5*           \\
ESO\,362-G018   &   05h19m35.8s&    -32d39m28s&53.4&259 &71&20.0        &1.5*           \\
UGC\,3478&  06h32m47.2s& +63d40m25S&57.0&276            &77&21.18       &1.2                \\ 
Mrk\,6          &   06h52m12.2s&    74d25m38s&82.9&402  &63&20.76       &1.5*           \\
Mrk\,79         &   07h42m32.8s&    49d48m35s&97.1&471  &37&20.0        &1.2*           \\
NGC\,2992		&	09h45m42.0s&	-14d19m35s&35.6&173 &62&21.72       &1i/2.0*	     \\ 
NGC\,3227		&	10h23m30.6s&	+19d51m56s&21.2&103 &57&20.95       &1.5/comp$^{i,j}$		        \\
NGC\,3516       &   11h06m47.4s&    +72d34m07s&43.1&209  &44&20.0       &1.5                \\		
NGC\,3783		&	11h39m01.8s&	-37d44m19s&37.7&183  &37&20.49      &1.5/1.0*              \\
NGC\,3786 & 11h39m42.5s& +31d54m33s&45.5&221            &65&22.36         &1.8                     \\ 
UGC\,6728       &	11h45m16.0s&	+79d40m53s&33.2&161  &56&20.00        &1.2*                  \\
NGC\,3982         &   11h56m28.1s& +55d07m30s&22.7&110      &29&23.34$^b$ &2.0              \\
NGC\,4051		&	12h03m09.6s&	+44d31m53s&13.5&65  	&40&20.0      &1n/1.5*        \\
NGC\,4138		&	12h09m29.9s&	+43d41m06s&18.4&89  	&61&22.89     &1.9	      \\
NGC\,4151		&	12h10m32.5s&	+39d24m21s&20.8&101 	&26&22.71     &1.5	      \\
NGC\,4235       &   12h17m09.9s&    +07d11m29s&40.8&198   &90&21.26       &1.2              \\	
NGC\,4253  (Mrk\,766)&12h18m26.5s&  29d48m46s&63.2&306    &44&20.32       &1.5*             \\
NGC\,4388$\dagger$&	12h25m46.7s&	+12d39m41s&17.0&82  	&78&23.52     &1h/2.0*        \\
NGC\,4395		&	12h25m48.9s&	+33d32m48s&4.0&19   	&90&21.04      &1.8/1.9*         \\
NGC\,4501		&	12h31m59.3s&    +14d25m13s&14.6&71   &63&$<$21.3$^c$   &2.0                \\	
NGC\,4593       &   12h39m39.4s&    -05d20m39s&43.1&209  &40&20.0          &1.0                 \\	
NGC\,4639       &   12h42m52.5s&    +13d15m25s&14.5&70   &51&$<$20.00$^c$  &1.8               \\
ESO\,323-G77    &   13h06m26.2s&    -40d24m52s&62.1&301 &60&22.81          &1.2             \\
NGC\,5033       &   13h13m27.5s&    +36d35m38s&18.0&87   &65&20.00         &1.8             \\
NGC\,5135       &   13h25m44.0s&    -29d50m02s&60.2&292  &24&$>$24.00$^d$  &2.0/comp$^k$             \\	
MCG-06-30-015 	&	13h35m53.8s&	-34d17m44s&27.8&135 &59&20.85          &1.2*              \\
IC\,4329A       &   13h49m19.3s&    -30d18m34s&82.0&397 &66&21.52          &1.2             \\	
NGC\,5347       &   13h53m17.8s&    +33d29m27s&41.7&202  &45&$>$24.00$^e$  &2.0             \\	
NGC\,5506		&	14h13m14.8s&	-03d12m26s&31.2&151 	&90&22.44      &1i/1.9*          \\
NGC\,5548       &   14h17m59.6s&    +25d08m13s&82.5&400  &33&20.69         &1.5         \\	
NGC\,5953       &   15h34m32.3s&    +15d11m38s&35.7&173  &44&...           &2.0/LINER$^l$         \\
IRAS\,18325-5926 (Fairall49)&18h36m58.3s &59d24m09s&92.1 &50&22.03         &446&1.8     \\
NGC\,6814		&	19h42m40.7s&	-10d19m23s&26.7&130 &20&20.97          &1.5		 \\
NGC\,6860  &   20h08m47.1s&    -61d06m00s&68.0&330       &58&21.08         &1.5/1.0*    \\
NGC\,6890  &   20h18m18.1s&     -44d48m24s&37.2&180     &38&21.00$^b$      &1.9         \\
NGC\,7130       &   21h48m19.5s&   -34d57m05s&72.6& 340    & 34 &24.22     & 2.0/comp$^k$               \\
NGC\,7314		&	22h35m46.1s& -26d03m02s&21.7&105    &65&21.6           &1h/1.9*	   \\
NGC\,7469       &   23h03m15.6s&   +08d52m26s&69.9&339       &53&20.53     &1.5/1.2*/comp$^{n,m}$       \\
NGC\,7496       &   23h09m47.3s&   -43d25m40s&24.1&117   &53&22.70$^f$     &2.0/comp$^l$             \\
NGC\,7582		&	23h18m23.5s&-42d22m14s&22.9&111     	&62&24.15      &1i/2.0*/comp$^l$      \\
NGC\,7590       &   23h18m54.8s&   -42d14m21s&22.9&111   &69&$<$20.96$^g$  &2.0/comp$^l$            \\
\hline
\end{tabular}						 
\caption{Seyfert sample. Right ascension (R.A.), declination (Dec.). The luminosity distance and spatial scale were calculated using a cosmology with H$_0$=70 km~s$^{-1}$~Mpc$^{-1}$, $\Omega_m$=0.3, and $\Omega_{\Lambda}$=0.7, and a velocity-field corrected using the \citet{Mould00} model, which includes the influence of the Virgo cluster, the Great Attractor, and the Shapley supercluster. $\dagger$ This galaxy is part of the Virgo Cluster \citep{Binggeli85}. Galaxy inclination were taken from \citet{Whittle92} when available, otherwise from HyperLeda Database. N$_{\rm H}^{\rm X-rays}$ corresponds to the foreground extinction due to the host galaxy. The N$_{\rm H}^{\rm X-ray}$ were taken from \citet{Ricci17} when possible otherwise are indicated by superscripts. N$_{\rm H}^{\rm X-ray}$ references: a) \citet{Nemmen06}; b) \citet{Brightman11}; c) \citet{Cappi06}; d) \citet{Guainazzi05}; e) \citet{Risaliti99}; f) \citet{Shu07}; g) \citet{Bassani99}. Spectral type were taken from \citet{Veron06} when possible, otherwise from \citet{Tueller2008} which are indicated by * symbol. The only exceptions are NGC\,3786, NGC\,3982, NGC\,4639, Fairall\,49 and UGC\,3478 classifications which were taken from \citet{Maiolino95}. Note that LINERb classification means LINER with broad Balmer lines, and the i, h and n subscripts correspond to Sy1 with detected broad Paschen lines (in the infrared spectra), narrow-lines and broad polarized Balmer lines, respectively. AGN/SF composites, based on their UV emission and/or emission line ratios, are denoted by "comp". References: h) \citet{Sosa-Brito01}; i) \citet{Gonzalez-Delgado97}; j) \citet{Ardila03}; k) \citet{Gonzalez-Delgado01}; l) \citet{Yuan10}; m) \citet{Wilson91}; n) \citet{Diaz-Santos07}.} 

\label{tab1}
\end{table*}


\section{Sample selection}
\label{sample_selection}

\subsection{Seyfert sample}

\begin{figure}
\centering
\par{
\includegraphics[width=9.22cm]{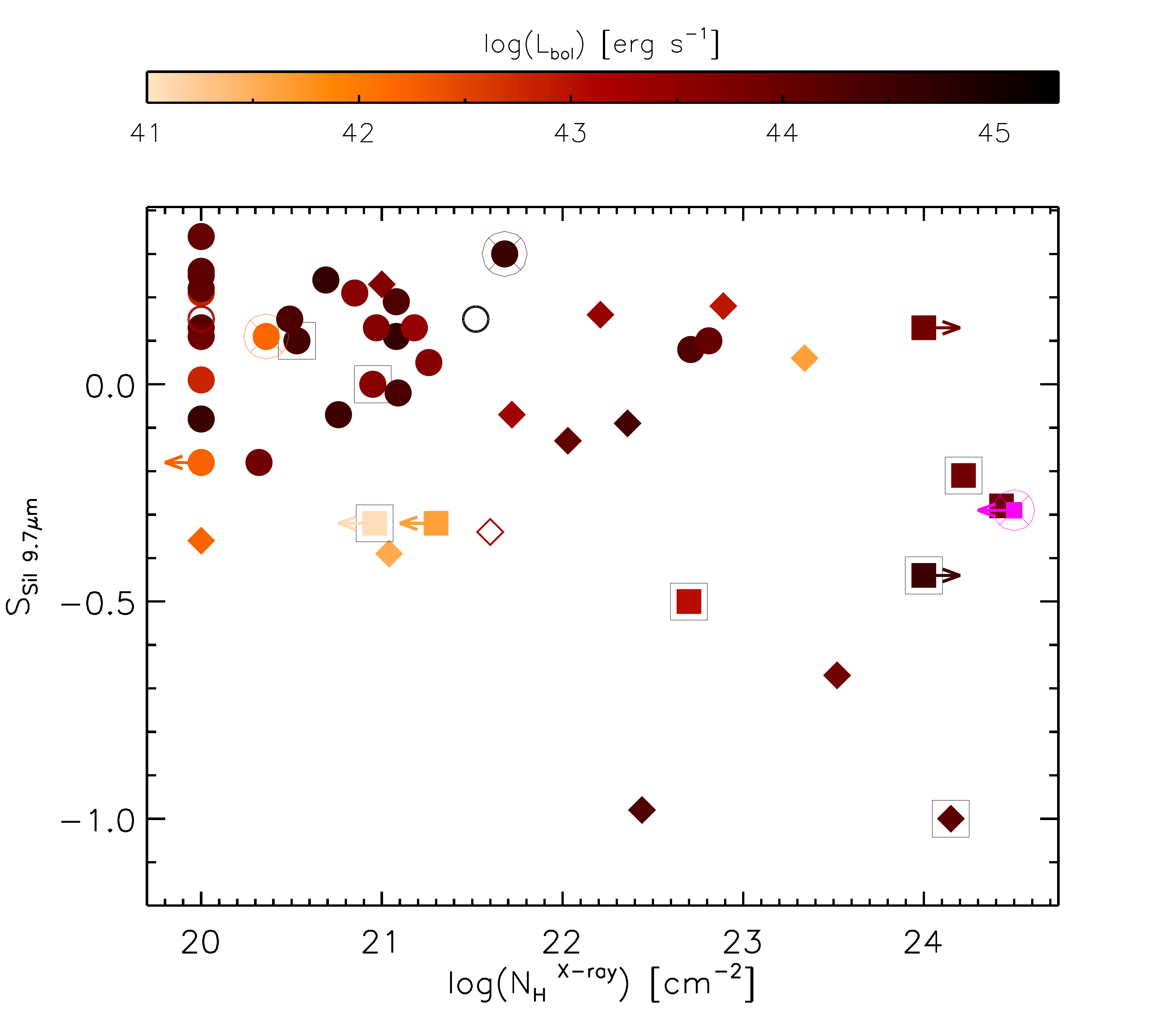}
\par}
\caption{The strength of the 9.7~$\mu$m silicate feature (see Section \ref{results}) vs. the X-ray hydrogen column density (see Table \ref{tab1} for references). The silicate strength is computed as S$_{\rm Sil}=$ln(f$_{9.7}/$f$_{\rm cont}$). Color-coded symbols represent the bolometric luminosity. The magenta data point correspond to the only galaxy with no X-ray flux detected (i.e. NGC\,5953). Filled circles, diamond and squares correspond with Sy1, Sy1.8/1.9 and Sy2, respectively. Note that open black squares and circles with X represent galaxies that have been also classified as AGN/SF composite and LINER, respectively. Finally, open circles and diamond correspond with Sy1 and Sy1.8/1.9 without any PAH bands detected within the 11.3~$\mu$m PAH complex (11.23 and 11.33~$\mu$m features) in the Spitzer/IRS spectra, respectively. Note that all Sy2 galaxies show the 11.3~$\mu$m PAH feature in the Spitzer/IRS spectra. }
\label{silicates}
\end{figure}
The main aim of this work is to study the properties of the PAH molecules located in the central/extended regions of AGN and compare them with those in star-forming galaxies. Therefore, we first focus on building a representative sample of Sy galaxies sampling the different Seyfert properties. For the Sy galaxy sample we used the ultra-hard 14--195~keV band of the nine-month catalogue, which is observed with Swift/BAT \citep{Tueller2008}. The catalogue is far less sensitive to the effects of obscuration than optical or softer X-ray wavelengths, making this AGN selection one of the least biased for N$_{\rm H}$ $<$10$^{24}$~cm$^{-2}$ to date (see e.g. \citealt{Winter2009,Winter2010,Weaver2010,Ichikawa2012,Ricci15,Ueda15}). We note that this catalogue covers 80 per cent of the sky with a flux threshold of 3.5$\times$10$^{-11}$~erg~cm$^{-2}$~s$^{-1}$ in the 14--195~keV band and covers one-third of the sky near the ecliptic poles at 2.5$\times$10$^{-11}$~erg~cm$^{-2}$~s$^{-1}$. The Swift/BAT sources were selected based on detection at 4.8$\sigma$ or higher. Therefore, the catalog is flux-limited in the 14--195~keV X-ray band and consists of 153 sources. We note that we have excluded from this work broad-absorption line (BAL) quasars because their inner region geometry is thought to be significantly different from standard AGN.

Since in this work we are interested in the study of the effect of the AGN in the central and extended emission of Seyfert galaxies, we mainly focused on type-1 Seyfert galaxies (i.e. Sy1, Sy1.2 and Sy1.5; hereafter Sy1). To this end, we first selected all the Sy1 from the parent Swift/BAT sample (69 sources). From these we culled sources with luminosity distances D$_{\rm L}<$100~Mpc (29 sources). Finally,  we cross-correlated these sources with the Spitzer Heritage Archive (SHA)\footnote{https://sha.ipac.caltech.edu/applications/Spitzer/SHA/} to select those with available Spitzer/Infrared Spectrograph (IRS) MIR spectra covering the  5--35~$\mu$m range (i.e. observed in the short- and long-low modules) with the IRS instrument (\citealt{Houck04}). The final Swift/BAT Sy1 sample comprises 23 objects (excluding NGC\,7213 which exhibits very strong silicate emission). 

However, the above mentioned flux threshold of the Swift/BAT catalog excludes relatively low-luminosity Seyfert galaxies, which are present in the local Universe. Therefore, to counteract this effect, we augmented our original sample with Seyfert galaxies from the RSA sample \citep{Maiolino95}. This results in five more type-1 Seyfert galaxies with D$_{\rm L}<$100~Mpc (namely, ESO\,323-G77, NGC\,1097, NGC\,1566, NGC\,4235 \& UGC\,3478). Therefore, our final sample contains 28 Sy1 galaxies. 

Although our focus is on Sy1 galaxies we are also interested in sampling a wide range of Seyfert properties. Thus,
we added  11 Sy1.8/1.9 and 11 Sy2 galaxies from the RSA sample (13 objects) and the Swift/BAT catalogue (9 objects). These additional Seyferts have similar luminosities, inclinations and satisfy the same distance limit as the original Sy1 sample.

Taking into account the spatial scale probed by Spitzer/IRS ($\sim$4--6\arcsec) and the distance limit selected, this allows us to study the inner central~kpc of these galaxies. The final sample used in this paper comprises 50 local Seyfert galaxies. The majority of the observations (36/50) employed in this work were observed in the low-resolution (R$\sim$60-120) IRS spectral mapping mode, and the rest of the data were observed in the staring mode (see Section \ref{observations}.) 

In Figure \ref{silicates} we present the range of X-ray hydrogen column densities, bolometric luminosities and strengths of the 9.7~$\mu$m silicate feature for our sample. Note that column densities of N$_{\rm H}^{\rm X-ray}=$10$^{20}$~cm$^{-2}$ correspond to the lowest values reported by \citet{Ricci17}. This shows that our selected AGN sample probes well the observed ranges for the various types of Seyfert galaxies. The sample sorted by right ascension is shown in Table 1.

\begin{figure}
\centering
\par{
\includegraphics[width=8.5 cm]{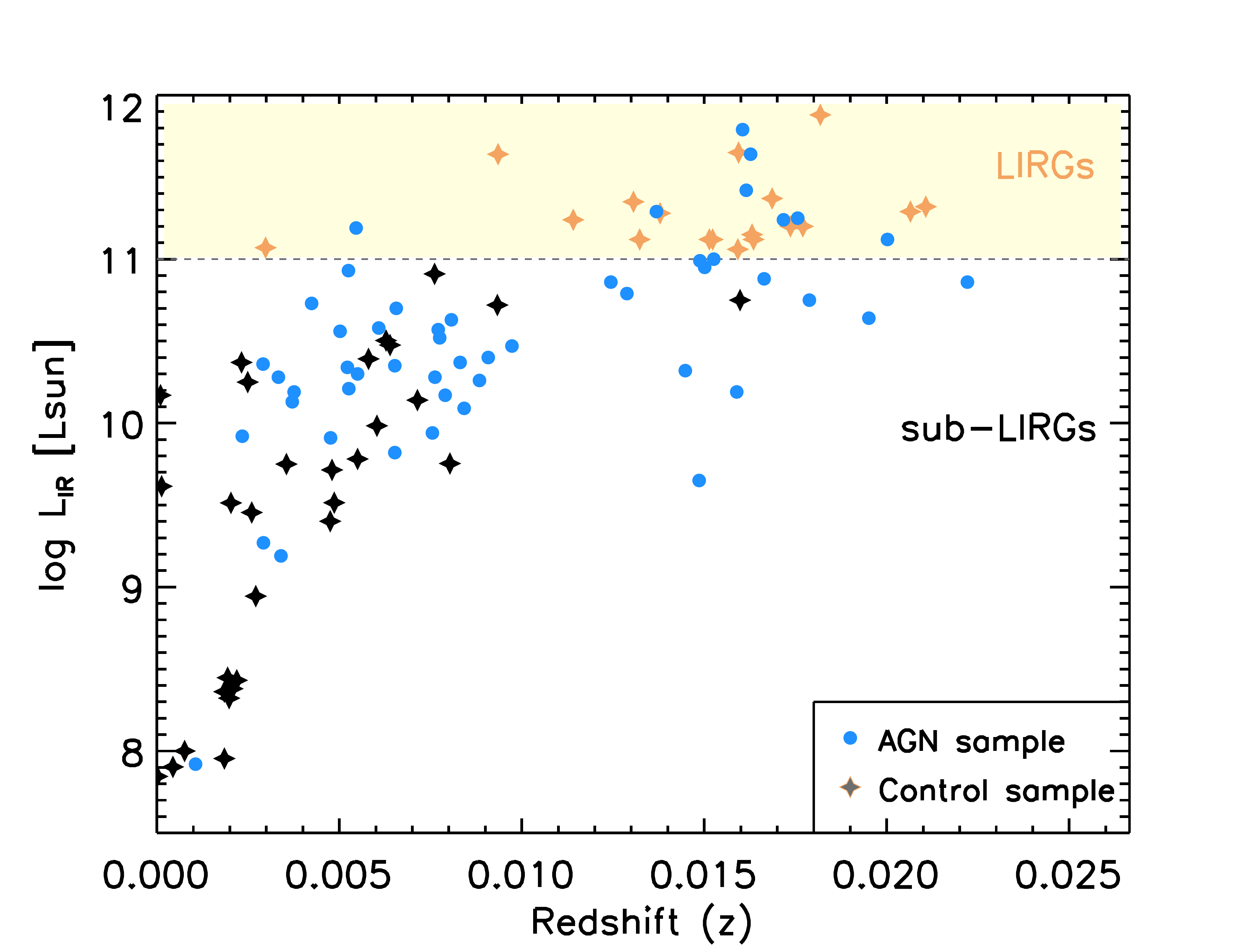}
\par}
\caption{Total IR luminosity vs. redshift for all the sources used in this work. Filled stars and blue circles correspond to the star-forming control sample and AGN sample. Black stars and orange stars indicate \hii galaxies (sub-LIRGs) and LIRGs. The shaded yellow region corresponds to the LIRG luminosity region.}
\label{agn_control_sample}
\end{figure}

In summary, our sample covers an AGN luminosity range of log(L${_{\textrm{bol}}^{14-195~\textrm{keV}}}$)$\sim$41.0--45.5~erg~s$^{-1}$ and luminosity distances D$_{\rm L}<$100~Mpc. We note that there are no statistically significant differences between the distances for the various Sy groups according to the Kolmogorov-Smirnov (KS) test. As expected, we find that the 92\% of Sy1 galaxies in our sample have lower values of X-ray hydrogen column densities (N$_{\rm H}^{\rm X-ray}<$10$^{22}$~cm$^{-2}$), in comparison with the 60\% and 20\% for Sy1.8/1.9 and Sy2 galaxies, respectively. We find that this difference is statistically significant using the Fisher's exact test.

\subsection{Star-forming control sample}
\label{control_sample}
For the present study we also selected a control sample of star-forming galaxies that span similar redshifts and total IR luminosities as the Seyfert sample (see Fig. \ref{agn_control_sample}). We selected luminous infrared galaxies (LIRGs) and \hii galaxies from the Great Observatories All-sky LIRG Survey (GOALS; \citealt{Armus09}), \citet{Brandl06} and the Spitzer Infrared Nearby Galaxies Survey (SINGS; \citealt{Kennicutt03}).

The GOALS sample consist of a complete subset of the Infrared Astronomical Satellite (IRAS) Revised Bright Galaxy Sample (RBGS; \citealt{Sanders03}). The LIRGs within GOALS cover different interaction stages and nuclear spectral types. For our study we selected 21 LIRGs from the GOALS sample that have Spitzer/IRS observations and are not interacting or classified as AGN and/or LINER. To select \hii galaxies, we used the SINGS sub-sample (23 sources) presented in \citet{Smith07a}, which was observed with Spitzer/IRS. 

We also use an additional sample of 22 star-forming galaxies observed with Spitzer/IRS  \citep{Brandl06}. By removing those sources classified as AGN/SF composite in \citet{Brandl06}, we add one LIRG and 7 \hii galaxies (sub-LIRGs; log (L$_{\rm IR}$)$<$11~L$_{\sun}$) to the control sample. Therefore, the final star-forming control sample consists of 22 LIRGs and 30 \hii galaxies. In the following, we consider the star-forming control sample as the combination of these two samples (i.e. 52 galaxies).

\begin{figure}
\centering
\par{
\includegraphics[width=6.82cm, angle=90]{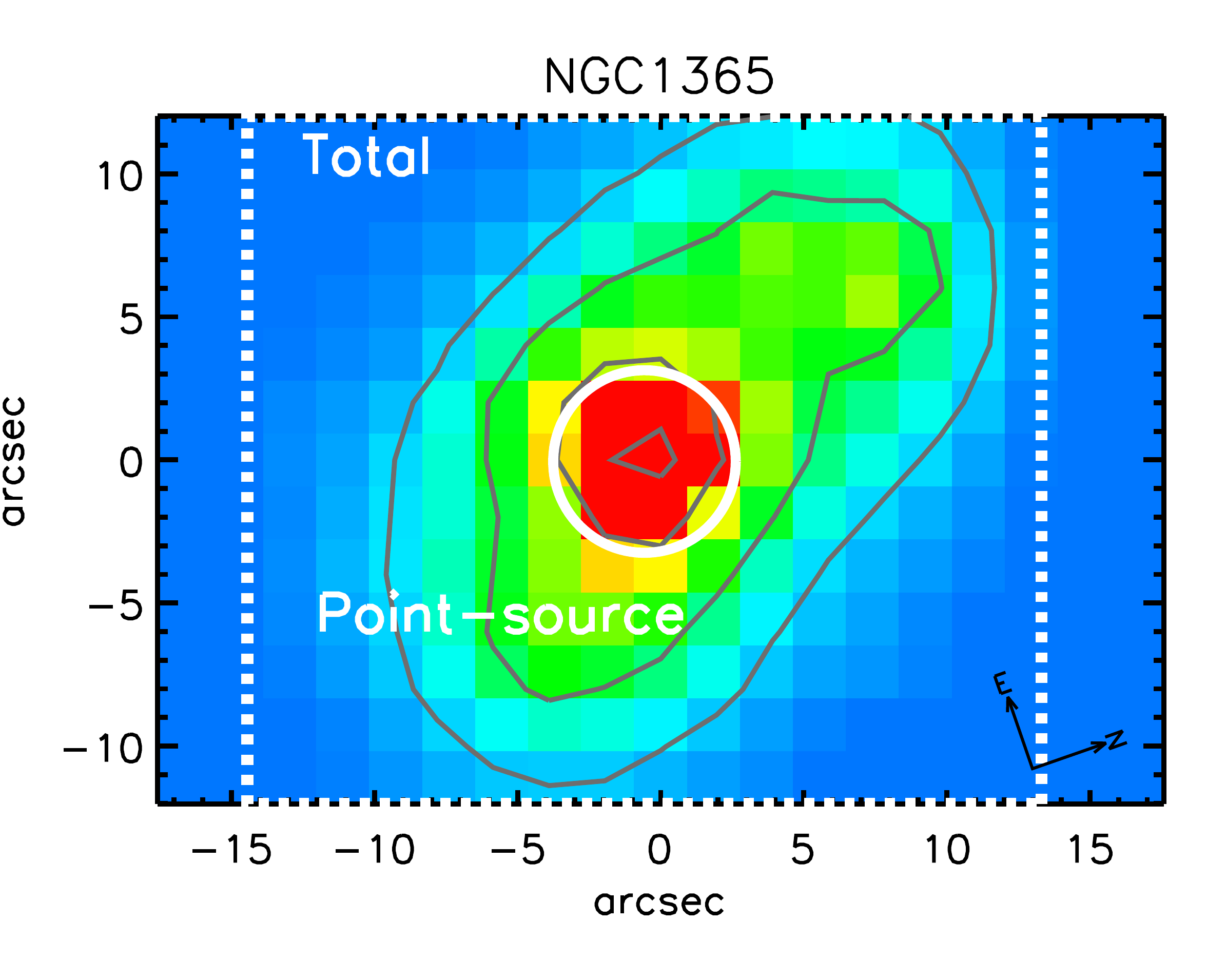}
\par}
\caption{Schema of the different apertures used to calculate the central and extended emission of our sample. The apertures are overplotted on the 10~$\mu$m Spitzer/IRS map (using 1~$\mu$m window) of NGC\,1365. The white dashed region and solid circle correspond to the apertures used to calculate the fluxes of the total galaxy emission and the central point-like component that is subtracted from the former to remove the AGN contribution, respectively. The extended emission spectrum is calculated by subtracting the central emission from the total emission.}
\label{apertures}
\end{figure}

\section{Observations}
\label{observations}
In what follows we describe the IR observations used in this work.

\begin{table*}
\centering
\begin{tabular}{lccccr}
\hline
Name &	PAH$\lambda$6.2$\mu$m & 	PAH$\lambda$7.7$\mu$m (complex) & 	PAH$\lambda$11.3$\mu$m (complex) & AGN MIR & S$_{\rm Sil}$\\
&&&&contribution (\%)&\\

 (1)&(2)&(3)&(4)&(5)& (6)\\	
\hline
{\textit{Mrk\,352}}$\dagger$          & \ldots & \ldots &  \ldots & 73$\pm^{18}_{25}$ & 0.11$\pm$0.40 \\
NGC\,931          & 38$\pm2$ & 185$\pm26$ &  63$\pm3$ & 89$\pm^{5}_{5}$ & -0.02$\pm$0.16\\
NGC\,1097         & 86$\pm1$ & 401$\pm8$ &  211$\pm1$ & 32$\pm^{7}_{8}$ & 0.11$\pm$0.40\\
NGC\,1275         & 64$\pm2$ & 268$\pm17$ &  167$\pm3$ & 97$\pm^{2}_{2}$ & 0.30$\pm$0.12\\
NGC\,1365		  & 561$\pm3$ & 2111$\pm37$ &  648$\pm5$ & 61$\pm^{7}_{10}$ & 0.16$\pm$0.23\\
{\textit{ESO\,548-G081}}$\dagger$     & \ldots & \ldots &  \ldots & 90$\pm^{10}_{14}$ & 0.25$\pm$0.18\\
NGC\,1566         & 87$\pm1$ & 333$\pm10$ &  138$\pm2$ & 49$\pm^{10}_{13}$ & 0.21$\pm$0.25\\
NGC\,1667         & 29$\pm1$ & $<$91 &  59$\pm1$ & 19$\pm^{12}_{7}$ & -0.28$\pm$0.54\\
{\textit{MCG-01-13-025}}$\dagger$     & \ldots & \ldots &  \ldots & 93$\pm^{7}_{10}$ & 0.34$\pm$0.19\\
4U0517$+$17       & 56$\pm6$ & $<$339 &  $<$69 & 79$\pm^{6}_{5}$ & 0.11$\pm$0.26\\
ESO\,362-G018     & 47$\pm2$ & 194$\pm17$ &  $81\pm2$ & 79$\pm^{5}_{13}$ & 0.22$\pm$0.19\\
UGC\,3478$\dagger$         & 39$\pm3$ & $<$157 &  61$\pm7$ &71$\pm^{10}_{13}$ &0.13$\pm$0.27\\
Mrk\,6            & 23$\pm2$ & $123\pm26$ &  53$\pm2$ & 88$\pm^{9}_{10}$ & 0.15$\pm$0.17\\
Mrk\,79           & 24$\pm2$ & $<$155 &  45$\pm3$ & 91$\pm^{6}_{8}$ & 0.14$\pm$0.17\\
NGC\,2992		  & 144$\pm2$ & 660$\pm23$ &  197$\pm3$ & 71$\pm^{9}_{7}$ & 0.07$\pm$0.23\\
NGC\,3227	      & 332$\pm13$ & 1695$\pm124$ &  423$\pm42$ & 53$\pm^{10}_{11}$ & 0.00$\pm$0.28\\
NGC\,3516         & 20$\pm2$ & $<$98 &  $45\pm3$ & 95$\pm^{5}_{9}$ & 0.13$\pm$0.18\\
{\textit{NGC\,3783}}$\dagger$		  & \ldots & \ldots &  \ldots & 73$\pm^{11}_{4}$ & 0.15$\pm$0.24\\
NGC\,3786$\dagger$         & 72$\pm2$ & $<$232 &  109$\pm4$ & 42$\pm^{12}_{10}$ & -0.09$\pm$0.30\\
{\textit{UGC\,6728}}$\dagger$         & \ldots & \ldots &  \ldots & 97$\pm^{3}_{13}$ &0.15$\pm$0.24\\
NGC\,3982         & 28$\pm1$ & 148$\pm11$ &  50$\pm1$ & 40$\pm^{10}_{10}$ & 0.06$\pm$0.26\\
NGC\,4051         & 97$\pm3$ & 552$\pm30$ &  188$\pm4$ & 80$\pm^{6}_{7}$ & 0.22$\pm$0.15\\
NGC\,4138		  & 18$\pm1$ & $<$58 &  47$\pm1$ & 37$\pm^{10}_{10}$ & 0.18$\pm$0.26\\
{\textit{NGC\,4151}}		  & \ldots & \ldots &  \ldots & 97$\pm^{3}_{6}$ & 0.08$\pm$0.17\\
{\textit{NGC\,4235}}$\dagger$         & \ldots & \ldots &  \ldots & 77$\pm^{17}_{23}$ & 0.05$\pm$0.43 \\
NGC\,4253         & 86$\pm2$ & 495$\pm27$ &  115$\pm3$ & 87$\pm^{5}_{9}$ & 0.08$\pm$0.20\\
NGC\,4388         & 105$\pm6$ & $<$975 &  191$\pm6$ & 65$\pm^{7}_{6}$ & -0.67$\pm$0.16\\
NGC\,4395$\dagger$		  & 2$\pm1$ & $<$37 &  4$\pm1$ & 61$\pm^{15}_{11}$ & -0.39$\pm$0.26\\
NGC\,4501		  & 10$\pm1$ & 37$\pm2$ &  40$\pm1$ & 8$\pm^{6}_{6}$ &-0.32$\pm$0.68\\
NGC\,4593         & 31$\pm2$ & 180$\pm28$ &  77$\pm3$ & 91$\pm^{4}_{9}$ & 0.26$\pm$0.14\\
NGC\,4639$\dagger$         & 7$\pm1$ & $<$24 &  17$\pm1$ & 35$\pm^{18}_{16}$ & -0.18$\pm$0.65\\
ESO\,323-G77$\dagger$      & 170$\pm16$ & $<$734 &  247$\pm28$ & 69$\pm^{11}_{11}$ & 0.10$\pm$0.24\\
NGC\,5033         & 58$\pm1$ & 248$\pm10$ &  156$\pm1$ & 14$\pm^{9}_{6}$ & -0.36$\pm$0.75\\
NGC\,5135         & 519$\pm1$ & 2187$\pm15$ &  591$\pm2$ & 14$\pm^{10}_{11}$ & -0.44$\pm$0.76\\
MCG-06-30-015 	  & 32$\pm2$ & $<$192 &  70$\pm3$ & 94$\pm^{5}_{7}$ & 0.21$\pm$0.15\\
{\textit{IC\,4329A}}         & \ldots & \ldots &  \ldots & 92$\pm^{7}_{9}$ &0.15$\pm$0.18\\
NGC\,5347         & 35$\pm2$ & $<$234 &  76$\pm3$ & 83$\pm^{8}_{7}$ & 0.13$\pm$0.21 \\
NGC\,5506		  & 242$\pm3$ & $<$2897 &  396$\pm5$ &60$\pm^{9}_{8}$ &-0.98$\pm$0.29\\
NGC\,5548         & 27$\pm2$ & 121$\pm20$ &  $54\pm3$ & 88$\pm^{4}_{3}$ & 0.24$\pm$0.15\\
NGC\,5953         & 253$\pm1$ & 905$\pm19$ &  326$\pm3$ & 9$\pm^{8}_{5}$ & -0.29$\pm$0.69\\
IRAS\,18325-5926$\dagger$  & 113$\pm8$ & 698$\pm83$ &  180$\pm15$ &67$\pm^{8}_{8}$ & -0.13$\pm$0.21 \\
{\textit{NGC\,6814}}$\dagger$		  & \ldots & \ldots &  \ldots & 83$\pm^{10}_{13}$ & 0.13$\pm$0.24\\
{\textit{NGC\,6860}}         & \ldots & \ldots &  \ldots &82$\pm^{10}_{12}$ & 0.19$\pm$0.16 \\
NGC\,6890         & 36$\pm2$ & $<$175 &  62$\pm2$ &82$\pm^{5}_{6}$ & 0.23$\pm$0.16\\
NGC\,7130         & 236$\pm1$ & 1096$\pm9$ & 296$\pm13$ & 45$\pm^{9}_{9}$ & -0.21$\pm$0.42\\
NGC\,7314$\dagger$		  & 21$\pm5$ & $<$129 &  $<$64 &84$\pm^{5}_{8}$ &-0.34$\pm$0.29\\
NGC\,7469         & 834$\pm3$ & 3181$\pm47$ &  844$\pm6$ &45$\pm^{9}_{7}$ & 0.10$\pm$0.28\\
NGC\,7496         & 189$\pm3$ & 646$\pm32$ &  196$\pm5$ & 39$\pm^{4}_{4}$ & -0.41$\pm$0.28\\                   
NGC\,7582		  & 1382$\pm2$ & 6641$\pm51$ &  1258$\pm5$ & 25$\pm^{9}_{9}$ & -1.00$\pm$0.41\\
NGC\,7590         & 32$\pm1$ & 125$\pm11$ &  49$\pm1$ & 14$\pm^{11}_{11}$ & -0.32$\pm$0.73\\
\hline
\end{tabular}						 
\caption{Seyfert galaxy PAH measurements with PAHFIT using the Spitzer/IRS spectra. $\dagger$ These galaxies were observed using the staring mode, whereas the remaining galaxies have spectral mapping data. Columns 2--4 correspond to the fluxes and 1$\sigma$ errors, which are in units of 10$^{-14}$~erg$^{-1}$~cm$^{-2}$, of the 6.2, 7.7, and 11.3$\mu$m PAH bands. Column 5 indicates the fractional contribution of the AGN component to the MIR spectrum derived from the spectral decomposition of the Spitzer/IRS spectra (see Section \ref{fagn}). Column 6 corresponds to the silicate strength. We note that $<$ corresponds to 3$\sigma$ upper limits. Sources in italics have unsatisfactory fits of their spectral features and have been ignored from further analysis.}
\label{tab3}
\end{table*}

\begin{table*}
\centering
\begin{tabular}{lcccccc}
\hline
Name               &	[Ar\,II]$\lambda$6.99 &	[S\,IV]$\lambda$10.51          &	[Ne\,II]$\lambda$12.81 &	[Ne\,V]$\lambda$14.32&	[Ne\,III]$\lambda$15.56 &	[O\,IV]$\lambda$25.89 \\
\hline
E$_{\rm ion}$ (eV) &15.8 & 34.8 & 21.6 & 97.1 & 41.0& 54.9\\
\hline
{\textit{Mrk\,352}}$\dagger$       &   \ldots                   & \ldots                       &\ldots                       & \ldots                       & \ldots                        & \ldots  \\
NGC\,931                & $<$19.3                     &   156.3$\pm5.4$               &   38.6$\pm3.5$       & 78.1$\pm12.1$                 &   111.2$\pm6.2$      &   287.7$\pm7.3$\\    
NGC\,1097               & $<$0.9                      &   36.5$\pm1.7$       &   15.9$\pm1.5$                & $<$44.4                       &   17.4$\pm3.3$                &   31.6$\pm6.4$\\
NGC\,1275         	    &   113.6$\pm4.2$             &    114.7$\pm4.7$              &    348.1$\pm3.2$              & $<$8.7                        &    149.9$\pm6.6$              &    29.1$\pm7.6$   \\
NGC\,1365		        &   286.0$\pm6.7$             &     208.6$\pm7.3$    &      820.6$\pm5.4$   & $<$209.7                      &     157.9$\pm13.3$  &    672.1$\pm17.4$  \\
{\textit{ESO\,548-G081}}$\dagger$  &   \ldots                   & \ldots                      &  \ldots        & \ldots                       & \ldots                  &    \ldots\\
NGC\,1566               &   33.2$\pm1.9$              &   30.1$\pm1.8$       &   84.2$\pm1.4$                & $<$29.2                       & $<$59.0                       & $<$79.8  \\
NGC\,1667               &   39.9$\pm2.2$              &     18.7$\pm2.2$     &      75.5$\pm1.8$    & $<$45.5                       &     3.3$\pm4.2$      & 23.1$\pm7.1$\\
{\textit{MCG-01-13-025}}$\dagger$  &   \ldots              &   \ldots       &   \ldots                & \ldots                        &  \ldots                 & \ldots \\
4U0517$+$17             &   52.4$\pm16.6$             &   135.6$\pm32.6$    &   126.8$\pm32.2$    & 75.9$\pm10.2$                 &   141.8$\pm5.8$      &   269.9$\pm13.0$\\
ESO\,362-G018           &   11.7$\pm3.1$              &   38.6$\pm3.2$       &   64.9$\pm2.2$       & $<$19.8                       &   39.1$\pm5.2$       &   73.2$\pm6.0$\\
UGC\,3478$\dagger$      &  $<$65.7                    &    92.1$\pm9.9$      &  77.3$\pm8.5$        & 87.8$\pm6.6$                  &   127.5$\pm6.4$      &   402.9$\pm2.8$\\
Mrk\,6                  &   73.9$\pm4.1$              &   161.4$\pm4.6$      &   179.0$\pm3.3$      & $<$43.5                       &   262.2$\pm4.5$      &   253.1$\pm5.1$\\
Mrk\,79                 &   45.6$\pm4.4$              & 116.6$\pm5.0$        &   60.2$\pm3.3$       & 60.3$\pm13.7$       &   135.2$\pm5.1$               &   304.6$\pm6.3$\\
NGC\,2992		        &   161.9$\pm5.0$             &   255.3$\pm5.6$      &   376.2$\pm4.0$      & 107.9$\pm15.2$      &   377.8$\pm7.4$      &   660.6$\pm8.7$\\ 
NGC\,3227	            &   252.2$\pm37.7$            &   245.0$\pm43.1$    &   613.1$\pm61.8$    & 158.0$\pm20.0$      &   430.8$\pm8.4$               &   480.0$\pm9.1$\\
NGC\,3516               &   14.9$\pm4.4$              &   127.5$\pm5.4$      &   46.3$\pm3.3$       & 41.9$\pm11.8$       &   116.1$\pm6.0$      &   402.5$\pm6.3$\\
{\textit{NGC\,3783}}$\dagger$		&   \ldots             &   \ldots    &   \ldots    & \ldots      &   \ldots    &   \ldots\\
NGC\,3786$\dagger$      &   50.8$\pm5.3$              &   56.0$\pm6.1$       &   118.9$\pm5.0$      & $<$124.6                      & 106.2$\pm4.3$                 &    132.1$\pm2.4$\\
{\textit{UGC\,6728}}$\dagger$      &     \ldots                 &    \ldots                    &    \ldots                    & \ldots                       &   \ldots                     &    \ldots\\
NGC\,3982               &   14.2$\pm2.6$              & 24.5$\pm2.5$                  &   72.7$\pm1.9$       & $<$40.8                       &      24.4$\pm3.9$    & $<34.0$ \\
NGC\,4051               &   31.3$\pm5.8$              &   59.7$\pm6.5$       &   124.8$\pm4.2$      & $<$84.9                       &   64.16$\pm6.4$      &   218.4$\pm7.2$\\                        
NGC\,4138		        &   15.6$\pm2.7$              &   11.1$\pm1.2$       &   35.9$\pm1.1$       & $<$8.2                        &   32.6$\pm1.4$                & 30.2$\pm2.5$\\
{\textit{NGC\,4151}}		        &   \ldots            &   \ldots   &   \ldots     & \ldots      &   \ldots   &   \ldots\\
{\textit{NGC\,4235}}$\dagger$      &   \ldots                   &    \ldots                    &    \ldots      & \ldots                       &    \ldots     &     \ldots  \\
NGC\,4253               &   51.0$\pm4.6$              &   147.8$\pm5.3$      &   162.2$\pm3.7$               & 78.7$\pm13.9$       &   116.9$\pm8.1$      &   302.6$\pm8.0$\\                         
NGC\,4388               &   243.3$\pm6.0$             &     577.5$\pm6.8$    &     569.2$\pm4.5$    & 458.7$\pm20.4$      & 925.6$\pm15.9$      & 2473.0$\pm11.3$\\
NGC\,4395$\dagger$		& 12.5$\pm2.1$                &    20.8$\pm2.7$               &    36.3$\pm3.1$      & 13.5$\pm1.3$         &     28.7$\pm0.6$              &    18.1$\pm2.4$\\
NGC\,4501		        &   19.8$\pm0.4$              &     6.9$\pm0.4$      &    40.5$\pm0.3$               & $<$5.0                        &      14.0$\pm0.8$    &     8.6$\pm1.5$\\
NGC\,4593               &   $<$26.5                   & 47.6$\pm5.7$         &   46.1$\pm3.6$                & $<$50.9                       &   31.6$\pm7.0$       &  120.8$\pm7.4$\\
NGC\,4639$\dagger$      &   $<$7.8                    & $<$7.3                        &     10.3$\pm1.5$     & $<$6.7                        &     4.6$\pm1.2$      & $<$4.3 \\
ESO\,323-G77$\dagger$   &   $<$144.1                  &    122.1$\pm32.0$   &   249.1$\pm26.5$    & $<$237.3                      & 156.4$\pm14.6$      &    290.7$\pm30.1$\\
NGC\,5033               &   45.8$\pm1.8$              &    18.2$\pm1.8$      &    151.2$\pm1.5$              & 11.6$\pm2.9$         &   41.5$\pm1.2$       & 41.3$\pm2.1$\\
NGC\,5135               &   354.7$\pm2.1$             &     219.9$\pm2.7$    & 1006.0$\pm1.9$       & 137.3$\pm8.8$        & 346.5$\pm2.9$        & 455.8$\pm5.0$\\
MCG-06-30-015 	        &   19.4$\pm5.0$              &   98.1$\pm5.4$       &   35.1$\pm3.3$       & $<$70.9                       &   $<$40.2                     &   131.5$\pm6.6$\\
{\textit{IC\,4329A}}               &    \ldots                 &   \ldots                  & \ldots                      & \ldots      &    \ldots     &   \ldots\\
NGC\,5347               &   27.1$\pm4.1$              &   56.9$\pm4.4$       &    45.8$\pm3.0$               & $<$61.3                       & $<$12.5                       & 47.2$\pm6.5$\\
NGC\,5506		        &   340.6$\pm7.6$             &  804.4$\pm8.5$       &   706.5$\pm6.4$      & 494.9$\pm22.9$      &   845.1$\pm9.7$      &   2069.0$\pm11.3$\\
NGC\,5548               &   27.0$\pm4.1$              &   32.0$\pm4.2$       &   59.0$\pm3.3$                & $<$47.5                       & 46.2$\pm5.5$         &    58.7$\pm6.6$\\
NGC\,5953               &   127.3$\pm3.7$             &  40.0$\pm3.5$        &     437.3$\pm3.5$    & $<$62.6                       & 88.8$\pm5.7$         &  84.5$\pm8.1$\\
IRAS\,18325-5926$\dagger$ &  106.1$\pm35.6$           &   166.5$\pm15.8$              &    359.4$\pm20.3$   & 291.9$\pm13.2$      & 365.2$\pm10.9$      & 450.2$\pm9.0$\\
{\textit{NGC\,6814}}$\dagger$		&   \ldots                 & \ldots       &   \ldots     & \ldots      &    \ldots   & \ldots\\                        
{\textit{NGC\,6860}}               &   \ldots              &   \ldots       &   \ldots       & \ldots                       &   \ldots       &   \ldots\\
NGC\,6890               &   14.6$\pm3.4$              &   36.8$\pm3.6$       &   74.5$\pm2.4$       & $<$49.4                       &   35.0$\pm5.5$       &   69.3$\pm6.5$\\
NGC\,7130               & 226.6$\pm1.5$               & 94.1$\pm1.5$ & 580.60$\pm1.6$& 52.0$\pm6.4$& 166.6$\pm3.8$& 91.8$\pm4.8$\\
NGC\,7314$\dagger$		&   $<$86.9                   &     163.3$\pm30.8$  & $<$219.9                      & 178.0$\pm8.3$        &  239.2$\pm5.2$       & 479.1$\pm5.5$\\
NGC\,7469               &    656.8$\pm6.9$            &   226.50$\pm7.5$     &   1569.0$\pm6.8$              & $<$85.3                       &   151.6$\pm12.8$    &   278.0$\pm15.0$\\                         
NGC\,7496               &   106.4$\pm3.8$             & 47.5$\pm3.7$         &  346.2$\pm4.4$       & $<$40.1                       & $<$31.8                       &   47.5$\pm13.8$\\
NGC\,7582		        &  851.6$\pm4.9$              &     388.4$\pm6.4$    &     1956.0$\pm5.2$   & 407.7$\pm34.1$      &  620.4$\pm17.6$     & 1668.0$\pm20.1$\\
NGC\,7590               &   86.4$\pm8.1$              &     $<$37.7                   &     218.7$\pm6.1$    & $<$42.9                       &    16.6$\pm3.7$      & $<$26.6\\
\hline
\end{tabular}						 
\caption{Seyfert galaxy emission line measurements with PAHFIT using the Spitzer/IRS spectra. $\dagger$ These galaxies were observed using the staring mode, whereas the remaining galaxies have spectral mapping data. The fluxes and 1$\sigma$ errors are in units of 10$^{-15}$~erg$^{-1}$~cm$^{-2}$. We note that $<$ corresponds to 3$\sigma$ upper limits. Sources in italics have unsatisfactory fits of their spectral features and have been ignored from further analysis. }
\label{tab4}
\end{table*}

\begin{figure*}
\centering
\par{
\includegraphics[width=7.62cm]{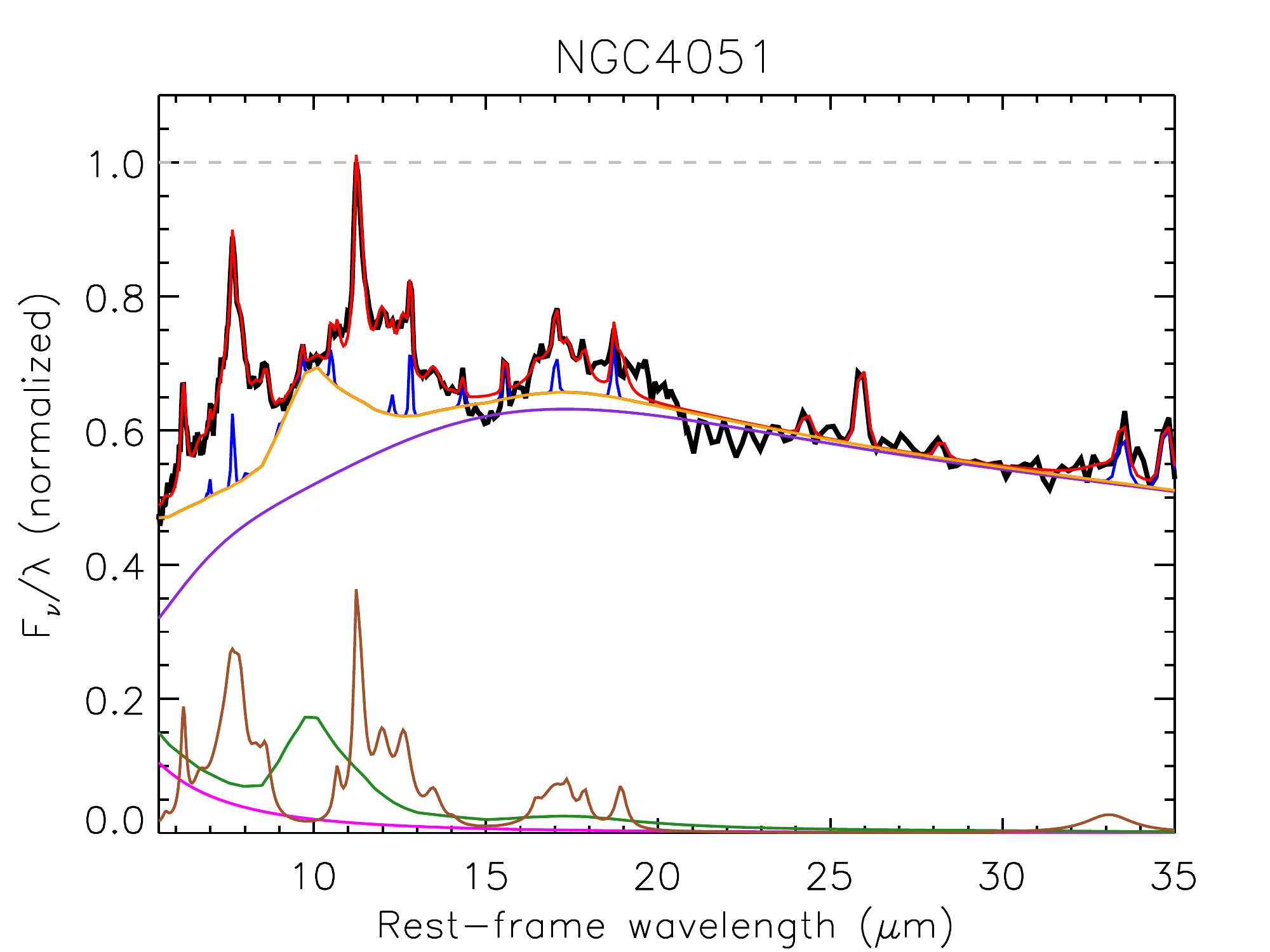}
\includegraphics[width=7.62cm]{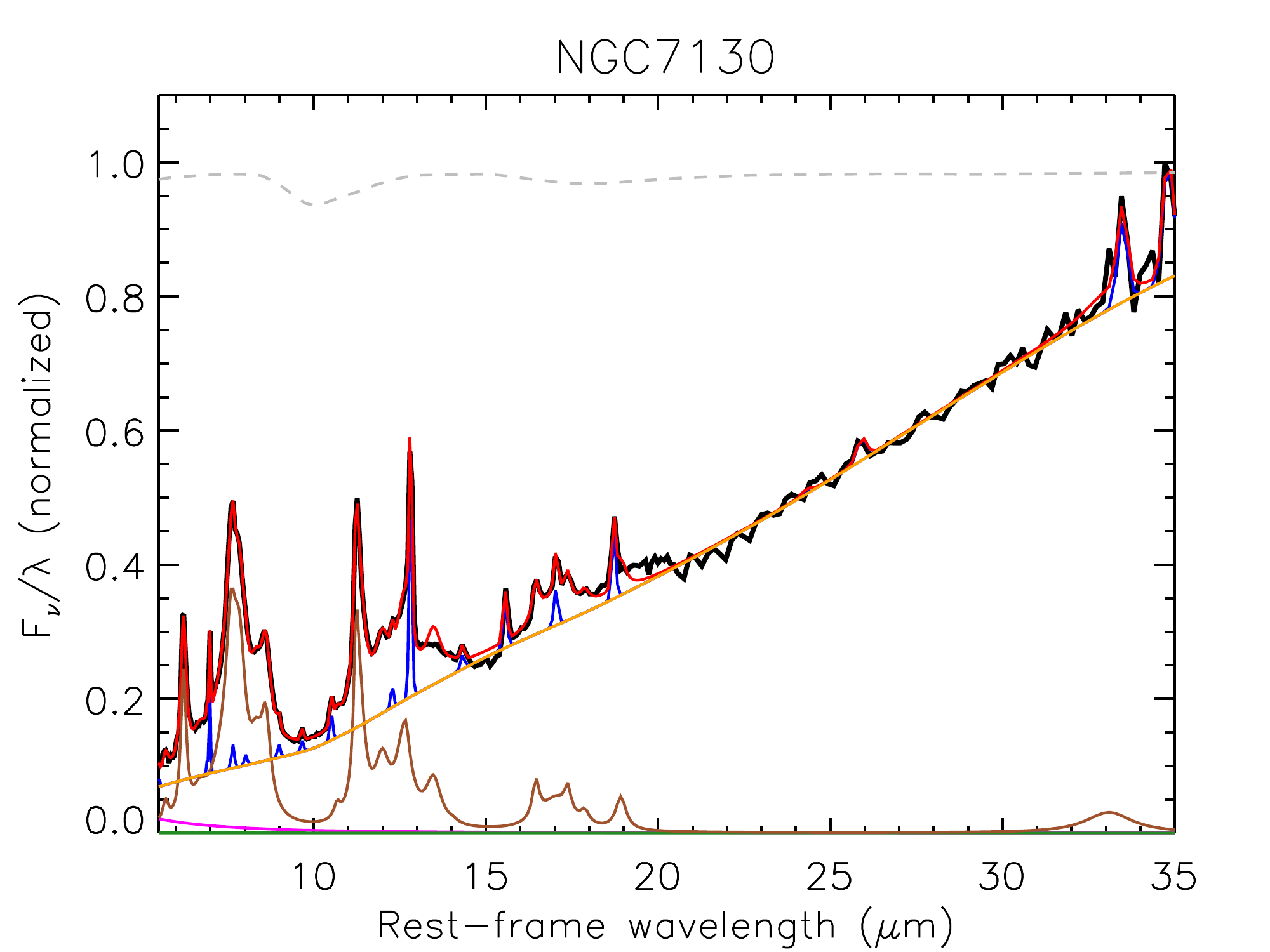}
\includegraphics[width=7.62cm]{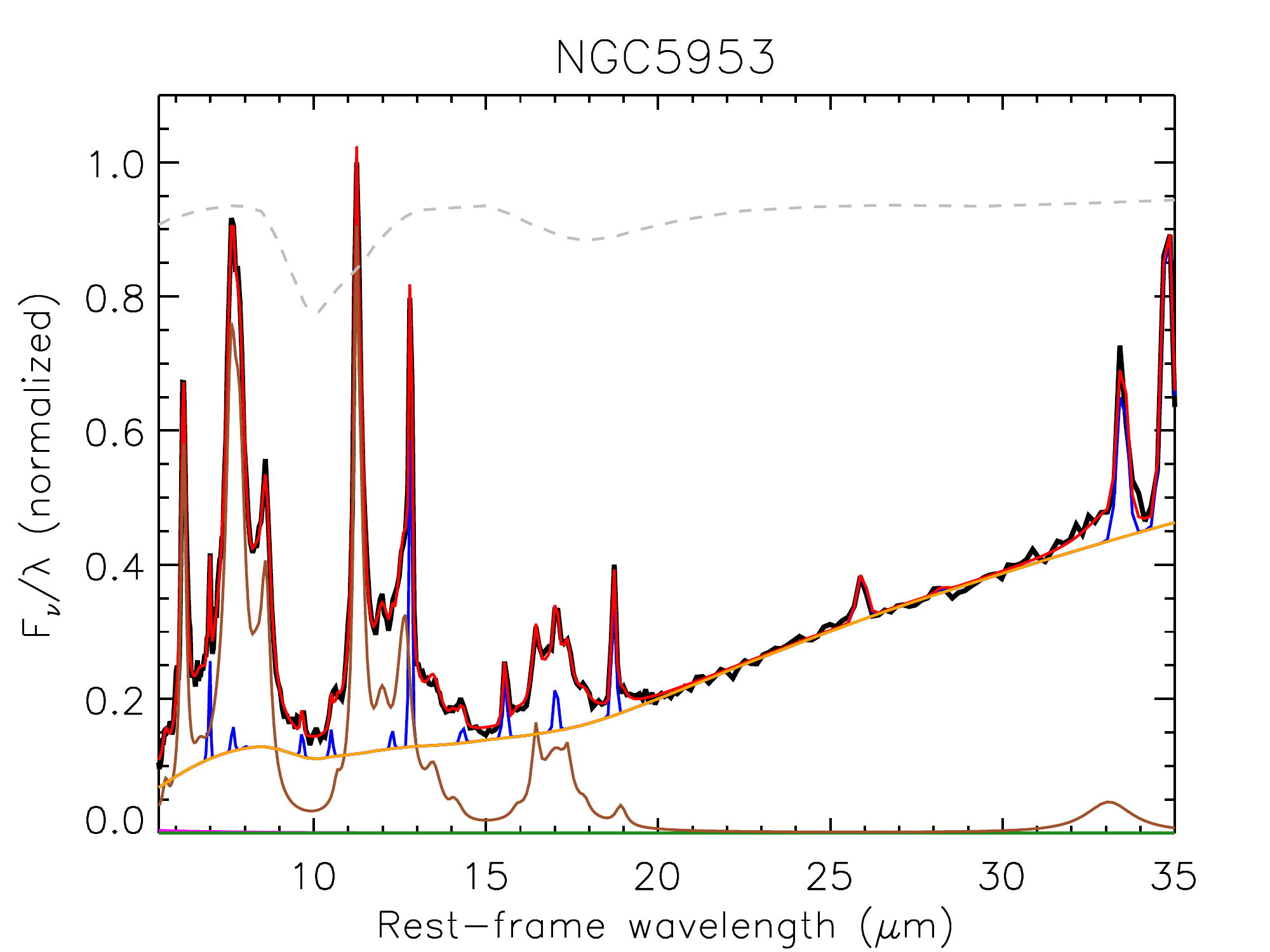}
\includegraphics[width=7.62cm]{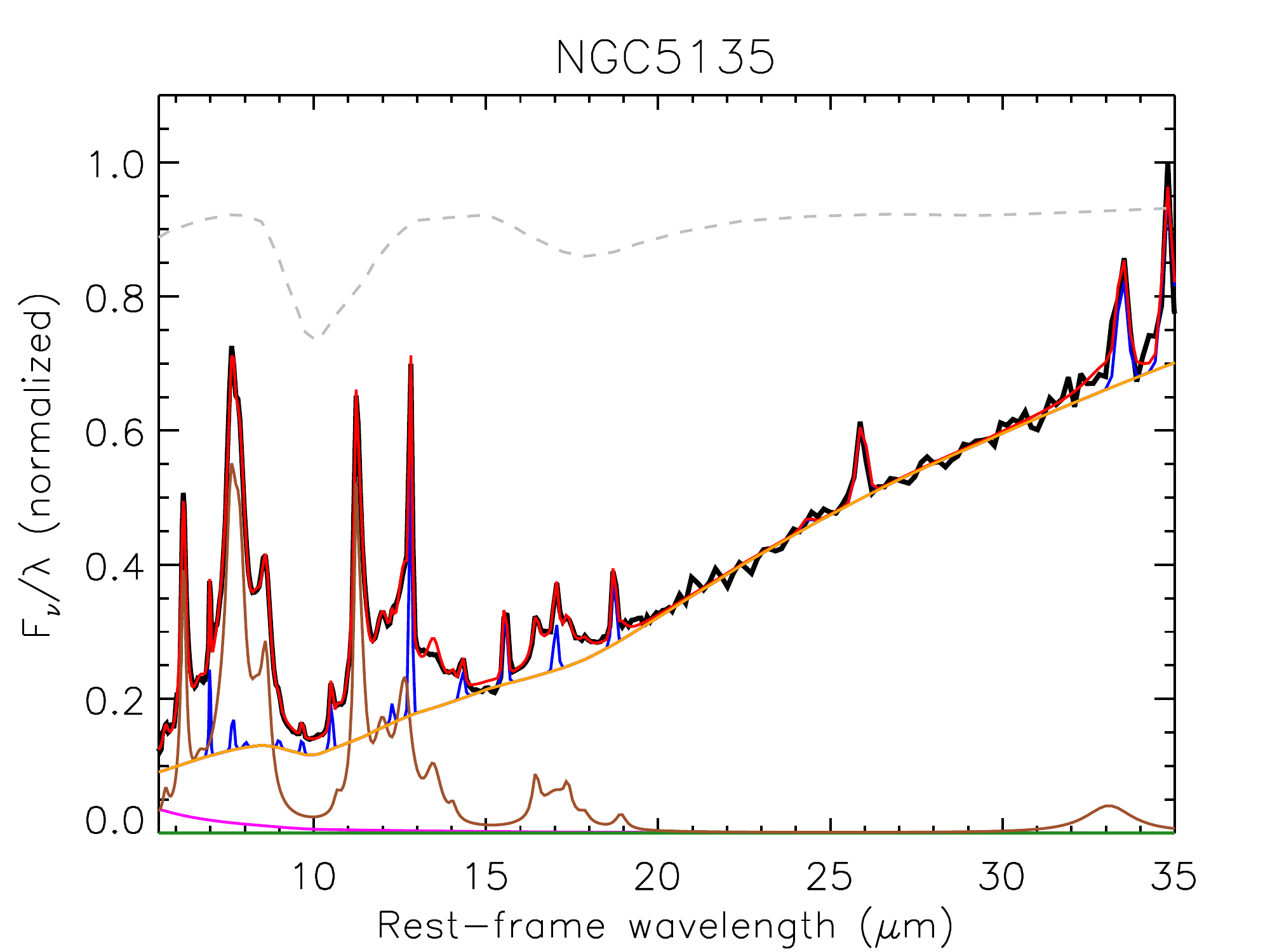}
\par}
\caption{Examples of the MIR spectral modelling using PAHFIT. The Spitzer/IRS rest-frame spectra and model fits correspond to the black and red solid lines. We show the dust continuum (purple solid lines), stellar continuum (pink solid lines), silicate feature in emission (green solid line), and their sum (orange solid line). The grey dashed line indicates the extinction profile. We also show the fitted PAH features (brown solid lines) and emission lines (blue solid lines).}
\label{pahfit}
\end{figure*}

\subsection{Archival data}

\subsubsection{Spitzer Space Telescope MIR spectra}
The majority of the Seyfert galaxies (36/50) in our AGN sample and 2 LIRGs from our control sample of star-forming galaxies with extended MIR emission (see Section \ref{control_sample}) were observed in the spectral mapping mode. We used the low (R$\sim$60--120) resolution IRS modules: Short-Low (SL1; 7.4--14.5~$\mu$m, SL2; 5.2--7.7~$\mu$m) and Long-Low (LL1; 19.5--38.0~$\mu$m, LL2; 14.0--21.3~$\mu$m). We downloaded the Basic Calibrated Data (BCD) from the SHA archive, which are processed by the Spitzer pipeline.

In the case of galaxies with spectral mapping observations available, we used CUBISM \citep{Smith07b} to perform sky background subtraction, rejection of rogue pixels, flux calibration and estimates of the statistical uncertainty at each wavelength and, finally, to combine the various slit observations and build the spectral cubes. 
 
Using these spectral cubes and assuming that the AGN emission is unresolved, we extracted 1-D central spectra treating the AGN as a point-like source. For the extraction we used square apertures of 5.9\arcsec$\times$5.9\arcsec~and 17.3\arcsec$\times$17.3\arcsec~for the SL and LL modules, respectively. Then we applied an aperture correction to account for slit losses using a standard star, which is equivalent to a point source extraction and provides the best signal-to-noise ratio (S/N). For those galaxies showing extended emission in the spectral mapping data, we extracted a second spectrum using an aperture containing the total flux (see Fig. \ref{apertures}). Then, for these galaxies, we derived the extended emission spectrum by subtracting the central spectrum (point-like extraction) from the total one to remove the contribution of the point-like source.

Note that 2 LIRGs (NGC\,3110 and NGC\,6701) within our star-forming control sample have available Spitzer spectral mapping observations and they show extended MIR emission \citep{Pereira10}. Thus, we also derived their extended emission and central spectra using the method described above.

For the star-forming galaxies selected from \citet{Brandl06} and the remainder of the AGN sample, we retrieved low-resolution MIR staring spectra from the Cornell Atlas of Spitzer/IRS Source (CASSIS, version LR7; \citealt{Lebouteiller11}). The spectra were reduced with the CASSIS software, using the optimal extraction to get the best signal-to-noise ratio, which is equivalent to a point source extraction. Finally for comparison, we also retrieved low-resolution MIR staring spectra from CASSIS of six planetary nebulae (PNe) selected from \citet{Stanghellini07}, these are discussed in Section \ref{pahband}.

Note that for both spectral mapping and staring point-like extractions we only needed to apply a small offset to stitch together the different modules (SL/LL; $\lesssim$10\%). 

\subsubsection{Spitzer Space Telescope MIR imaging}
We compiled MIR images of the AGN sample obtained with the Infrared Array Camera (IRAC; \citealt{Fazio04}) using the 3.6 and 8~$\mu$m channels (angular resolution $\sim$1.9\arcsec). The IRAC FOV is 5.2\arcmin ~$\times$~5.5\arcmin on the sky and its pixel scale is 1.2\arcsec. We downloaded the reduced and calibrated mosaiced data from the SHA archive. Note that these mosaics are re-sampled to a pixel size of 0.6\arcsec.

\subsection{Literature data}
The 12, 25, 60 and 100~$\mu$m IRAS fluxes were taken from the Infrared Astronomical Satellite (IRAS) Revised Bright Galaxy Sample (RBGS; \citealt{Sanders03}) and the IRAS Point Source Catalog{\footnote{https://irsa.ipac.caltech.edu/IRASdocs/surveys/psc.html}} (IRAS PSC; \citealt{Helou86}). Note that IRAS data were obtained with 0.6-m telescope, with a pixel size of 2\arcmin ~and an angular resolution ranging from 4 to 6\arcmin~. We also collected IR $\sim$3.4, 4.6 and 12~$\mu$m photometry from the All-Sky Data Release Point Source Catalog{\footnote{http://wise2.ipac.caltech.edu/docs/release/allsky/}}. These data were taken with the 0.4--m Wide-field Infrared Survey Explorer telescope (WISE; \citealt{Wright10}), which has an FoV of 47\arcmin~$\times$47\arcmin~on the sky and its pixel scale is 2.75\arcsec.

We retrieve 6.2, 7.7 and 11.3~$\mu$m PAH fluxes for the LIRGs from GOALS and for the \hii galaxies from SINGS. Note that these values were calculated using the same methodology as in our work (see Section \ref{PAHFIT}). We refer the reader to \citet{Smith07a} and \citet{Armus09} for details of the observations. The PAH$\lambda$3.3~$\mu$m fluxes for the samples of \hii galaxies and LIRGs were taken from \citet{Ichikawa14} and \citet{Inami18}, respectively, and, were observed using AKARI and an aperture to match the Spitzer/IRS slit.

Finally, we compiled PAH$\lambda$3.3~$\mu$m feature measurements available in the literature for our AGN sample (see Table \ref{tab2} in Appendix \ref{literature}). The 3.3~$\mu$m PAH band fluxes are both from ground- and space-based data (see Table \ref{tab2}). Ground-based spectra were obtained with the SpeX spectrograph \citep{Rayner03} on the 3.0--m NASA Infrared Telescope Facility (IRTF) and space-based spectra with the infrared camera (IRC) on-board the Japanese infrared astronomy satellite AKARI \citep{Murakami07,Onaka07} and the Short Wavelength Spectrometer (SWS) on board the Infrared Space Observatory (ISO; \citealt{Kessler96}). \citet{Smith07a} found that the fractional power of the PAH features to the IRAC 8~$\mu$m band is $\sim$70\% in SINGS galaxies. Therefore, we made the reasonable assumption that the emission of the various PAH features are co-spatial and can be traced by the IRAC 8~$\mu$m band. Following the methodology presented in \citet{McKinney21} we projected the Spitzer/IRS aperture and those used for deriving the PAH$\lambda$3.3~$\mu$m feature emission onto the IRAC 8~$\mu$m images. To estimate the aperture matching correction we used the ratio between the two apertures. A large fraction of the sample have IRAC 8~$\mu$m data, but we used the IRAC 3.6~$\mu$m band when there is no 8~$\mu$m images available or the central part of the image is saturated (i.e. NGC\,1365, NGC\,4593, MCG-06-30-015 and NGC\,7469)\footnote{Note that for those galaxies with both channel images available we find practically the same ratios using IRAC 3.6 and 8~$\mu$m data.}. Note that for NGC\,5506 and NGC\,7582 both IRAC 3.6 and 8~$\mu$m images are saturated, and thus we used the median aperture matching correction.

\section{MIR modelling}
\label{PAHFIT}
PAHFIT \citep{Smith07a} is a selection of routines to fit the MIR continuum and the dust emission features of the Spitzer/IRS MIR spectra. In brief, PAHFIT includes thermal continuum radiation from dust grains, fine-structure lines, H$_2$ lines, PAH features and the extinction profile which includes silicate in absorption. However, the original version of PAHFIT did not include silicates in emission, which is a key feature in type 1 AGN. Therefore, in the present work we use a modified version of the PAHFIT code that also includes the 10 and 18~$\mu$m silicate features in emission (see e.g. \citealt{Baum10,Diamond10,Gallimore10,LaMassa10,Dicken12,Dicken14,Sales14,Lyu18,Herrero20,Gleisinger20}). We follow the same methodology as presented in \citet{Gleisinger20} to obtain the fluxes of the different PAH bands and emission lines from the Spitzer/IRS MIR spectra of our sample of AGN and star-forming galaxies from \citet{Brandl06}.

This version of PAHFIT uses a Markov Chain Monte Carlo technique to evaluate the parameter uncertainties. It also makes use of the DREAM(Z) stepping algorithm \citep{braak08}, which performs self-adaptive steps through parameter space. The code employs the Metropolis likelihood criterion \citep{Metropolis53} to decide whether or not to keep the next set of parameters in the chain. This version of PAHFIT has been already tested in \citet{Gleisinger20} for a sample of AGN using Spitzer/IRS MIR spectra (see their Figure 1). We note that the Spitzer/IRS spectra of the majority of the galaxies in this sample have been previously studied (e.g. \citealt{Baum10,Diamond10,Gallimore10,LaMassa10,Lyu18,Herrero20}) by using the same method to extract the PAH feature strengths. 

The modified version of PAHFIT employed in this work uses the \citet{Ossenkopf92} cold dust model for the feature in emission, which peaks at $\sim$10.1~$\mu$m. We use the default model continuum dust temperatures and a stellar continuum component. Since Sy1 galaxies show evidence of the silicates feature in emission or nearly flat in the MIR spectra (see e.g. \citealt{Bernete19} and references therein), we use the silicates in emission only for fitting type 1 Seyfert galaxies. Because of the low spatial resolution afforded by Spitzer, the silicate features can produce an additional source of uncertainty in the fitted continuum. The latter can also affect the total measured PAH fluxes. However, we have consistently used the same method for fitting the underlying continuum of all the PAH bands, and thus these effects are mitigated when measuring PAH band ratios. Four sources in our sample (NGC\,4388, NGC\,5506, NGC\,7496 and NGC\,7582) show deep silicate absorption that can influence the derived PAH fluxes. We label these sources in all the figures to show that the observed trends are not in any way skewed by the presence of these galaxies. In the future, studies such as this, will benefit from the availability of higher angular resolution spectra for the entire MIR range that will be afforded by the James Webb Space Telescope. Such spectra will help to better constrain the various components of the continuum. Fig. \ref{pahfit} shows examples of the fitting procedure followed, whereas, Appendix \ref{pahfit_figs} shows the fits for the entire sample. Note that for some AGN-dominated systems in our sample (10/50) PAHFIT does not reproduce the region of the silicate features and, subsequently, we classify those as unsatisfactory fits of their spectral features (see Table \ref{tab3} and Appendix \ref{pahfit_figs}). 

Once we subtracted the continuum emission and the emission-line contribution to each MIR spectrum, we obtain the PAH spectrum for each galaxy. In Tables \ref{tab3} and \ref{tab4}, we list the integrated fluxes and corresponding errors of the MIR PAH features and emission lines, respectively. The list includes, [Ar\,II]$\lambda$6.98, [S\,IV]$\lambda$10.51, [Ne\,II]$\lambda$12.81, [Ne\,V]$\lambda$14.32, [Ne\,III]$\lambda$15.56, [O\,IV]$\lambda$25.91, PAH$\lambda$6.2, PAH$\lambda$7.7 complex ($\lambda$7.42, $\lambda$7.60 and $\lambda$7.85 features), PAH$\lambda$11.3 complex ($\lambda$11.23 and $\lambda$11.33 features) (all wavelengths given in ~$\mu$m). In the case of the galaxies for which particular emission lines are not detected, we calculated their 3$\sigma$ upper limits (see Tables \ref{tab3} and \ref{tab4}).

\section{AGN MIR fraction}
\label{fagn}

When considering the spatial scales probed by the Spitzer/IRS spectra of the sample galaxies ($\sim$kpc scales), we expect some degree of contribution from the host galaxy to the MIR spectra, in addition to the AGN. To estimate the relative contributions, we use the DEBLENDIRS routine \citep{Hernan-caballero15}. This is an IDL routine that decomposes MIR spectra using a linear combination of three spectral components: AGN, PAH and stellar emission. The
combination of components that best reproduce the IRS/Spitzer spectrum is obtained through $\chi^2$ minimization. Subsequently, DEBLENDIRS estimates the fractional contribution of each component to the total MIR spectrum.

The spectrum of each component is selected from extreme cases of galaxies dominated by AGN, PAH, or stellar emission, using a large library of Spitzer/IRS spectra. We have consistently avoided the use of the sources themselves as templates, since few of our galaxies are in the AGN library of DEBLENDIRS. Note that we cannot expand the spectral decomposition beyond 22~$\mu$m due to limitations in the library. In Table \ref{tab3} we list the AGN MIR contributions and corresponding errors derived from the spectral decomposition of our sample. The median values of the fractional contribution of the AGN to the MIR Spitzer/IRS spectra for Sy1, Sy1.8/1.9 and Sy2 are 83$\pm$17, 61$\pm$24 and 25$\pm$25, respectively. These estimates are in good agreement with previous works using samples of Sy galaxies (e.g. \citealt{Bernete16}).

\section{PAH emission and hardness of the radiation field}
\label{results}

The spectral features of the PAH molecules are directly related to their molecular size and structure, charge state and the radiation field to which the molecules are exposed. Indeed, PAH ratios can be used as tracers of the size of the PAH molecule (e.g. 6.2/7.7 and 11.3/3.3), the PAH ionization fraction (11.3/7.7 and 11.3/6.2) and the hardness of the radiation field (e.g. 11.3/3.3, 11.3/7.7 and 11.3/6.2) (e.g. \citealt{Jourdain90,Draine01, Galliano08,Sales10,Maragkoudakis20,Draine20,Rigopoulou20}). In what follows, we infer the hardness of the radiation field based on key observables while we discuss how PAH band ratios enable us to determine the properties of the PAH molecules in Section \ref{pahdiagram}.

Earlier studies have routinely used PAH emission to estimate the SFR in the circumnuclear regions of AGN (see e.g. \citealt{Diamond12,Esquej14}). However, limited knowledge of the effect of the radiation field on these molecules is a shortcoming of this approach. Often, AGN-dominated systems do not show strong PAH features in their nuclear IR spectra. Even in AGN that show these features, they often have low equivalent width compared to those observed in star-forming galaxies. Therefore, it has been proposed that PAH molecules are destroyed by the very hard radiation field of the AGN (e.g. \citealt{Aitken85,Roche91,Voit92,Siebenmorgen04}) or that these features are diluted by the strong AGN continuum (e.g. \citealt{Herrero14,Almeida14,bernete15}). However, other studies did not find dilution on the 11.3~$\mu$m feature using radial profiles of subarcsecond resolution spectra \citep{Esparza-Arredondo18}. In addition, AGN-dominated systems that show PAH features could be explained by a shielding effect of the PAH due to H$_2$ molecules when the column density is high (e.g. \citealt{Aitken85,Voit92,Herrero20}) or by AGN excitation of these molecules (e.g. \citealt{Jensen17}).

Therefore, it is interesting to investigate the link between PAH emission and the hardness of the radiation field to better understand the effect of very hard radiation fields on PAH molecules which can potentially dramatically influence the PAH properties. With this aim, we compare the relative intensities of the different PAH bands with the hardness of the radiation field of various environments such as powerful AGN, relatively weak AGN and star-forming galaxies. To do so, we compile IR photometry (IRAS and WISE), together with the MIR spectra, for our sample of AGN and a control sample of star-forming galaxies. 

\subsection{Dust components and the hardness of the radiation field}

The NIR-to-MIR range is well suited to probe the various components of the dust emission produced by the AGN heating. The NIR emission is mainly related to the accretion disk and very hot dust (close to the sublimation temperature; T$_{\rm sub}\sim$2000~K). The NIR emission also contains the contribution of the stellar emission from the host galaxy, which can be particularly significant in Sy2 galaxies (e.g. \citealt{Herrero96}, \citealt{Riffel06}). On the contrary, the MIR emission comes from warm dust mainly heated by the AGN, but there is also a significant contribution from young stars. These young stars heat the dust in star-forming regions that can even dominate the total MIR emission of galaxies (e.g. \citealt{Horst08,Siebenmorgen08}). 

We use IRAS and WISE colours (e.g. \citealt{Grijp87,Low88, Stern12, Mateos12, Assef13}), to determine the dust temperature in both of our samples. Powerful heating sources produce warmer IR colors allowing us  to distinguish them from cooler sources. We find that F$_{25~\rm \mu m}$/F$_{60~\rm \mu m}$ ratio is even better in selecting AGN-dominated systems than the WISE colours according to the AGN contribution to the MIR Spitzer/IRS spectra (see Section \ref{fagn}). Therefore, we use the F$_{25~\rm \mu m}$/F$_{60~\rm \mu m}$ ratio as a proxy for the dust temperature. Note that due to the large IRAS apertures this ratio is representative of the integrated properties of the galaxies (and may dilute the real impact of the AGN). Thus, we also estimate the AGN contribution to the MIR Spitzer/IRS point-like spectra (see also column 5 of Table \ref{tab3} in Section \ref{PAHFIT}). See Section \ref{fagn} for further details on the spectral decomposition.

The hardness of the radiation field can also be traced by MIR fine-structure emission line ratios. The [Ne\,III]/[Ne\,II] ratio is used as a robust indicator of the hardness of the radiation field in the surroundings of massive young stars (e.g. \citealt{Thornley00}), and in AGN (e.g. \citealt{Groves06}). However, a recent study reported that [Ne\,II] emission line can be significantly contaminated by the AGN-dominated systems using a sample of unobscured type 1 AGN \citep{Bernete17}. 
Because of that, and as a sanity check, we also employ other ionization ratios involving emission lines such as [O\,IV] and [Ar\,II] (see Appendix \ref{optical}). Finally, we employ the optical [O\,III]/[O\,II] ratio (see Appendix \ref{optical}) which has been known to be a good indicator of the hardness of the radiation field in AGN (e.g. \citealt{Kewley06}), although this ratio can be affected by extinction.

\begin{figure*}
\centering
\par{
\includegraphics[width=8.1cm]{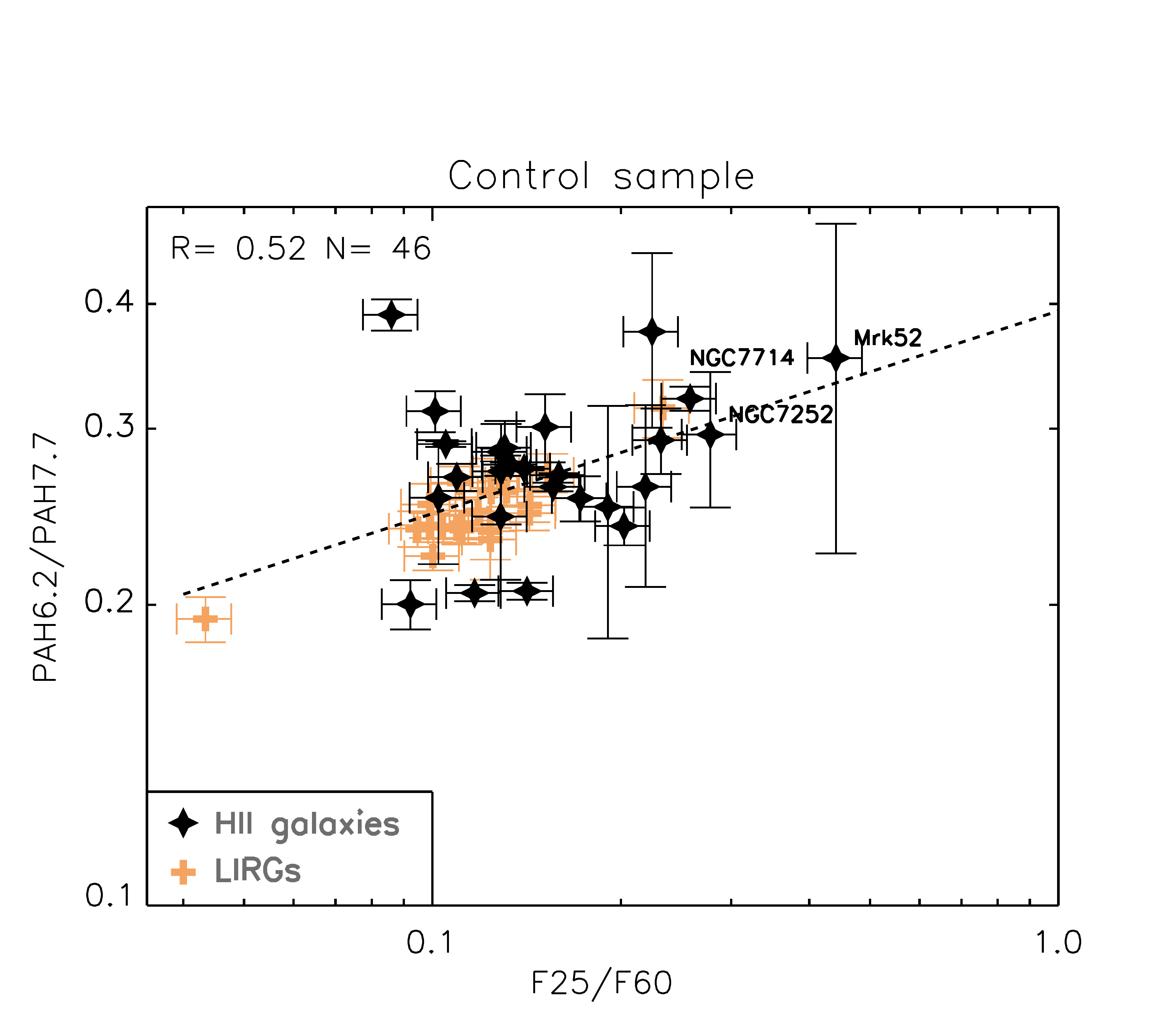}
\includegraphics[width=8.1cm]{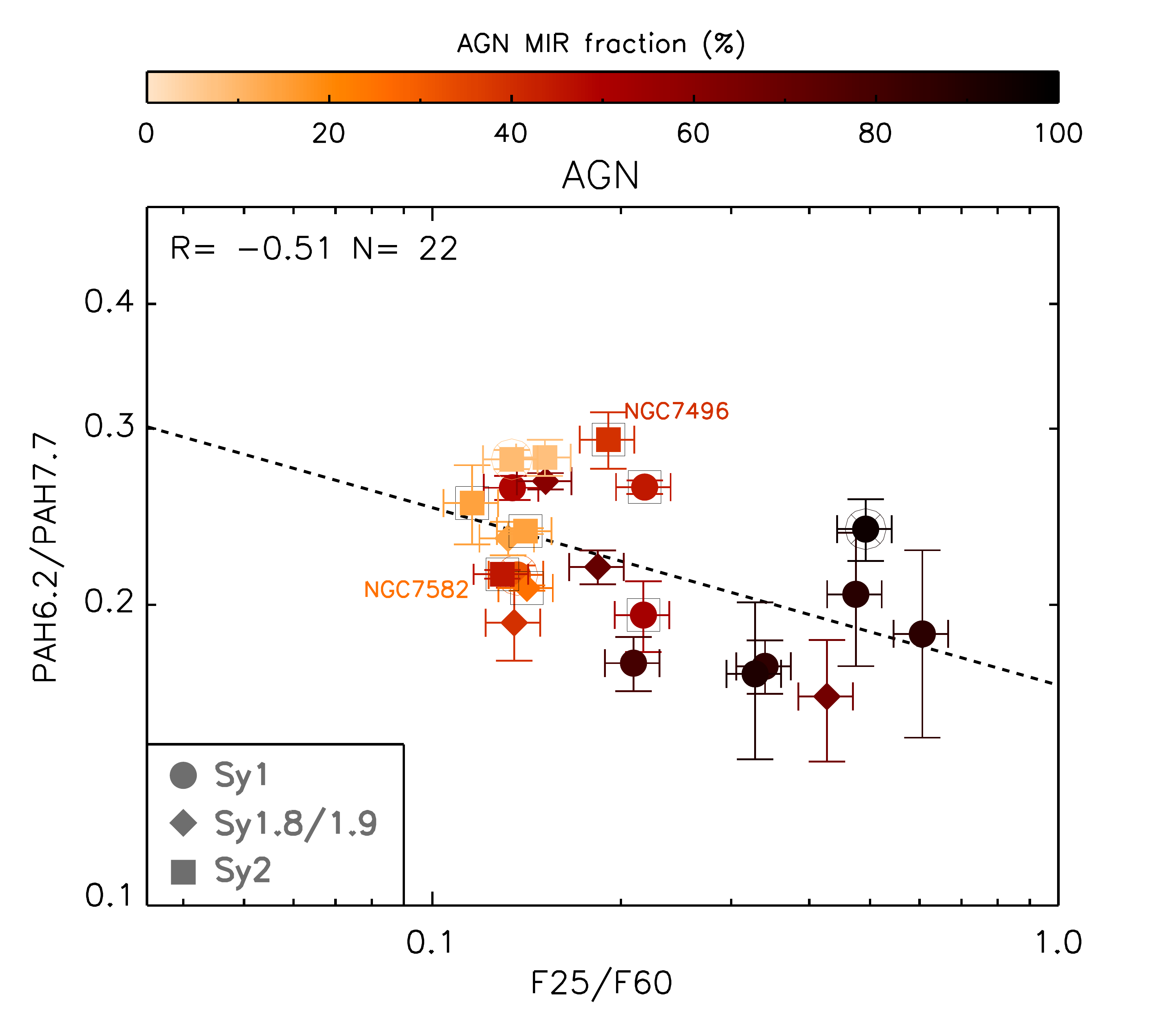}
\par}
\caption{6.2/7.7 PAH ratio vs. F$_{25~\rm \mu m}$/F$_{60~\rm \mu m}$ IRAS color for those sources with measured PAH fluxes. Left panel: control sample of star-forming galaxies. Black stars and brown crosses correspond to \hii galaxies and LIRGs. Right panel: Seyfert galaxies. Filled circles, diamond and squares correspond to Sy1, Sy1.8/1.9 and Sy2, respectively. Note that open squares and circles represent galaxies that have been also classified as AGN/SF composite and LINER, respectively. The black dashed lines in both panels are linear fits (see Table \ref{tabcor}). Labelled sources correspond to Seyfert galaxies with deep silicates in absorption. }
\label{iras67}
\end{figure*}

\begin{figure*}
\centering
\par{
\includegraphics[width=8.1cm]{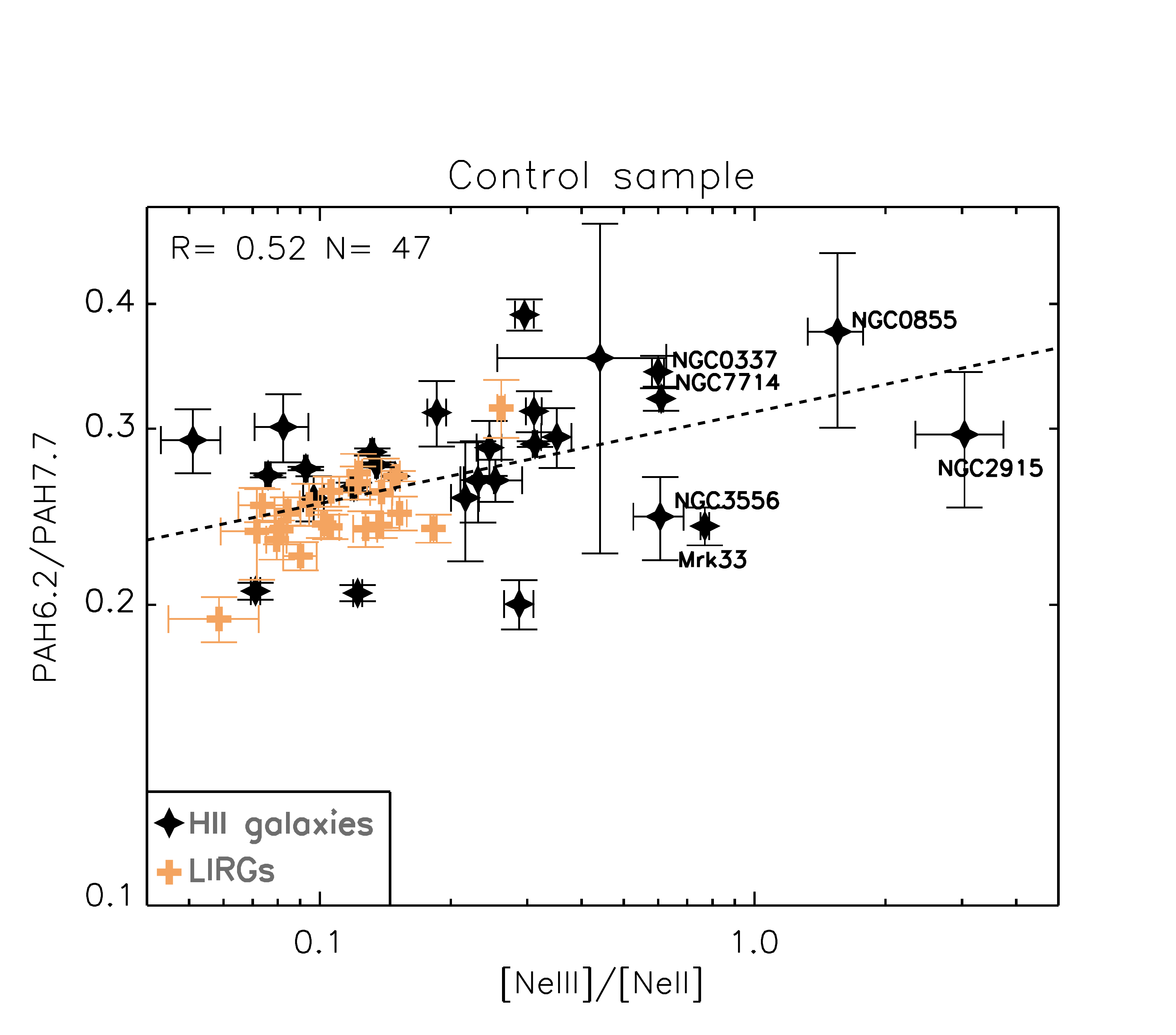}
\includegraphics[width=8.1cm]{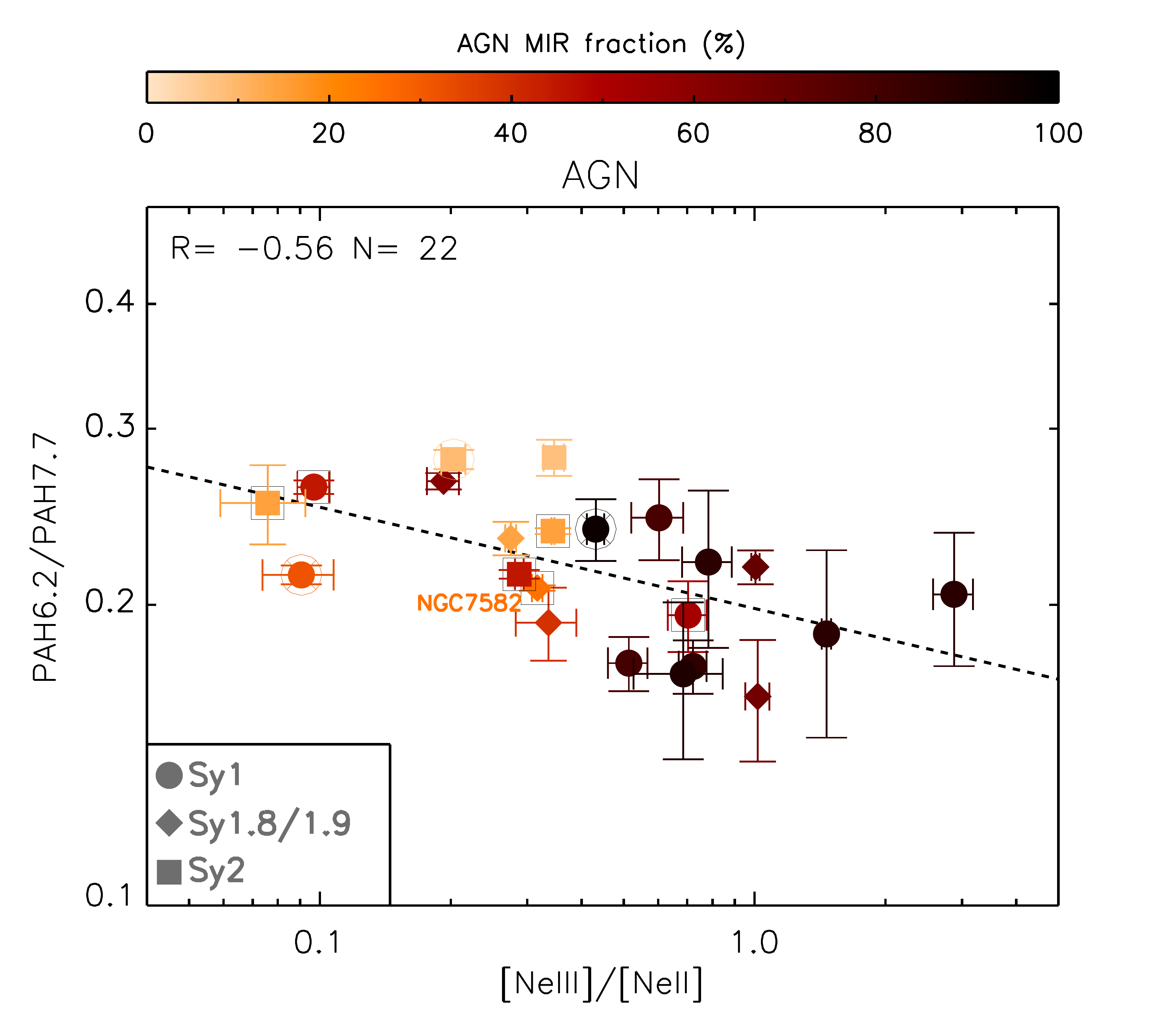}
\par}
\caption{Same as Fig. \ref{iras67} but using the [Ne\,III]/[Ne\,II] ratio instead of the F$_{25~\rm \mu m}$/F$_{60~\rm \mu m}$ ratio.}
\label{ne67}
\end{figure*}

\begin{figure*}
\centering
\par{
\includegraphics[width=8.1cm]{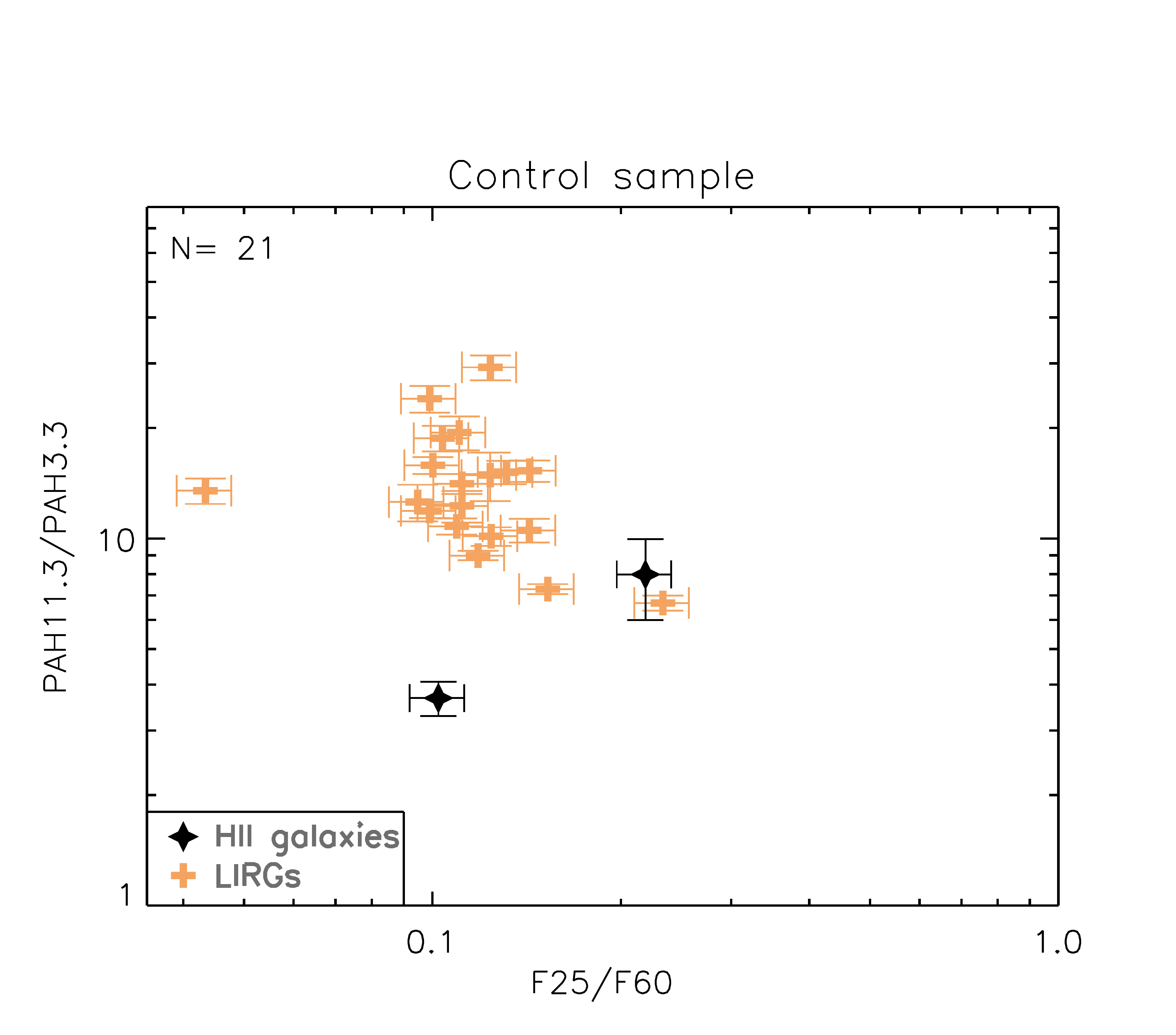}
\includegraphics[width=8.1cm]{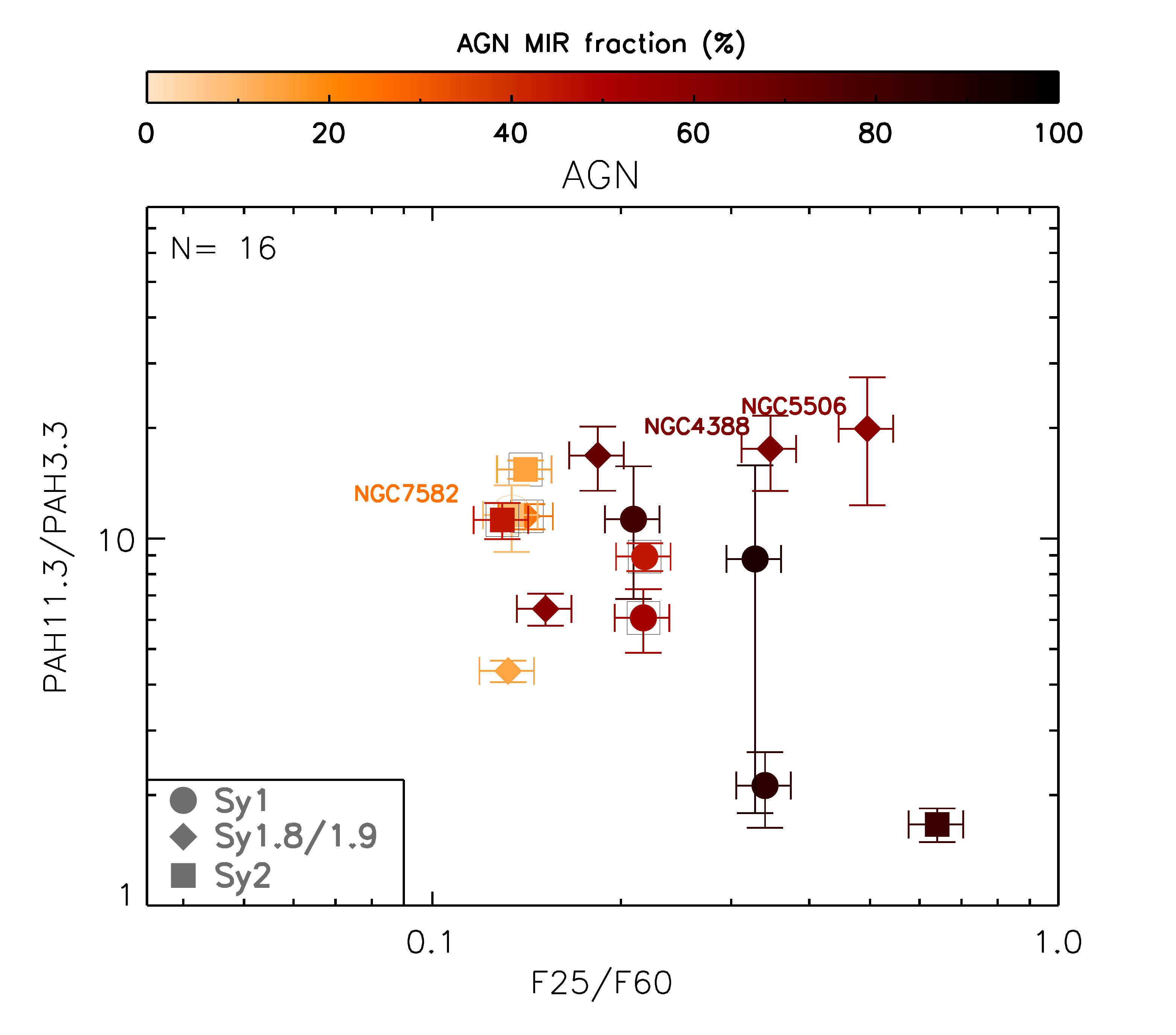}
\par}
\caption{11.3/3.3 PAH ratio vs. F$_{25~\rm \mu m}$/F$_{60~\rm \mu m}$ IRAS color for those sources with measured PAH fluxes. Left panel: control sample of star-forming galaxies. Black stars and brown crosses correspond to \hii galaxies and LIRGs. Right panel: Seyfert galaxies. Filled circles, diamond and squares correspond with Sy1, Sy1.8/1.9 and Sy2, respectively. Note that open squares and circles represent galaxies that have been also classified as AGN/SF composite and LINER, respectively. Labelled sources correspond to Seyfert galaxies with deep silicates in absorption.}
\label{iras113}
\end{figure*}

\begin{figure*}
\centering
\par{
\includegraphics[width=8.1cm]{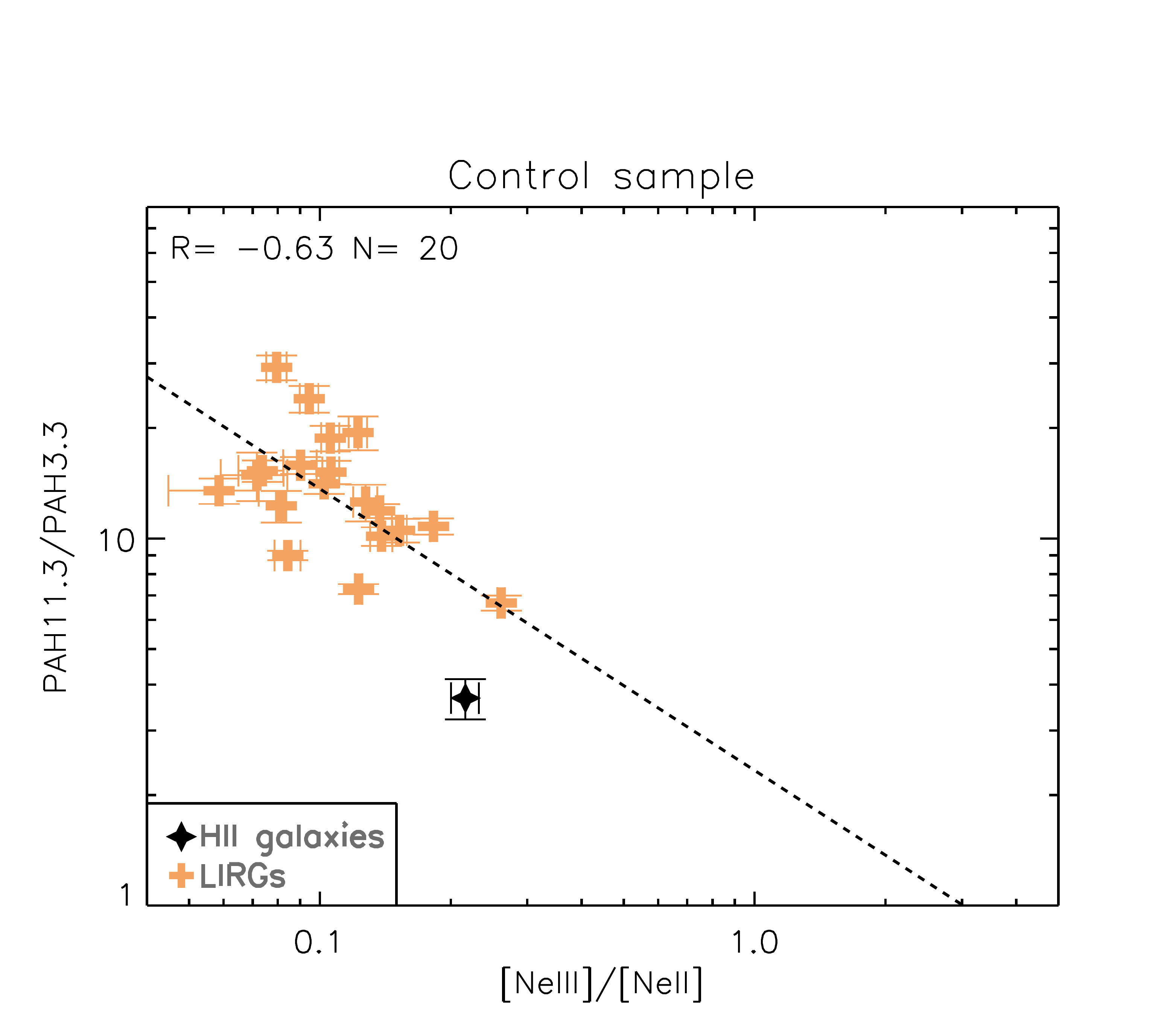}
\includegraphics[width=8.1cm]{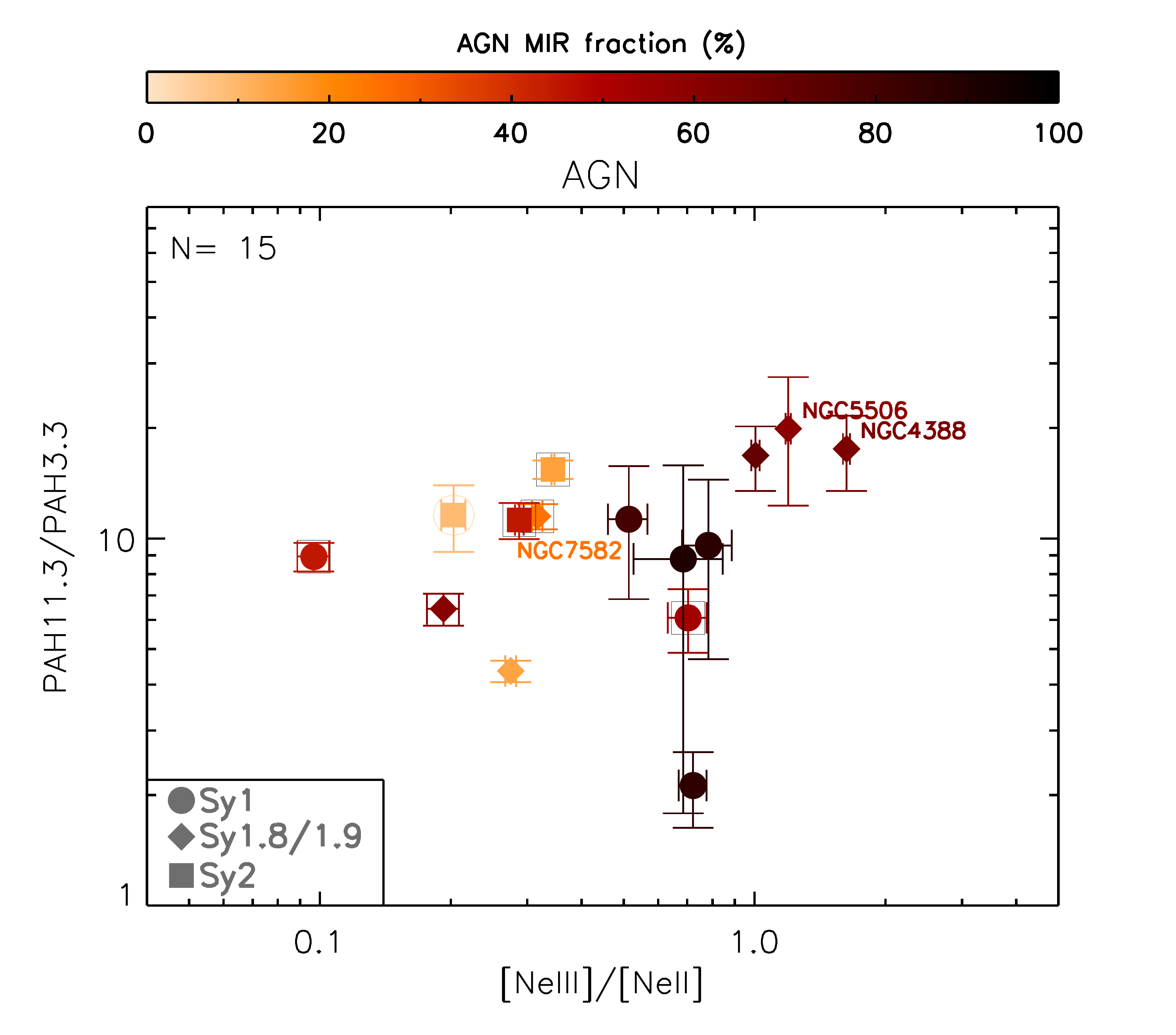}
\par}
\caption{Same as Fig. \ref{iras113} but using the [Ne\,III]/[Ne\,II] ratio instead of the F$_{25~\rm \mu m}$/F$_{60~\rm \mu m}$ ratio. The black dashed line in the left panel is a linear fit (see Table \ref{tabcor}). }
\label{ne113}
\end{figure*}

\begin{figure*}
\centering
\par{
\includegraphics[width=8.1cm]{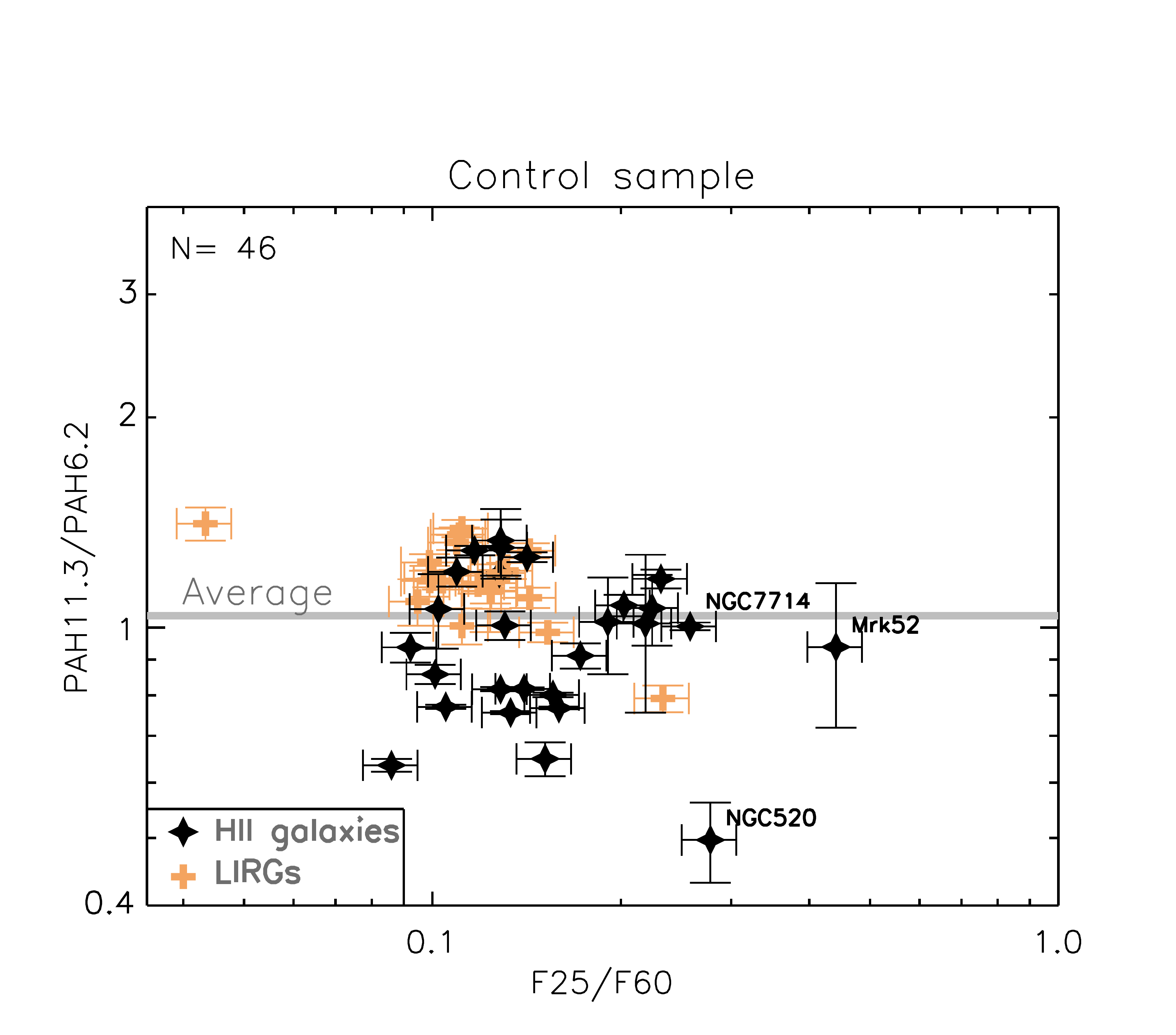}
\includegraphics[width=8.1cm]{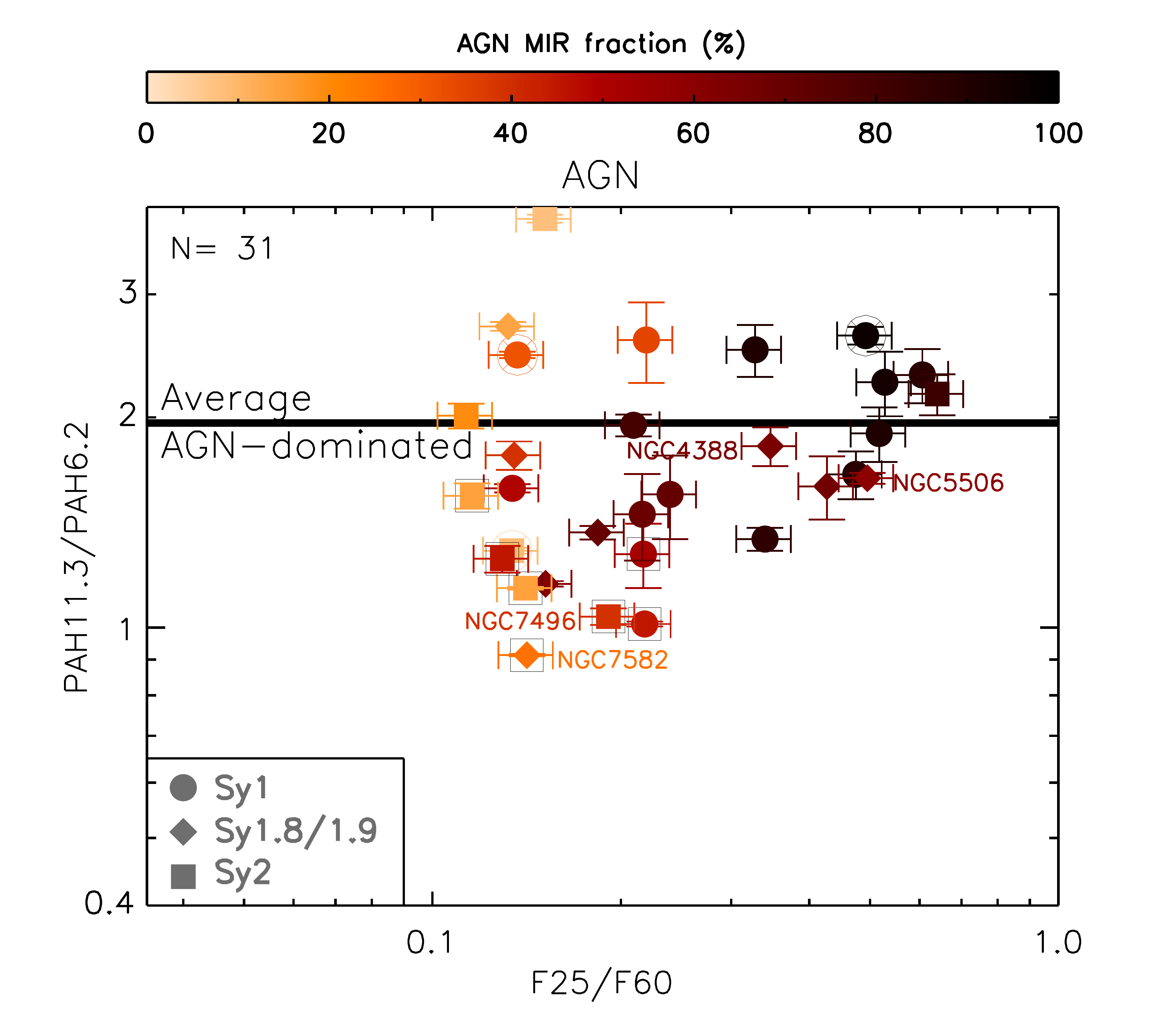}
\par}
\caption{11.3/6.2 PAH ratio vs. F$_{25~\rm \mu m}$/F$_{60~\rm \mu m}$ IRAS color for those sources with measured PAH fluxes. Left panel: control sample of star-forming galaxies. Black stars and brown crosses correspond to \hii galaxies and LIRGs. Right panel: Seyfert galaxies. Filled circles, diamond and squares correspond with Sy1, Sy1.8/1.9 and Sy2, respectively. Note that open squares and circles represent galaxies that have been also classified as AGN/SF composite and LINER, respectively. The grey and orange horizontal solid lines show the average 11.3/6.2 PAH ratios of star-forming galaxies and AGN-dominated systems (AGN fraction $>$70\%), respectively. Labelled sources correspond to Seyfert galaxies with deep silicates in absorption.}
\label{iras116}
\end{figure*}

\begin{figure*}
\centering
\par{
\includegraphics[width=7.62cm]{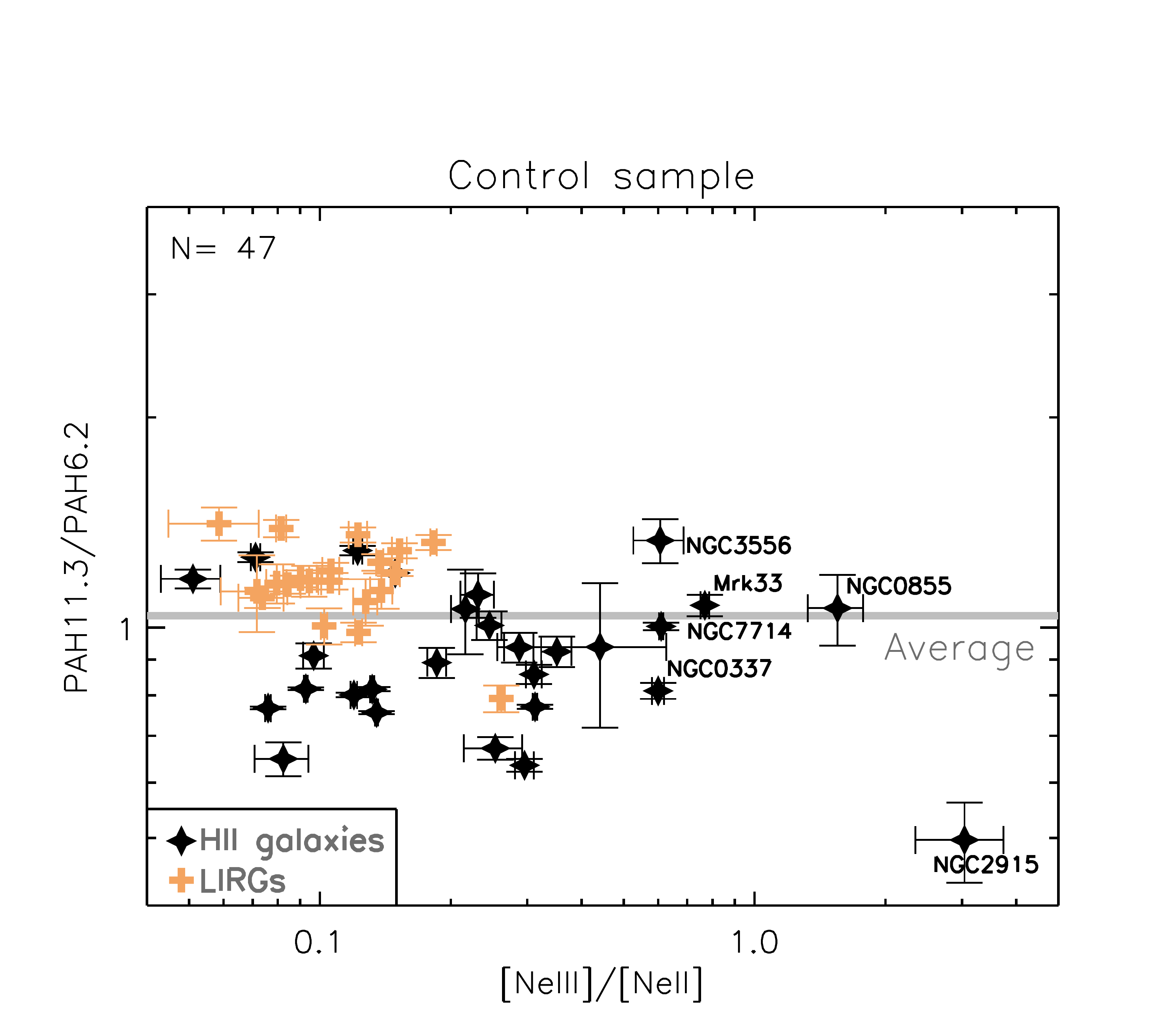}
\includegraphics[width=7.62cm]{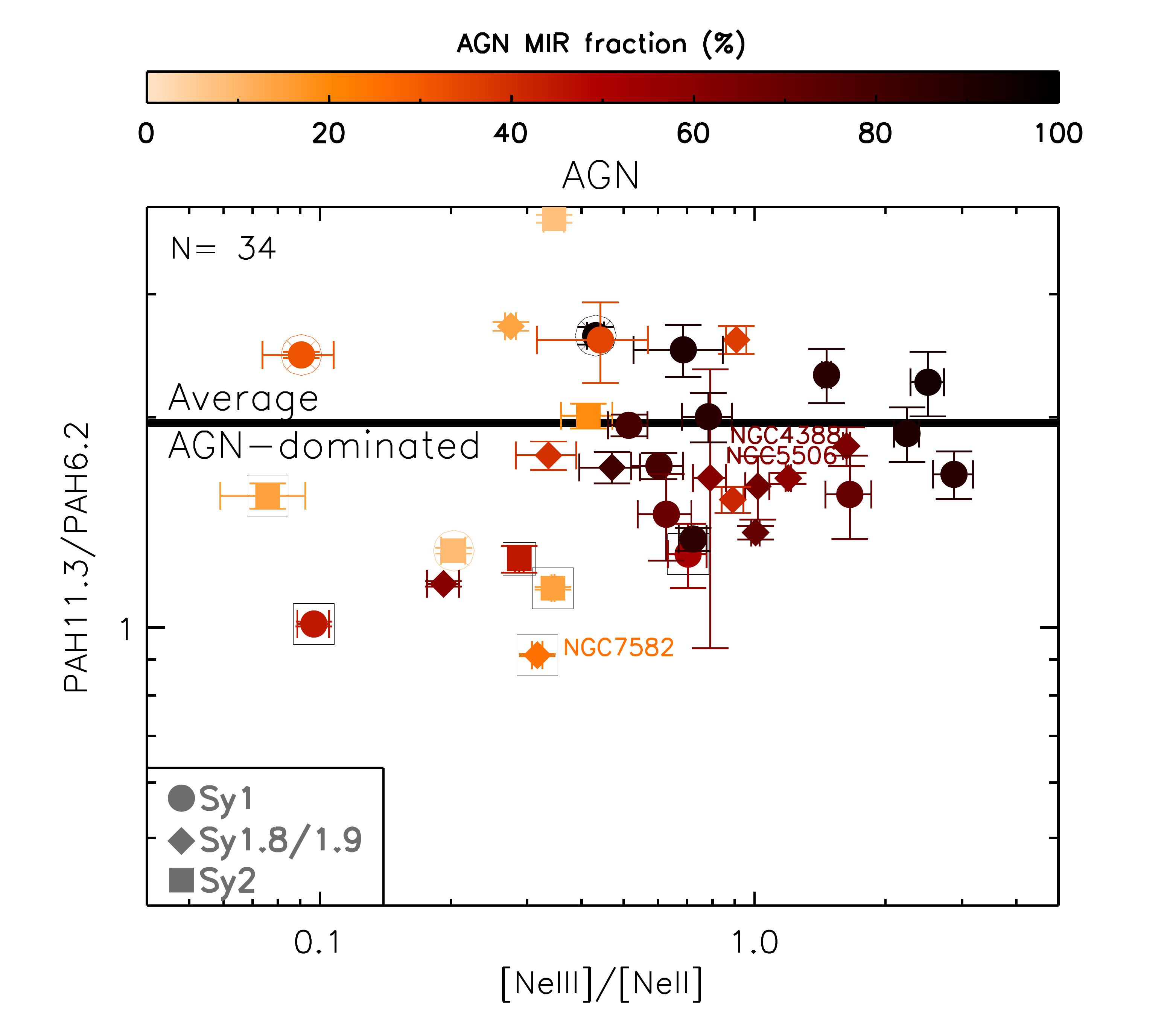}
\par}
\caption{Same as Fig. \ref{iras116} but using the [Ne\,III]/[Ne\,II] ratio instead of the F$_{25~\rm \mu m}$/F$_{60~\rm \mu m}$ ratio.}
\label{ne116}
\end{figure*}

\subsection{The effect of the hardness of the radiation field to PAH band ratios}
\label{ratios_sec}

In Figs. \ref{iras67} and \ref{ne67} we present the relationship of
the widely used PAH molecular size indicator 6.2/7.7 PAH ratio (e.g. \citealt{Jourdain90,Draine01,Sales10,Draine20,Rigopoulou20}) 
with the dust temperature (parameterised by F$_{25~\rm \mu m}$/F$_{60~\rm \mu m}$) and the hardness of the radiation field by using the [Ne\,III]/[Ne\,II] ratio. We compare the 6.2/7.7 PAH ratio values of Seyfert galaxies to those of the star-forming galaxies in the control sample using the KS test, and, find that they are statistically different (see Table \ref{tabks}). The same holds when using LIRGs or \hii galaxies instead of the entire star-forming control sample. However, the 6.2/7.7 PAH ratio values of those sources classified as AGN/SF composite (based on their UV emission and/or emission line ratios; see Table \ref{tab1}) and the star-forming galaxies in the control sample are similar according to the KS test, since the null hypothesis cannot be rejected (see Table \ref{tabks}). In general, we find larger values of the 6.2/7.7 PAH ratio for harder radiation fields in star-forming galaxies compared with AGN-dominated systems. We also find that AGN show a gradual decrease in the 6.2/7.7 PAH ratio with harder radiation fields, whereas the control sample of star-forming galaxies shows a marked increase of this ratio with the hardness of the radiation field (see black dashed lines in Figs. \ref{iras67} and \ref{ne67} and Table \ref{tabcor}). We discuss this issue  further in Section \ref{pahdiagram}. 
\begin{table}
\scriptsize
\centering
\begin{tabular}{lcccc}
\hline
PAH ratio & Samples   &	D-statistics & p-vaule & Number\\
 &    &	 &  & of sources\\
 (1)&(2)&(3)&(4) & (5)\\	
\hline
{\bf{6.2/7.7}} & {\bf{AGN vs. Control}} & 0.493 & $<$0.05 & 24 vs. 46 \\            
{\bf{6.2/7.7}} & {\bf{AGN vs. LIRGs}} & 0.492 & $<$0.05 & 24 vs. 20 \\ 
{\bf{6.2/7.7}} & {\bf{AGN vs. \hii galaxies}} & 0.554 & $<$0.05 & 24 vs. 26 \\ 
6.2/7.7 & AGN/SF comp. vs. Control & 0.492 & 0.176 & 7 vs. 46\\

{\bf{11.3/7.7}} & {\bf{AGN vs. Control}} & 0.496 & $<$0.05 & 24 vs. 46 \\           
{\bf{11.3/7.7}} & {\bf{AGN vs. LIRGs}} & 0.533 & $<$0.05 & 24 vs. 20 \\           
{\bf{11.3/7.7}} & {\bf{AGN vs. \hii galaxies}} & 0.468 & $<$0.05 & 24 vs. 26 \\
11.3/7.7 & AGN/SF comp. vs. Control & 0.193 & 0.960 & 7 vs. 46\\

{\bf{11.3/6.2}} & {\bf{AGN vs. Control}} & 0.734 & $<$0.05 & 38 vs. 46 \\           
{\bf{11.3/6.2}} & {\bf{AGN vs. LIRGs}} & 0.737 & $<$0.05 & 38 vs. 20 \\           
{\bf{11.3/6.2}} & {\bf{AGN vs. \hii galaxies}} & 0.789 & $<$0.05  & 38 vs. 26\\         
11.3/6.2 & AGN/SF comp. vs. Control & 0.193 & 0.960 & 7 vs. 46\\
11.3/3.3 & AGN vs. Control & 0.349 & 0.148 & 18 vs. 21\\
11.3/3.3 & AGN/SF comp. vs. Control & 0.371 & 0.526 & 5 vs. 21\\
\hline
{\bf{11.3/7.7 vs. 6.2/7.7}} & {\bf{AGN vs. Control}} & 0.586 & $<$0.05& 24 vs. 46\\
{\bf{11.3/7.7 vs. 6.2/7.7}} & {\bf{AGN vs. LIRGs}} & 0.596 & $<$0.05& 24 vs. 20\\
{\bf{11.3/7.7 vs. 6.2/7.7}} & {\bf{AGN vs. \hii galaxies}} & 0.596 & $<$0.05& 24 vs. 26\\

{\bf{11.3/7.7 vs. 6.2/7.7}} & {\bf{AGN-dominated vs. Control}} & 0.802 & $<$0.05 & 9 vs. 46\\
{\bf{11.3/7.7 vs. 6.2/7.7}} & {\bf{AGN-dominated vs. LIRGs}} & 0.783 & $<$0.05 & 9 vs. 20\\
{\bf{11.3/7.7 vs. 6.2/7.7}} & {\bf{AGN-dominated vs. \hii galaxies}} & 0.868 & $<$0.05 & 9 vs. 26\\
11.3/7.7 vs. 6.2/7.7 & AGN/SF comp. vs. Control & 0.365 & 0.331 & 7 vs. 46\\
11.3/7.7 vs. 6.2/7.7 & AGN/SF comp. vs. LIRGs & 0.364 & 0.400 & 7 vs. 20\\
11.3/7.7 vs. 6.2/7.7 & AGN/SF comp. vs. \hii galaxies & 0.446 & 0.160 & 7 vs. 26\\
{\bf{11.3/7.7 vs. 6.2/7.7}} & {\bf{AGN-central vs. AGN-extended}} & 0.627 & $<$0.05 & 15 vs. 18\\
11.3/7.7 vs. 11.3/3.3 & AGN vs. Control & 0.417 & 0.103 & 14 vs. 22\\
\hline
\end{tabular}						 
\caption{The Kolmogorov-Smirnov test probability. In bold we indicate distributions that can be considered as statistically significant different (i.e. p-value$<$0.05).}
\label{tabks}
\end{table}

\begin{table}
\scriptsize
\centering
\begin{tabular}{lccccc}
\hline
X & Y & Sample   & N & R & p-vaule\\
(1)&(2)&(3)&(4) & (5) & (6)\\	
\hline
{\bf{F25/F60}} & {\bf{PAH6.2/PAH7.7}} & {\bf{Control}} & 46 & 0.52 & $<$0.05\\
{\bf{F25/F60}} & {\bf{PAH6.2/PAH7.7}} & {\bf{LIRGs}} & 20 & 0.87 & $<$0.05\\ 
F25/F60 & PAH6.2/PAH7.7 & \hii galaxies & 26 & 0.30 & 0.130\\ 
F25/F60 & PAH11.3/PAH3.3 & Control & 21 & -0.30 & 0.194\\
F25/F60 & PAH11.3/PAH3.3 & LIRGs & 19 & -0.36 & 0.127\\ 
{\bf{[Ne\,III]/[Ne\,II]}} & {\bf{PAH6.2/PAH7.7}} & {\bf{Control}} & 46 & 0.52 & $<$0.05\\
{\bf{[Ne\,III]/[Ne\,II]}} & {\bf{PAH6.2/PAH7.7}} & {\bf{LIRGs}} & 20 & 0.70 & $<$0.05 \\ 
{[Ne\,III]/[Ne\,II]} & PAH6.2/PAH7.7 & \hii galaxies & 26 & 0.36 & 0.07\\ 
{\bf{[Ne\,III]/[Ne\,II]}} & {\bf{PAH11.3/PAH3.3}} & {\bf{Control}} & 20 & -0.63 & $<$0.05\\
{\bf{[Ne\,III]/[Ne\,II]}} & {\bf{PAH11.3/PAH3.3}} & {\bf{LIRGs}} & 19 & -0.53 & $<$0.05 \\ 
\hline
{\bf{F25/F60}} & {\bf{PAH6.2/PAH7.7}} & {\bf{AGN}} & 22 & -0.51 & $<$0.05\\ 
{\bf{[Ne\,III]/[Ne\,II]}} & {\bf{PAH6.2/PAH7.7}} & {\bf{AGN}} & 22 & -0.56 & $<$0.05\\ 
\hline
\end{tabular}						 
\caption{Correlation properties. Columns 1 and 2 correspond to the abscissa and ordinate of the linear fit, respectively. Columns 3 and 4 list the samples considered and the number of galaxies included, respectively. Columns 5 and 6 correspond to the Pearson's correlation coefficient (R) and the null probability (p-value). In bold we indicate fits that can be considered statistically significant (i.e. p-value$<$0.05).}
\label{tabcor}
\end{table}

In the case of the 11.3/3.3 PAH ratio, star-forming galaxies in the control sample have comparable 11.3/3.3 PAH ratios to the AGN sample (see Figs. \ref{iras113} and \ref{ne113}). Indeed, the
11.3/3.3 PAH ratio distributions of AGN and star-forming galaxies are similar according to the KS test (see Table \ref{tabks}). For the control sample of star-forming galaxies, we did not find a strong relationship between the 11.3/3.3 PAH ratio and the dust temperature of the galaxy (see left panel of Fig. \ref{iras113} and Table \ref{tabcor}). However, we find that the harder the radiation field (traced by the [Ne\,II]/[Ne\,III]) the lower the 11.3/3.3 PAH ratio (see left panel of Fig. \ref{ne113} and Table \ref{tabcor}).

These findings are in agreement with the results reported by \citet{Mori12}. They showed that the 11.3/3.3 PAH ratio is mostly controlled by the incident radiation field in molecular clouds and photodissociation regions in the Large Magellanic Cloud. This also confirms the predicted enhancement of the 3.3~$\mu$m PAH band compare with other PAH bands with the increase of the hardness of the radiation field by theoretical models \citep{Rigopoulou20}.

On the other hand, for the AGN sample we find an almost flat relation between the 11.3/3.3 and the [NeIII]/[NeII] ratio (see right panel of Figs. \ref{iras113} and \ref{ne113} and Fig. \ref{ratio_oiiioii} in Appendix \ref{optical}, but also Section \ref{pahdiagram} for more details on this). It is worth noting that only a small fraction of the Sy galaxies in our sample have measurements of the PAH$\lambda$3.3~$\mu$m feature. Furthermore, the mixture of instruments (with different sensitivities), together with the different extinction level for 3.3 and 11.3~$\mu$m can be also significantly affecting this ratio.

We next examine whether there is any relationship of PAH ratios related to the ionization fraction of the molecules such as 11.3/7.7 and 11.3/6.2 ( e.g. \citealt{Draine01, Galliano08,Draine20,Rigopoulou20}) with the dust temperature (F$_{25~\rm \mu m}$/F$_{60~\rm \mu m}$; see Fig. \ref{iras116}) and the [Ne\,III]/[Ne\,II] ratio (see Fig. \ref{ne116}). Using the KS test and comparing the 11.3/7.7 and 11.3/6.2 PAH ratio distributions of both samples, we find statistically significant differences between Seyfert and star-forming galaxies (see Table \ref{tabks}). These differences persist when using LIRGs or \hii galaxies only instead of the entire star-forming control sample. In contrast, the 11.3/7.7 and 11.3/6.2 PAH ratio distributions of star-forming galaxies and those sources classified as AGN/SF composite are similar according to the KS test (see Table \ref{tabks}). 
We also find that the 11.3/7.7 and 11.3/6.2 PAH ratios show similar trends, but there is more scatter when using the 7.7~$\mu$m PAH band. This is related to the fact that this PAH band complex consisting of the 7.42, 7.60 and 7.85 ~$\mu$m features results in an upper limit due to the low detection rate of the 7.42 ~$\mu$m in a significant part of our sample (see Section \ref{PAHFIT}). Therefore, hereafter we discuss only the trends found for 11.3/6.2.

Figures \ref{iras116} and \ref{ne116} show that Sy galaxies with relatively low contribution of the AGN ($<$40\%) have a large scatter for this ratio. However, those sources classified as AGN/SF composites are found in the region where  the star-forming galaxies of the control sample are located. In contrast, the 11.3/6.2 values of galaxies in the control sample are more concentrated around  $\sim$0.7-1.4 indicating that in star-forming galaxies this ratio does not depend on the dust temperature (see left panel of Fig. \ref{iras116}) or the hardness of the radiation field as traced by the [Ne\,II]/[Ne\,III] ratio (see left panel of Fig. \ref{ne116}). It is worth stressing  that our findings are independent of the use of an IR or optical indicator for the hardness of the radiation field (see Fig. \ref{ratio_oiiioii} in Appendix \ref{optical}). This result is in agreement with those reported by \citet{Brandl06} and \citet{Smith07a} using star-forming galaxies. 

We also find that AGN dominated systems have an average 11.3/6.2 PAH ratio ($\sim$2) larger than that of the star-forming galaxies (average 11.3/6.2 PAH ratio $\sim$1; see Fig. \ref{iras116} and \ref{ne116}). The latter is in good agreement with those reported by \citet{Smith07a} (see also \citealt{Wu10,Diamond10}). \citet{Smith07a} also found larger values of the 11.3/7.7 PAH ratio for weak AGN in comparison with \hii galaxies in the SINGs sample. Therefore, using a sample including AGN-dominated systems the present work confirms and extends the result of \citet{Smith07a} based on the SINGs galaxies. \citet{Smith07a} interpreted this as the signature of the small PAH destruction by the hard radiation field of the AGN even in the relatively large aperture probed by Spitzer. Other explanations include, for example, the ionization fraction (see Section \ref{pahdiagram}).

Overall our findings suggest that Sy galaxies with a strong contribution of the AGN have smaller values of 6.2/7.7 PAH ratio and larger values of the 11.3/6.2 PAH ratios (same holds for the 11.3/7.7 PAH ratio but with more scatter). In the case of the 11.3/3.3 PAH ratio, only the galaxies in the control sample show a clear correlation with the hardness of the radiation field.

\section{PAH diagnostic diagrams}
\label{pahdiagram}
In this section we investigate the relative strengths of the PAH bands based on new model grids generated by using theoretically computed PAH spectra based on Density Functional Theory.  We refer the reader to \citet{Rigopoulou20} for more details on how these grids are constructed.

\subsection{The 6.2/7.7 vs. 11.3/7.7 grid}
\label{67_vs_117} 

First we examine the 11.3/7.7 vs. 6.2/7.7 PAH ratio diagram for the AGN sample and the control sample of star-forming galaxies studied here. For clarity, in the case of AGN, we exclude objects with upper limits in any of the 6.2, 7.7 and 11.3~$\mu$m PAH bands. However, in Appendix \ref{PAH_ratio_upper_limits} we show those sources with PAH upper limits and note that they generally show the same trend as those sources with detected PAH features.

\begin{figure*}
\centering
\par{
\includegraphics[width=15.72cm]{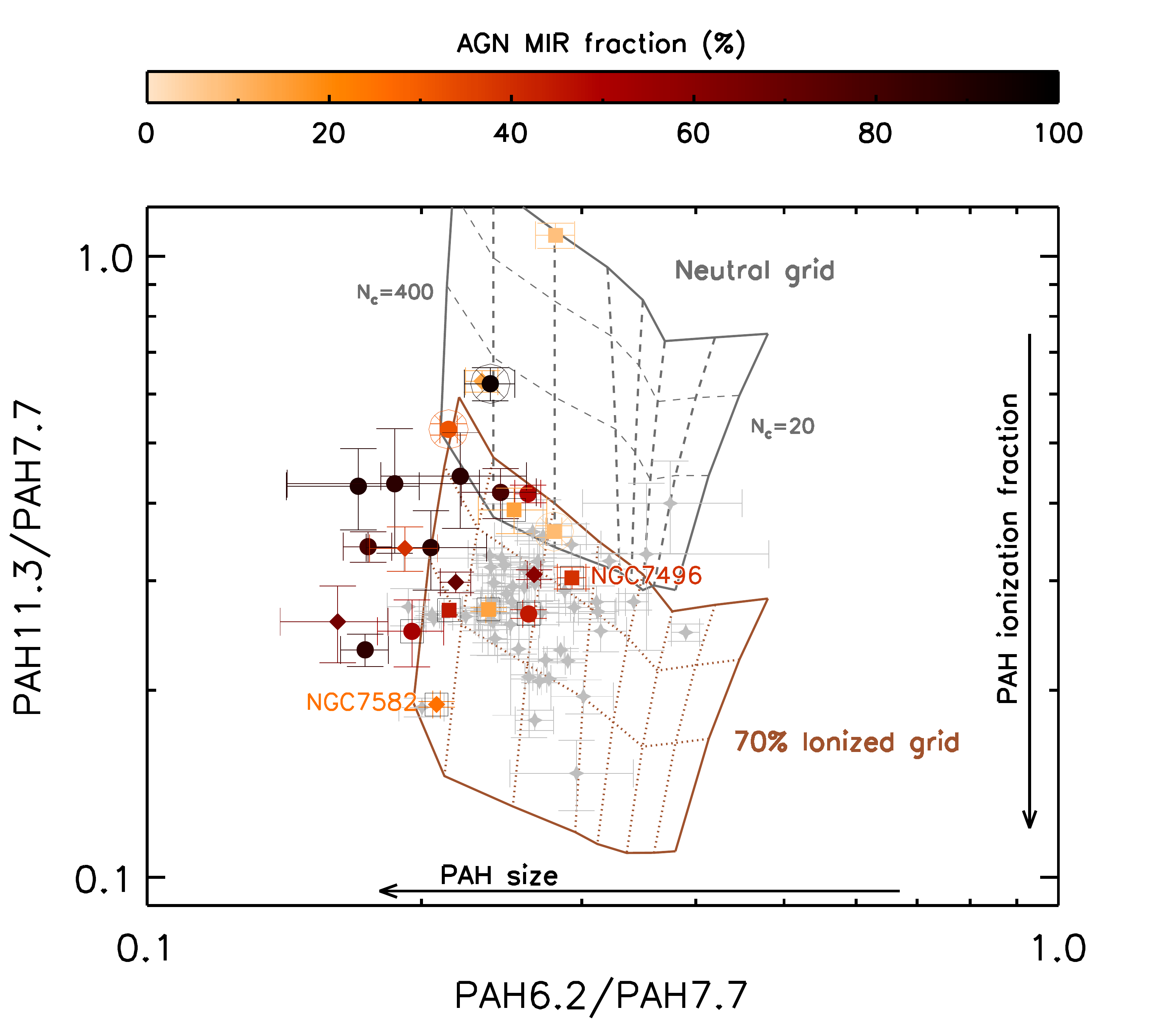}
\par}
\caption{Relative strengths of the 6.2, 7.7 and 11.3~$\mu$m PAH features for those AGN with measured PAH fluxes in these bands (24 sources). Filled color code circles, diamond and squares correspond with the AGN fraction of Sy1, Sy1.8/1.9 and Sy2, respectively. Note that open squares and circles represent galaxies that have been also classified as AGN/SF composite and LINER, respectively. Grey filled stars correspond to star-forming control sample (46 sources). The grey grid corresponds to neutral PAHs ranging from small PAHs (N$_{\rm c}$=20; right side of the grid) to large PAH molecules (N$_{\rm c}$=400; left side of the grid). The top and bottom tracks correspond to molecules exposed to a radiation field of ISRF and 10$^3\times$ISRF, respectively. Dashed grey lines correspond to intermediate number of carbons. The brown grid corresponds to a mixture of 70\% ionized - 30\% neutral PAH molecules exposed to the same radiation fields and for the same number of carbons as in the neutral grid. Dotted brown lines correspond to intermediate number of carbons. Labelled sources correspond to Seyfert galaxies with deep silicates in absorption.}
\label{ratio_diagram_detections}
\end{figure*}

Figure \ref{ratio_diagram_detections} shows a large scatter in the values of the measured 6.2/7.7 PAH ratios for the AGN sample suggesting that the mean size of PAHs, and hence the size of the dust grains, differs widely from source to source (compared to the star-forming galaxies). Indeed, weak AGN or those classified as AGN/SF composites tend to cluster in a narrower part of the diagram closer to the location of star-forming galaxies. Furthermore, most of the measured PAH band ratios in weak AGN and star-forming galaxies are well reproduced by the model grids with PAH molecules showing a $\sim$70\% ionization fraction, medium-to-large molecular size (100$<$N$_{\rm c}$ $<$300) and exposed to a relatively strong radiation field (see Fig. \ref{ratio_diagram_detections}).

Concerning the strength of the AGN, we find that AGN-dominated systems are generally located in the leftmost corner of the grid or even outside of the range of the grid (with measured 6.2/7.7 PAH ratios smaller than 0.25; see Fig. \ref{ratio_diagram_detections} and also \citealt{Herrero20}). The fact that a fraction of AGN-dominated systems lies beyond the values predicted by the current grid suggests that perhaps larger PAH molecules (N$_{\rm c}>$400), are required to reproduce their PAH ratio values. The outcome of the comparison between the observed PAH ratios and the grid suggests that in AGN-dominated systems the PAH emission is consistent with originating in large PAH molecules (N$_{\rm c}>$300-400) (see also Figs. \ref{iras67} and \ref{ne67} in Section \ref{ratios_sec}).

Following the same method as in \citet{Fasano87} (see also \citealt{Peacock83}), we use a two-dimensional extension of the KS test (hereafter 2-D KS test) to compare the PAH ratio distributions employed in the diagnostic diagrams for the various samples used in this study. We find that the values of the  11.3/7.7--6.2/7.7 PAH ratios for the AGN are different to those of LIRGs and \hii galaxies with statistical significance (see Table \ref{tabks}). Note that the difference increases when we use AGN-dominated systems only (see Table \ref{tabks}). However, 
we find that the 11.3/7.7--6.2/7.7 PAH ratios of sources classified as AGN/SF composites and those of star-forming galaxies are similar according to the 2-D KS test (see Table \ref{tabks}).

Extinction can also have a significant impact on PAH band ratios. This is particularly important for the 11.3$\mu$m where this broad PAH feature can be affected by the absorption band (but also emission band in Sy1) from silicate grains which is centred at $\sim$9.8 $\mu$m (e.g. \citealt{Caballero20}). For instance, NGC\,7582, the Seyfert galaxy with the lowest 11.3/7.7 PAH ratio (and also 11.3/6.6 PAH ratio) is also the galaxy with the strongest silicates absorption (S$_{\rm sil}\sim$-1.0) that could attenuate the 11.3~$\mu$m feature. Nevertheless, the 6.2 to 7.7~$\mu$m PAH ratio is less affected by extinction because both bands are relatively close in wavelength and, therefore, their level of extinction is expected to be similar.

Figure \ref{ratio_diagram_detections} shows that the average 11.3/7.7 PAH ratio of AGN dominated systems is larger than for star-forming galaxies (see also Figs. \ref{iras116} and \ref{ne116}). As discussed in Section \ref{ratios_sec}, the previous is more clear for the 11.3/6.2 PAH ratio where the number of upper limits is lower than for the 11.3/7.7 PAH ratio. We remark that the 11.3/7.7 and 11.3/6.2 PAH ratios strongly depend on the ionization fraction but these ratios have a non-negligible dependence on the hardness of the radiation field \citep{Rigopoulou20}. As expected, we find that AGN-dominated systems have harder radiation fields than star-forming galaxies (see Section \ref{ratios_sec}). According to theoretical works (e.g. \citealt{Rigopoulou20}), higher values of the 11.3/7.7 and 11.3/6.2 PAH ratios indicate a larger fraction of neutral molecules and/or lower intensity of the underlying radiation field that the PAH molecules are exposed to. Therefore, the relatively high values of the 11.3/7.7 and 11.3/6.2 PAH ratios in AGN-dominated systems suggest larger fractions of neutral PAH molecules compared to star-forming galaxies. This issue is discussed further in Section \ref{3d_grid}.

\begin{figure*}
\centering
\par{
\includegraphics[width=15.72cm]{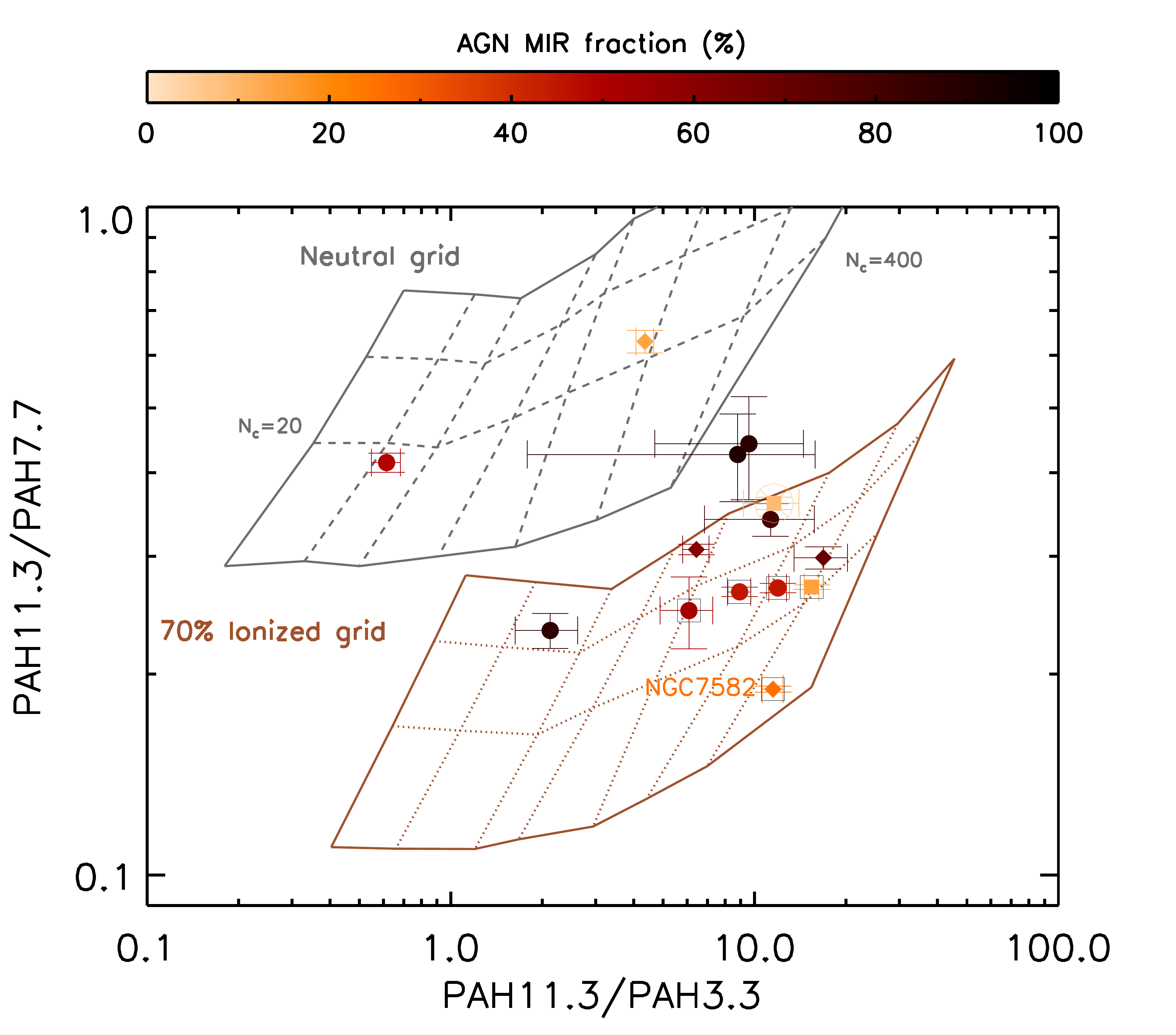}
\par}
\caption{Relative strengths of the 3.3, 7.7 and 11.3~$\mu$m PAH features for various astronomical sources with measured PAH fluxes in these bands (14 sources). Filled color code circles, diamond and squares correspond with the AGN fraction of Sy1, Sy1.8/1.9 and Sy2, respectively. Note that open squares and circles represent galaxies that have been also classified as AGN/SF composite and LINER, respectively. The grey grid corresponds to neutral PAHs ranging from small PAHs (N$_{\rm c}$=20; left side of the grid) to large PAH molecules (N$_{\rm c}$=400; right side of the grid). The top and bottom tracks correspond to molecules exposed to a radiation field of ISRF and 10$^3\times$ISRF, respectively. Dashed grey lines correspond to intermediate number of carbons. The brown grid corresponds to a mixture of 70\% ionized - 30\% neutral PAH molecules exposed to the same radiation fields and for the same number of carbons as in the neutral grid. Dotted brown lines correspond to intermediate number of carbons. Note that only a small fraction of the Sy galaxies in our sample have available the PAH$\lambda$3.3~$\mu$m feature detected. We labelled NGC\,7582, one of the galaxies in our sample with a relatively deep silicate feature.} 
\label{ratio_diagram_113}
\end{figure*}

\subsection{The 11.3/7.7 vs. 11.3/3.3 grid}
\label{117_vs_113}
\citet{Maragkoudakis20} found that the 11.3~$\mu$m and 3.3~$\mu$m bands can be used to trace the size of PAH molecules. \citet{Rigopoulou20} confirmed the use of the 11.3/3.3 PAH ratio as a tracer of PAH size but noted the strong dependence of the 3.3~$\mu$m PAH on the intensity of the radiation field (see their Figure 1). Indeed, \citet{Rigopoulou20} showed that while the peak intensity of all PAH features is dependant on the energy of the photon that is absorbed, the variation in the 11.3/3/3.3 PAH ratio is much higher than that of the 6.2/7.7 PAH ratio. In addition, \citet{Mori12} investigated molecular clouds and photodissociation regions in the Large Magellanic Cloud\footnote{Group A targets in \citet{Mori12} where the destruction of PAHs is expected to be inefficient.} and they found that the 11.3/3.3 PAH ratio is a useful tracer for the incident radiation field. This is also in good agreement with our result for the sample of star-forming galaxies, that the harder the radiation field the lower the 11.3/3.3 PAH ratio (see left panel of Fig. \ref{ne113}).

We next compare the observational data with the 11.3/7.7 vs. 11.3/3.3 theoretical grids (see Fig. \ref{ratio_diagram_113}). In agreement with the results presented in Section \ref{67_vs_117}, AGN tend to be located in the region of the harder radiation field and larger PAH molecules. We also find that 11.3/3.3--11.3/7.7 distributions of the AGN and control samples are similar according to the 2-D KS test (see Table \ref{tabks}). This is likely due to the similar values of the 11.3/3.3 PAH ratio of AGN compared to those of the star-forming galaxies (see also Section \ref{ratios_sec}). In addition, we note that AGN show a relatively flat relation with the various hardness tracers (see Section \ref{ratios_sec}). We remark that the 11.3/3.3 PAH can be affected by the hardness of the radiation field (low values for strong radiation fields) but also by the PAH molecular size (high values for large molecules), which could balance this ratio in AGN (see Section \ref{3d_grid} for further discussion). However, only a small fraction of the Sy galaxies in our sample have PAH$\lambda$3.3~$\mu$m feature measurements. The results presented here indicate that the 11.3/3.3 PAH ratio does not vary with the strength of the AGN, but we cannot rule out the possibility that it is due to the mixture of instruments (with different sensitivities) used for measuring the PAH$\lambda$3.3~$\mu$m fluxes. Note that the different extinction level for 3.3 and 11.3~$\mu$m can be also significantly affecting this ratio.

\begin{figure*}
\centering
\includegraphics[width=14.8cm]{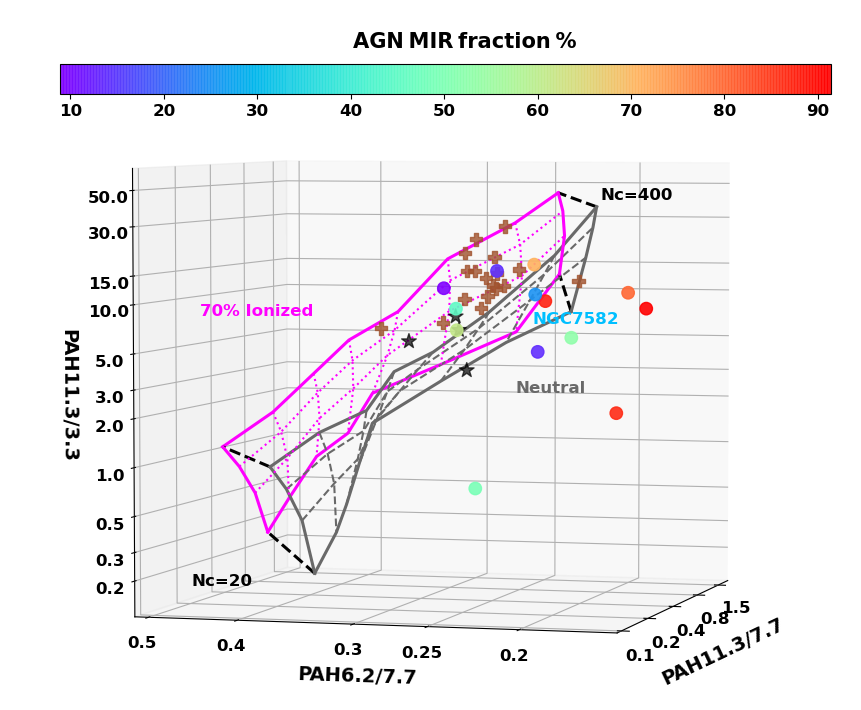}
\caption{3-D PAH diagnostic diagram. Relative strengths of the 3.3, 6.2, 7.7 and 11.3~$\mu$m PAH features for various astronomical sources with measured PAH fluxes in these bands. Color code circles correspond to central regions of AGN (14 sources). Black stars and brown crosses represent \hii galaxies and LIRGs (22 sources), respectively.
The grey grid corresponds to neutral PAHs ranging from small PAHs (N$_{\rm c}$=20; left side of the grid) to large PAH molecules (N$_{\rm c}$=400; right side of the grid). The top and bottom tracks correspond to molecules exposed to a radiation field of ISRF and 10$^3\times$ISRF, respectively. Dashed grey lines correspond to intermediate number of carbons. The magenta grid corresponds to a mixture of 70\% ionized - 30\% neutral PAH molecules exposed to the same radiation fields and for the same number of carbons as in the neutral grid. Dotted magenta lines correspond to intermediate number of carbons. The solid black line indicates the interception between both grids. Note that only a small fraction of the Sy galaxies in our sample have the PAH$\lambda$3.3~$\mu$m feature measurements. We labelled NGC\,7582, one of the galaxies in our sample with a relatively deep silicate feature.}
\label{ratio_diagram_3d}
\end{figure*}

\subsection{The 3-D PAH plane}
\label{3d_grid}

\citet{Rigopoulou20} presented a 3-D grid by combining the two 2-D diagrams discussed earlier (see also Fig. 9 of \citealt{Rigopoulou20}) to fully characterise the PAH molecular size, ionization fraction and hardness of the radiation field.

In Figure \ref{ratio_diagram_3d}, we plot the three PAH band ratios for the samples studied here on the PAH 3-D grid. The plot shows that the measured PAH ratios of \hii galaxies, LIRGs and relatively weak AGN mainly fall in the 70\% ionized grid plane (i.e. mixture of 70\% ionized - 30\% neutral PAH molecules) and intermediate values of the hardness of the radiation field.

From this plot it is clear that the measured PAH ratios of AGN-dominated systems are different to star-forming galaxies and relatively weak AGN. The measured PAH ratios of the former can be explained with a smaller fraction of small PAH molecules, and a smaller fraction of ionized PAHs (see also Sections \ref{ratios_sec} and \ref{67_vs_117}). As previously mentioned, the latter is in agreement with previous works using relatively weak AGN (e.g. \citealt{Smith07a,Diamond10}). 

However, assuming that PAH emission is produced by the UV radiation field, in principle, we would expect a higher ionization fraction for the harder UV radiation field of AGN. A possible explanation for the larger 11.3/7.7 (and 11.3/6.2) PAH ratio in AGN-dominated systems is the variation in the PAH molecular size. This could produce changes of short-wavelength features relative to the 11.3~$\mu$m feature~(e.g. \citealt{Draine01,Galliano08}). Alternatively, the 11.3/7.7 (and 11.3/6.2) PAH ratio can also be affected by hydrogenated PAH molecules. \citet{Reach00} proposed that hydrogenation, which produces an increase in the hydrogen-to-carbon ratio (H/C) in PAH molecules, could explain the extreme 11.3/7.7 ratios, like the one observed in SMB B1 No 1 (see also \citealt{Li20} for a review). In a forthcoming paper (Garc\'ia-Bernete et al., in prep) we will further investigate how the H/C ratio of PAH molecules affects their band ratios. Finally, the shielding of the PAH molecules by the molecular hydrogen can also be responsible for the increase of neutral species. We discuss this possibility further in Section \ref{central_extended}.

The results of the comparison with the PAH grid is in good agreement with our findings in Section \ref{ratios_sec}. Our findings suggest that AGN-dominated systems generally tend to have a greater fraction of larger molecules, which can be explained by the preferential destruction of small PAH molecules due to X-rays and/or shocks from the AGN (e.g. \citealt{Voit92,Siebenmorgen04}). However, star-forming galaxies show the opposite trend, the harder the radiation field the larger the value of the 6.2/7.7 PAH ratio (Fig. \ref{iras67} and \ref{ne67} in Section \ref{ratios_sec}). The trend can be explained by the ionization potential of PAH molecules\footnote{Note that the selective destruction of large PAH molecules is an unlikely explanation since large PAH molecules are expected to be more stable (e.g. \citealt{Allain96A}, \citealt{Bauschlicher08} and \citealt{Murga20}).}. Smaller PAH molecules have larger values of the ionization potential (see e.g. \citealt{Ruiterkamp05}). Consequently, a relatively harder environment could produce an enhancement of the 6.2~$\mu$m feature with respect to the 7.7~$\mu$m PAH in star-forming galaxies (assuming the same PAH molecular size).

We conclude that the presence of an AGN is likely to have a significant impact on the surrounding PAH population by destroying small PAH molecules 
and favouring large and neutral species. High angular resolution data of the entire $\sim$3-28~$\mu$m range with the James Webb Space
Telescope will enable a greater accuracy in the measured PAH ratios (see Section \ref{PAHFIT}) and elucidate the impact of the central AGN in the properties of the PAH molecules located in its vicinity.

\subsection{PAH ratios in central and extended AGN regions}
\label{central_extended}

As discussed in Section 3,
the majority of the Seyfert galaxies in our sample (36/50), were observed in the spectral mapping mode. To further investigate the impact of the AGN on the PAH ratios as a function of distance from the central source, we consider two distinct regions: the central emission and the diffuse/extended emission in the galaxy disk (see Fig. \ref{apertures}; see also \citealt{Diamond10}).  The central spectra are extracted assuming a point-like source (see Section \ref{observations}). For those AGN with extended emission present in the spectral mapping data (i.e. 50\% of the sub-sample with spectral mapping mode observations; 18/36 galaxies), we derive the extended emission spectra. These spectra are computed as the difference between the total minus the central (point-like source) extractions. For comparison, we also apply the same method to 2 LIRGs from our control sample of star-forming galaxies with extended MIR emission (see Fig. \ref{apertures}).

\begin{figure*}
\centering
\par{
\includegraphics[width=15.72cm]{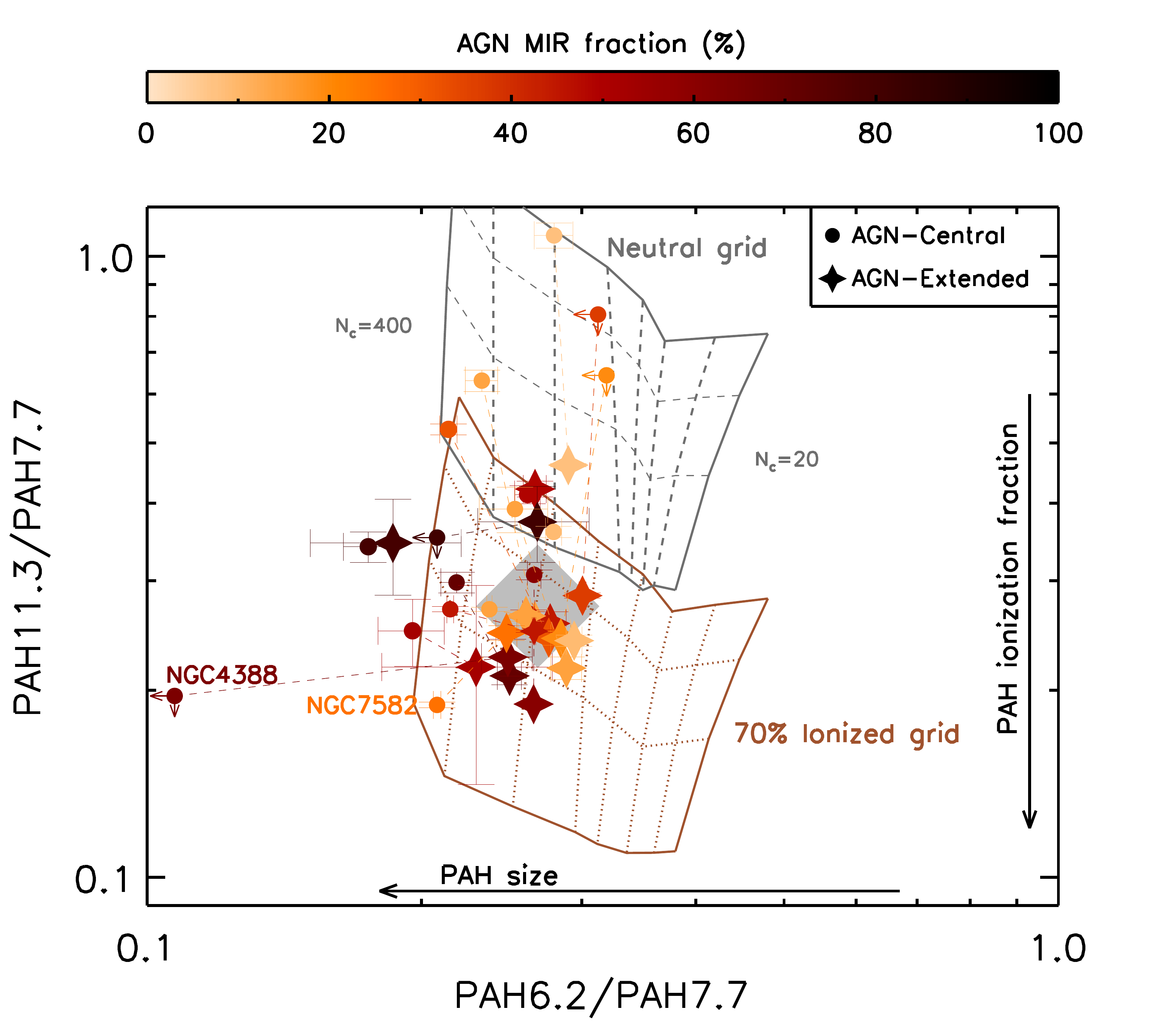}
\par}
\caption{Relative strengths of the 6.2, 7.7 and 11.3~$\mu$m PAH features for those AGN with extended emission (18 sources). Color code symbols indicate the AGN fraction of the source. Filled circles and stars correspond with the circumnuclear and extended emission, respectively. The model grids are the same as in Fig. \ref{ratio_diagram_detections}. Labelled sources correspond to Seyfert galaxies with deep silicates in absorption.}
\label{ratio_diagram_circumnuclear}
\end{figure*}

In Fig. \ref{ratio_diagram_circumnuclear} we present PAH band ratios for those AGN where it was possible to measure nuclear (central) and extended PAH emission (18 galaxies of the sample). Note that we use the 6.2/7.7 vs. 11.3/7.7 grid due to the lack of measurements of the PAH$\lambda$3.3~$\mu$m for the nuclear$/$extended regions (see Section \ref{117_vs_113}). We also compared the 6.2/7.7--11.3/7.7 PAH ratio distributions of the central and extended regions of Seyfert galaxies using the 2-D KS test and found statistically significant differences between them (see Table \ref{tabks}). The PAH properties in the extended regions (a few kpc scales) of Seyfert galaxies are similar to those seen in star-forming galaxies (see Fig. \ref{ratio_diagram_circumnuclear}). This is in good agreement with previous works comparing \hii galaxies and off-nuclear regions of Seyfert galaxies (e.g. \citealt{Diamond10}). Furthermore, \citet{Diamond10} also found larger values of the 11.3/7.7 PAH ratio for AGN in comparison with their off-nuclear regions and \hii galaxies. The PAH emission in the central regions of AGN appears to different, favouring larger and/or more neutral molecules. For the majority of the galaxies (15/18), the measured PAH ratios in the extended regions of Sys are consistent with PAH ionization fractions of~$\sim$50-70\% and smaller PAH sizes.

As previously discussed (see Section \ref{3d_grid}), the destruction of the small PAH molecules can have a significant effect on the 11.3/7.7 PAH ratio. For a given ionization, larger PAH molecules tend to emit more at longer wavelengths therefore large and neutral PAH molecules will produce larger 11.3/7.7 and smaller 6.2/7.7 PAH ratios (e.g. \citealt{Draine01,Li20,Rigopoulou20}). Therefore, PAH molecule destruction could result in an increase of the 11.3/7.7 PAH ratio, but also a change of the number of hydrogens per carbon in PAH molecules (e.g. \citealt{Li20}; see also Section \ref{3d_grid}). 

The distribution of the molecular hydrogen in AGN is also likely to influence PAH emission. Seyfert galaxies tend to have a large fraction of their molecular gas concentrated toward their centres (e.g. \citealt{Maiolino97}). As a result, the molecular gas column density would be smaller at larger distances from the central engine. The lower molecular gas column densities in the extended emission of these galaxies could, therefore, explain the observed increase in the fraction of ionized PAH molecules (e.g. \citealt{Herrero20}). 
In contrast, the higher fraction of large and neutral PAHs seen in the central regions of AGN (compared to their extended regions) points to the possible destruction of the smaller PAH molecules by the harsh environments present in the innermost regions of these galaxies.
The increase of neutral PAH molecules can be explained by the shielding effect 
(protection of PAH molecules by the molecular hydrogen present in these regions) due to the large concentration of molecular gas in the centres of Seyfert galaxies as suggested by (e.g. \citealt{Rigopoulou02,Herrero14,Herrero20}).

\begin{figure}
\centering
\par{
\includegraphics[width=8.72cm]{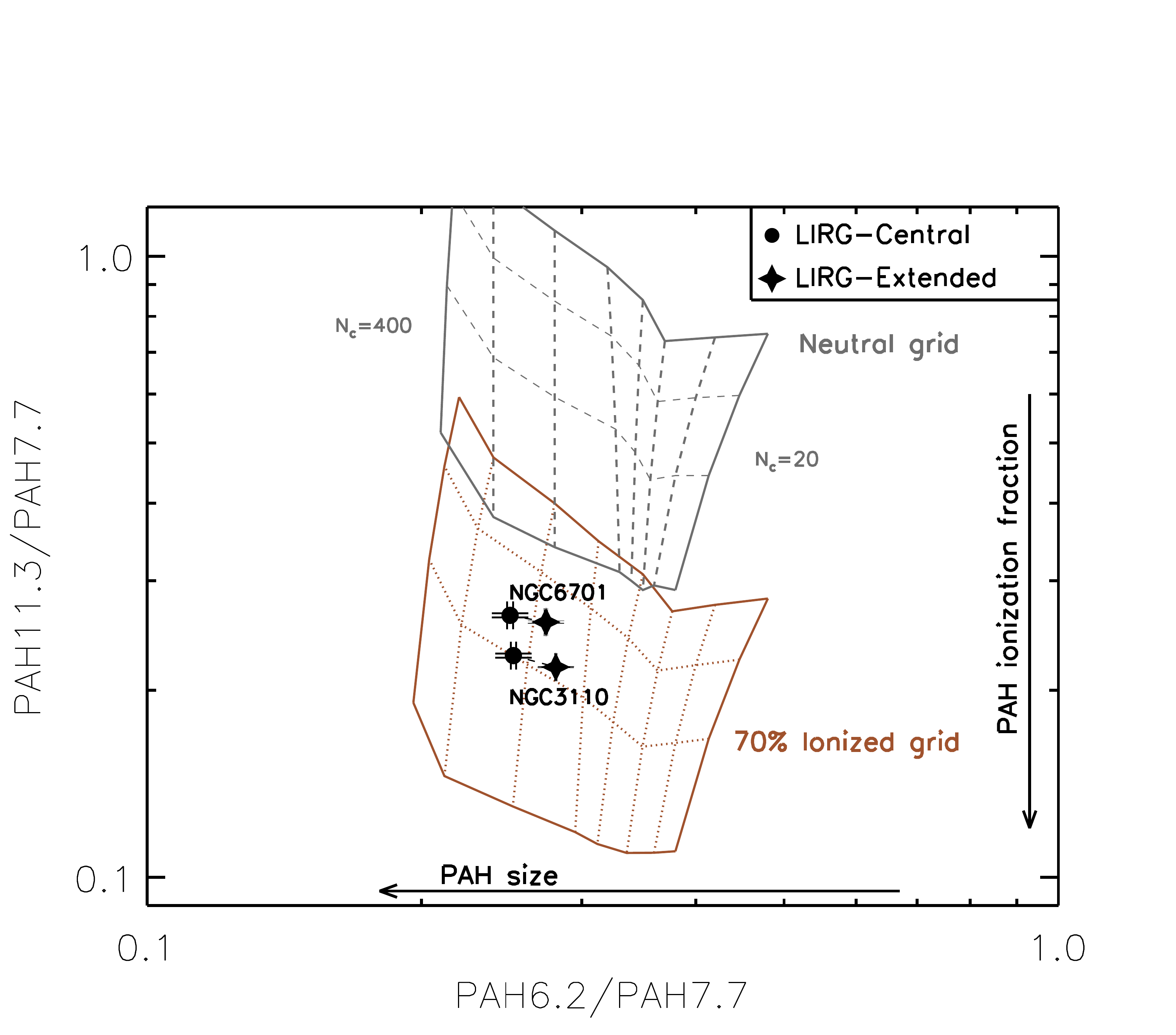}
\par}
\caption{Same as Fig. \ref{ratio_diagram_circumnuclear} but for the LIRG galaxies (i.e. NGC\,3310 and NGC\,6701). The model grids are the same as in Fig. \ref{ratio_diagram_detections}.}
\label{ratio_diagram_circumnuclear_lirgs}
\end{figure}

For comparison, we also investigate the central and extended PAH ratios of the 2 LIRGs that have no detectable AGN (see Fig. \ref{ratio_diagram_circumnuclear_lirgs}). We find rather small variations in their central and extended PAH 6.2/7.7 PAH ratios, but practically the same 11.3/7.7 PAH ratios indicating the same fraction of neutral vs. ionized PAH molecules is present in both central and extended regions. 

 In summary, our findings suggest that the radiation fields of AGN might be altering the molecular size distribution of PAH molecules in their central regions. The changes in the PAH properties due to the presence of the AGN are likely to happen close to the central engine, therefore, high angular observations of the innermost regions of AGN and star-forming galaxies are needed to further investigate this hypothesis.
 
\begin{figure*}
\centering
\includegraphics[width=17cm]{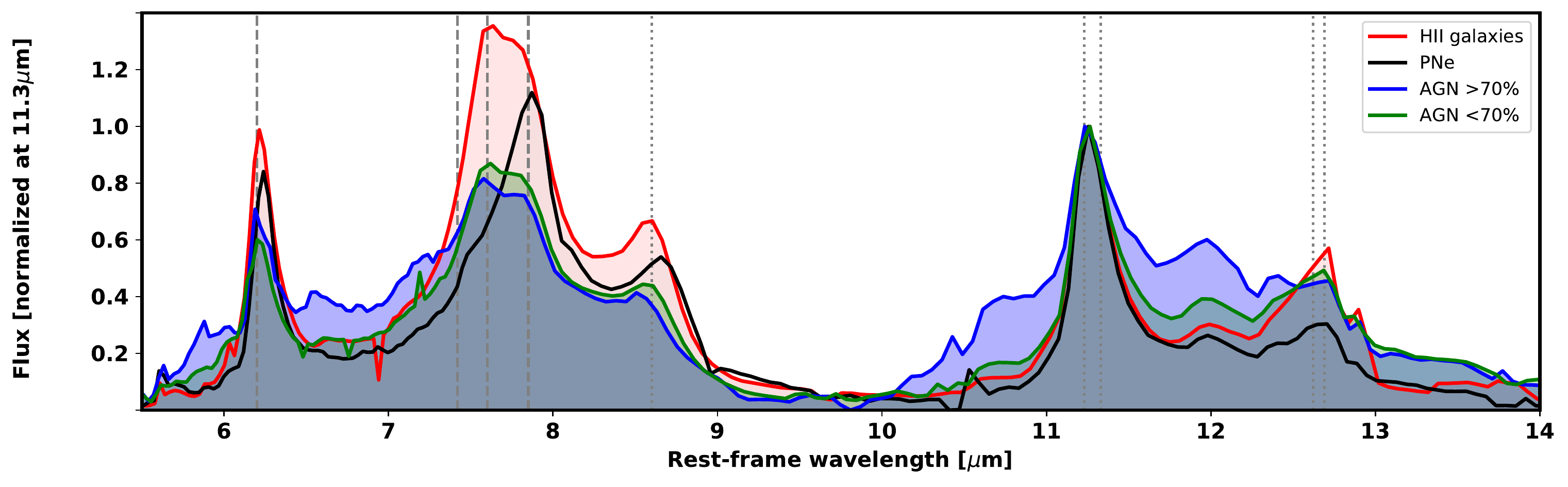}
\caption{Stacked PAH spectra. The PAH spectra of the various sources fitted in this work. Red, black, blue and green lines correspond to star-forming galaxies (\citealt{Brandl06}), planetary nebula, AGN-dominated Seyfert galaxies (i.e. AGN MIR contribution$>$70\%) and Seyfert galaxies with a relatively low contribution of the AGN (i.e. AGN MIR contribution$<$70\%). The grey vertical dashed lines correspond to the most important PAH bands (i.e. 6.2, 7.42, 7.60, 7.85, 8.61, 11.23, 11.33, 12.62 and 12.69~$\mu$m features).}
\label{stacked_spectra}
\end{figure*}

\section{Averaged properties of PAH spectral features}
\label{pahband}

In this section we compare {\it the average PAH band spectra} of AGN, star-forming galaxies (\citealt{Brandl06}) and those of local planetary nebulae in the Magellanic Clouds. 

For this comparison, we include PNe sources since it is well known that PAH band profiles in these objects vary significantly in comparison with other sources (e.g. \citealt{Peeters02,Shannon19}). In addition, due to the large aperture of Spitzer/IRS, local objects like PNe allow us to better isolate the effects of temperature or metallicity, because are less contaminated by the surrounding medium of the UV source than in galaxies. Therefore, these sources are more likely to show clear changes in the PAH bands for different temperatures/metallicity. By comparing the average PAH spectra, we find that the shape of 11.3$\mu$m is similar for PNe, weak AGN (i.e. AGN MIR contribution$<$70\%) and star-forming galaxies (black, green and red solid line of Fig. \ref{stacked_spectra}, respectively). In contrast, there is a marked difference in the shape and strength of the 7.7~$\mu$m PAH complex. In the case of the PNe the feature is narrower and shifted to longer wavelengths in comparison to SF and active galaxies. In a recent study, \citet{Shannon19} found a dependence of the 7.7~$\mu$m PAH band peak shift with PAH molecular size, charge, temperature and aliphatic and nitrogen content. However, the origin of the shift is yet unclear. 

The shape of the PAH averaged spectra of Seyfert galaxies with relatively low contribution of the AGN (AGN MIR contribution$<$70\%) and star-forming galaxies are similar. However, although star-forming galaxies have prominent 6.2 and 7.7~$\mu$m features, these features appear to be attenuated for both AGN subgroups (AGN MIR contribution$<$70\% and AGN MIR contribution$>$70\%). In agreement with the results presented in Section \ref{results}, the average PAH spectra of Seyfert galaxies with AGN MIR contribution$<$70\% and AGN MIR contribution$>$70\% show noticeable differences: AGN-dominated systems (AGN MIR contribution$>$70\%; blue solid line of Fig. \ref{stacked_spectra}) show broader PAH band profiles than those in weaker ones (AGN MIR contribution$<$70\%), star-forming galaxies or PNe. Therefore, to reproduce the broad PAH 11.3~$\mu$m band profile observed in AGN-dominated systems larger PAH molecules are needed. However, a significant fraction of small and ionized molecules would be necessary to enhance the 6.2 and 7.7~$\mu$m PAH bands to reach the same levels as those seen in star-forming galaxies. We refer the reader to \citet{Rigopoulou20} for further details on the theoretical PAH profiles. Summarising, the outcome of the comparisons of the averaged PAH spectra of various sources is in agreement with the findings from our earlier analysis presented in Section \ref{pahdiagram}.

\section{Conclusions}
\label{conclusions}

Using a sample of Seyfert galaxies (50 galaxies; D$_{\rm L}<$100~Mpc), we investigated the PAH emission originating in the central region of the AGN and compared the results with a control sample of star-forming galaxies (22 LIRGs and 30 \hii galaxies). In addition, we studied the difference between the central and extended PAH properties. To do so, we used newly developed PAH diagnostic grids derived from theoretical spectra. This allows disentangling between properties of the PAH molecules such as the size, ionization fraction and hardness of the underlying radiation field.  We compared the predicted and observed PAH ratios using Spitzer/IRS IR spectral data for the sample of Seyfert galaxies and a control sample of star-forming galaxies. The main results are as follows.

   \begin{enumerate}
   
\item Hardness of the radiation field. In agreement with theoretical predictions, we find that non-AGN sources have larger 11.3/3.3 PAH ratios for softer radiation fields as traced by the [Ne\,II]/[Ne\,III] ratio. In contrast, AGN tend to have a flat relation between 11.3/3.3 PAH ratios and the hardness of the radiation field. In general, we measure larger values of the 6.2/7.7 PAH ratio for harder radiation fields in star-forming galaxies compared with AGN-dominated systems.
   
\item PAH molecular size. We find that the 6.2/7.7 PAH ratio values of Seyfert galaxies and the star-forming galaxies are statistically different. In general, the measured 6.2/7.7 PAH ratio of AGN-dominated systems are smaller than those of star-forming galaxies. A fraction of these AGN-dominated systems lies beyond the range predicted by the grid suggesting that their PAH ratios would require the presence of large PAH molecules (N$_{\rm c}>$400).

\item PAH Ionization fraction. AGN-dominated systems have a higher average 11.3/6.2 (and 11.3/7.7) than star-forming galaxies. This indicates a larger fraction of neutral PAH molecules in AGN-dominated systems compared with the control sample of star-forming galaxies. This suggests that the nuclear molecular gas concentrated in AGN centres could be playing a role in protecting their nuclear PAH molecules (shielding), or that ionized PAH molecules can be destroyed more easily than neutral ones in hard environments.

\item Weak AGN or sources classified as AGN/SF composites and star-forming galaxies tend to cluster in a narrow part of 11.7/7.7 vs. 6.2/7.7 diagram. In general, they are well reproduced by the model grids with PAH molecules with $\sim$70\% ionization fraction, medium-to-large molecular size (100$<$N$_{\rm c}$ $<$300) and exposed to a relatively strong radiation field ($>$ ISRF).

\item We find that in the extended emission of AGN, the PAH ratios are closer to star-forming galaxies than their central regions. The measured PAH ratios of the extended emission of Seyfert galaxies can be explained with larger fractions of ionized and small PAH molecules than in their nuclear regions.

\item We find tentative evidence that Seyfert-like AGN-dominated systems can introduce changes in the PAH population that are significant even on kpc scales. This is particularly important for AGN-dominated systems which could potentially be responsible for the destruction of small PAH molecules, and affect the emission of large and neutral ones. Therefore, caution must be applied when using PAH bands as star-formation rate indicators because of the impact of extremely harsh environments on PAHs. \\

In the future, detailed studies such as this carried out across the whole NIR--MIR spectral range could be performed for samples of more distant and faint galaxies using the unprecedented spatial resolution (improvement of a factor of 10 with respect to Spitzer/IRS) and sensitivity that will be afforded by the James Webb Space Telescope. Such studies will enable the detailed study of PAH emission as a function of distance from the central AGN and investigate changes in the PAH population exposed to extremely harsh environments.

   \end{enumerate}

\section*{Acknowledgements}

We thank the anonymous referee for his$/$her useful comments.

IGB and DR acknowledge support from STFC through grant ST/S000488/1. DR and BK acknowledge support from the University of Oxford John Fell Fund. AAH acknowledges support from  PGC2018-094671-B-I00 (MCIU/AEI/FEDER,UE). AAH and MPS work was done under project  No. MDM-2017-0737 Unidad de Excelencia “Mar\'ia de Maeztu”- Centro de Astrobiolog\'ia (INTA-CSIC).
We thank Jack Gallimore for the modified version of the PAHFIT that include the silicate features, and for his useful comments. 

This work is based [in part] on archival data obtained with the Spitzer Space Telescope, which is operated by the Jet Propulsion Laboratory, California Institute of Technology under a contract with NASA. This research has also made use of the NASA/IPAC Extragalactic Database (NED), which is operated by the Jet Propulsion Laboratory, California Institute of Technology under a contract with NASA.

\section*{DATA AVAILABILITY}
The data underlying this article will be shared on reasonable request to the corresponding author.







\appendix
\begin{table*}
\centering
\begin{tabular}{lcccc}
\hline
Name &	PAH$\lambda$3.3$\mu$m & 	Aperture & 	Telescope& Reference \\
 (1)&(2)&(3)&(4)&(5)\\	
\hline
Mrk\,352          & ...& ...& ...&...\\
NGC\,931          & $<$11 & 2.5\arcsec$\times$0.8\arcsec& IRTF/SpeX& c    \\ 
NGC\,1097         & $<$1 &  0.45\arcsec$\times$0.45\arcsec& GNIRS& d  \\ 
NGC\,1275         & $<$20 & 4.38\arcsec$\times$4.38\arcsec& AKARI&b   \\
NGC\,1365		  & 124$\pm12$ & 4.4\arcsec radius aperture  &AKARI & e    \\
ESO\,548-G081     &  ...& ...& ...&...\\
NGC\,1566         & 610$\pm67$ & 13.5\arcsec radius aperture &ISO  &a \\  
NGC\,1667         & $<$2 & 3.0\arcsec$\times$1.6\arcsec& IRTF/SpeX& f\\
MCG-01-13-025     & ...& ...& ...&...\\
4U0517$+$17       &  ...& ...& ...&...\\
ESO\,362-G018     &  ...& ...& ...&...\\
UGC\,3478         &  ...& ...& ...&...\\
Mrk\,6            & $<$7 & 4.38\arcsec$\times$4.38\arcsec& AKARI&b   \\
Mrk\,79           & $<$7 & 4.38\arcsec$\times$4.38\arcsec& AKARI&b   \\
NGC\,2992		  & 9$\pm2$ & 4.38\arcsec$\times$4.38\arcsec& AKARI&b   \\
NGC\,3227	      & 7$\pm1$ & 1.8\arcsec$\times$0.8\arcsec & IRTF/SpeX& g\\
NGC\,3516         & $<$5& 4.38\arcsec$\times$4.38\arcsec& AKARI&b   \\
NGC\,3783		  & $<$12 & 4.38\arcsec$\times$4.38\arcsec& AKARI&b   \\
NGC\,3786         & $<$6 & 3.0\arcsec$\times$1.6\arcsec  & IRTF/SpeX& h\\
UGC\,6728         & $<$1 & 4.38\arcsec$\times$4.38\arcsec& AKARI&b   \\
NGC\,3982         & $<$6 & 2.5\arcsec$\times$0.8\arcsec  & IRTF/SpeX& c    \\ 
NGC\,4051         & 20$\pm8$ & 7.25\arcsec$\times$7.25\arcsec& AKARI& i   \\
NGC\,4138		  &  ...& ...& ...&...\\
NGC\,4151		  & 13$\pm9$ & 4.38\arcsec$\times$4.38\arcsec& AKARI&b   \\
NGC\,4235         & 3$\pm1$ &3.0\arcsec$\times$1.6\arcsec  & IRTF/SpeX& h\\
NGC\,4253         & 7$\pm2$ & 1.8\arcsec$\times$0.8\arcsec  & IRTF/SpeX& g\\
NGC\,4388         & 8$\pm2$ & 4.38\arcsec$\times$4.38\arcsec& AKARI&b   \\
NGC\,4395		  &  ...& ...& ...&...\\
NGC\,4501		  & $<$3 & 3.0\arcsec$\times$1.6\arcsec  & IRTF/SpeX& f\\
NGC\,4593         & 10$\pm9$ & 7.25\arcsec$\times$7.25\arcsec& AKARI& i   \\
NGC\,4639         &  ...& ...& ...&...\\
ESO\,323-G77      &  ...& ...& ...&...\\
NGC\,5033         & 358$\pm44$ & 13.5\arcsec radius aperture&ISO  &a \\  
NGC\,5135         & 79$\pm4$ & 7.3\arcsec radius aperture& AKARI& j\\
MCG-06-30-015 	  & 59$\pm41$ &13.5\arcsec radius aperture &ISO  &a \\  
IC\,4329A         & 17$\pm6$ & 4.38\arcsec$\times$4.38\arcsec& AKARI&b   \\
NGC\,5347         & 6$\pm1$ & 3.0\arcsec$\times$1.6\arcsec  & IRTF/SpeX& f\\  
NGC\,5506		  & 15$\pm6$ & 4.38\arcsec$\times$4.38\arcsec& AKARI&b   \\
NGC\,5548         & 4$\pm2$ & 4.38\arcsec$\times$4.38\arcsec& AKARI&b   \\
NGC\,5953         & 142$\pm39$ & 13.5\arcsec radius aperture&ISO  &a \\
IRAS\,18325-5926  &  ...& ...& ...&...\\                           
NGC\,6814		  &  ...& ...& ...&...\\
NGC\,6860         &  ...& ...& ...&...\\
NGC\,6890         &  ...& ...& ...&...\\
NGC\,7130         & 39$\pm4$ & 4.4\arcsec radius aperture  &AKARI & e    \\
NGC\,7314		  & 91$\pm32$ & 13.5\arcsec radius aperture&ISO  &a \\  
NGC\,7469         & 80$\pm7$ & 4.38\arcsec$\times$4.38\arcsec& AKARI&b   \\
NGC\,7496         &  ...& ...& ...&...\\
NGC\,7582		  & 84$\pm7$ & 4.38\arcsec$\times$4.38\arcsec& AKARI&b   \\ 
NGC\,7590         &  ...& ...& ...&...\\
\hline
\end{tabular}						 
\caption{Summary of 3.3~$\mu$m PAH fluxes. The fluxes and errors are in units of 10$^{-14}$~erg$^{-1}$~cm$^{-2}$. Note that we use ISO measurements as upper limits due to the large aperture extracted spectra. References: a) \citealt{Clavel00}; b) \citealt{Castro14}; c) \citealt{Oi10}; d) \citealt{Mason07}; e) \citealt{Inami18}; f) \citealt{ImanishiandAlonsoHerrero04}; g) \citealt{Ardila03}; h) \citealt{ImanishiandWada04}; i) \citealt{Kim19}; j) \citealt{Yamada13}. }
\label{tab2}
\end{table*}

\section{PAH$\lambda$3.3~microns feature measurements}
\label{literature}
In Table \ref{tab2}, we present the compiled PAH$\lambda$3.3~$\mu$m feature measurements available from the literature for our AGN sample.

\section{MIR spectral modelling using PAHFIT}
\label{pahfit_figs}
In Fig. \ref{appendix_pahfit_fig}, we show the MIR spectral modelling using PAHFIT.
Note that for some AGN-dominated sources with a strong contribution of the silicate in
emission PAHFIT does not produce satisfactory fits (namely ESO\,548-G081, IC\,4329A, MCG-01-13-025, Mrk\,352, NGC\,3783, NGC\,4151, NGC\,4235, NGC\,6860, NGC\,6814 and UGC\,6728).

\begin{figure*}
\centering
\par{
\includegraphics[width=7.62cm]{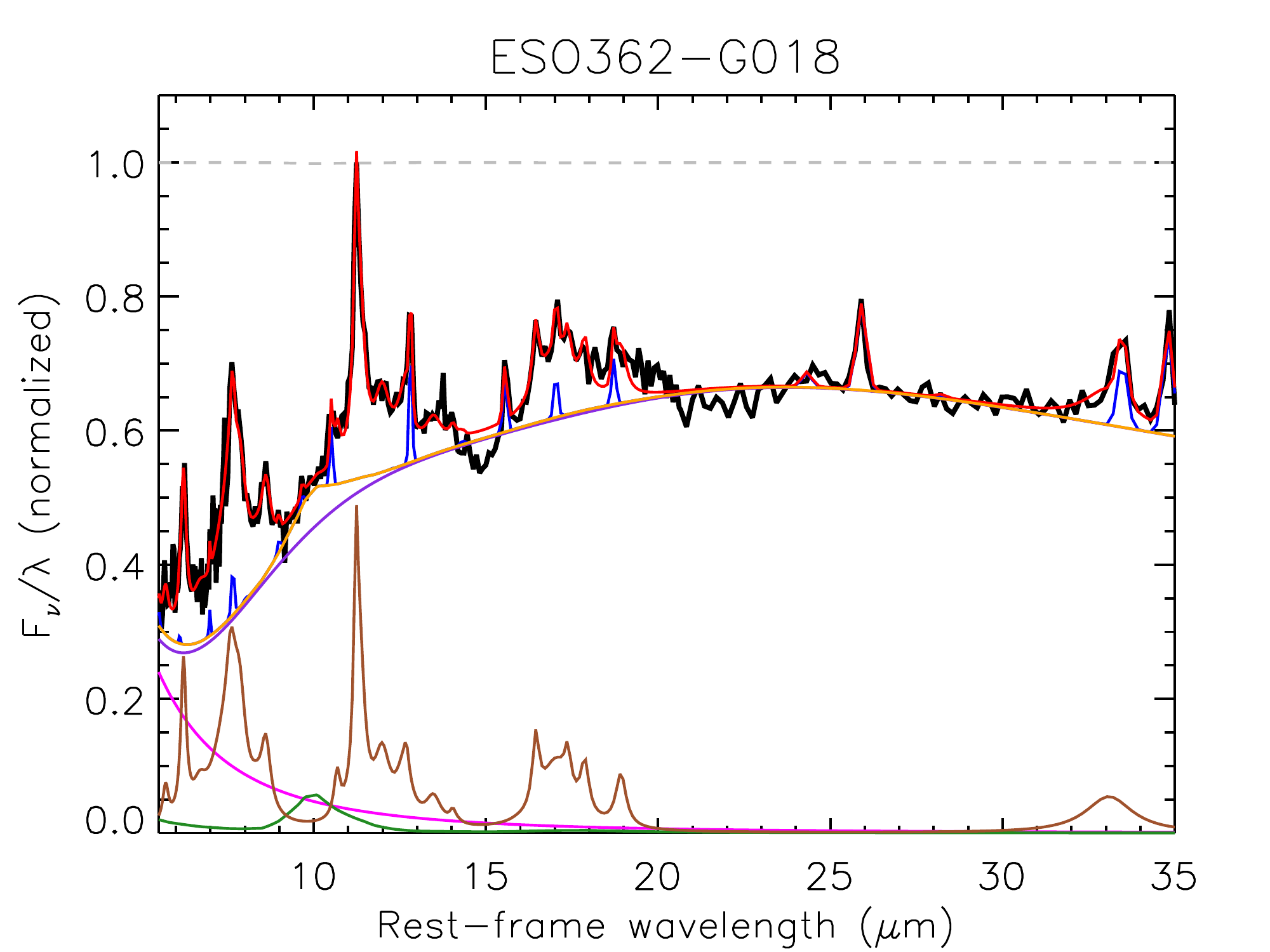}
\includegraphics[width=7.62cm]{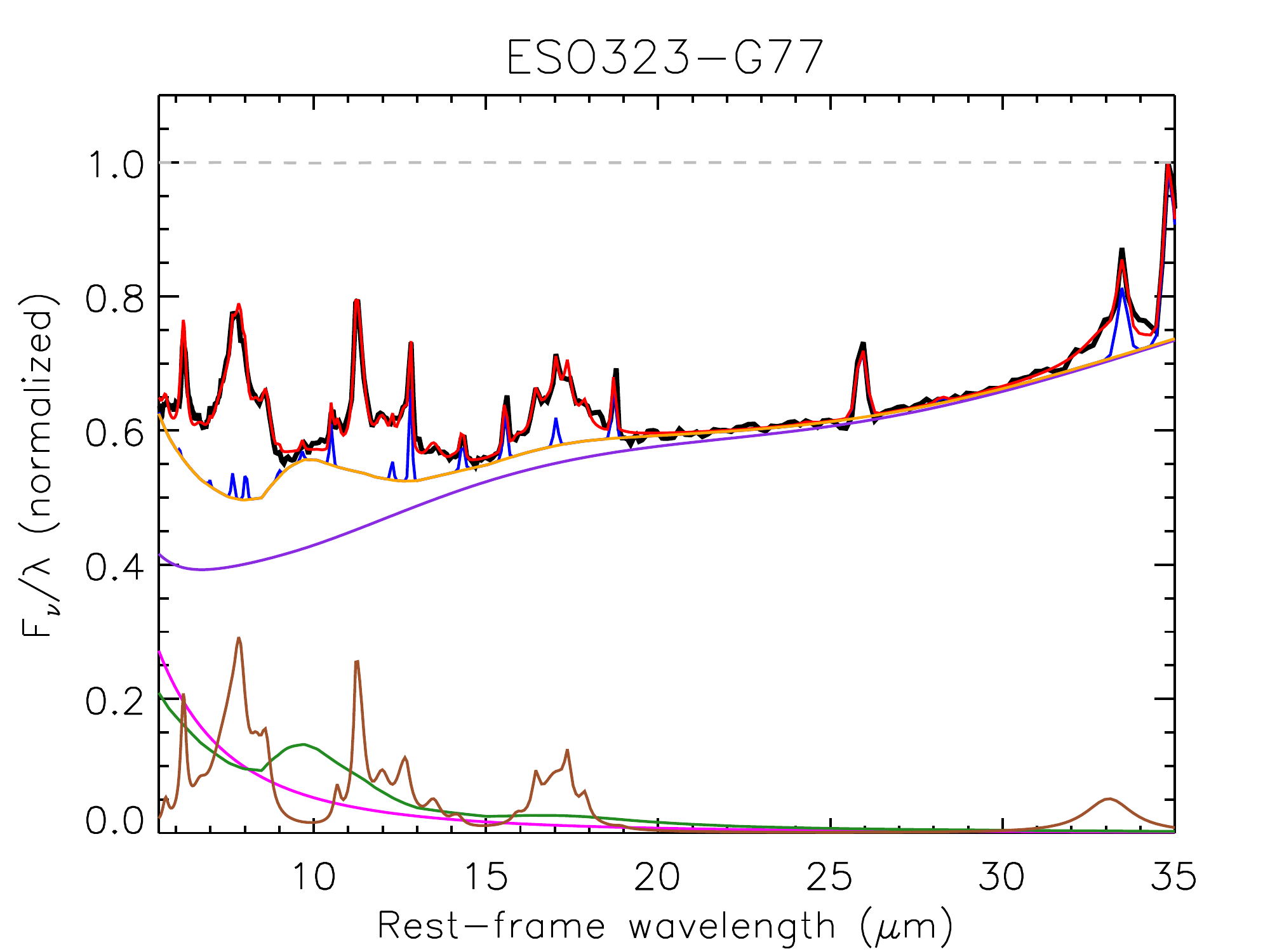}
\includegraphics[width=7.62cm]{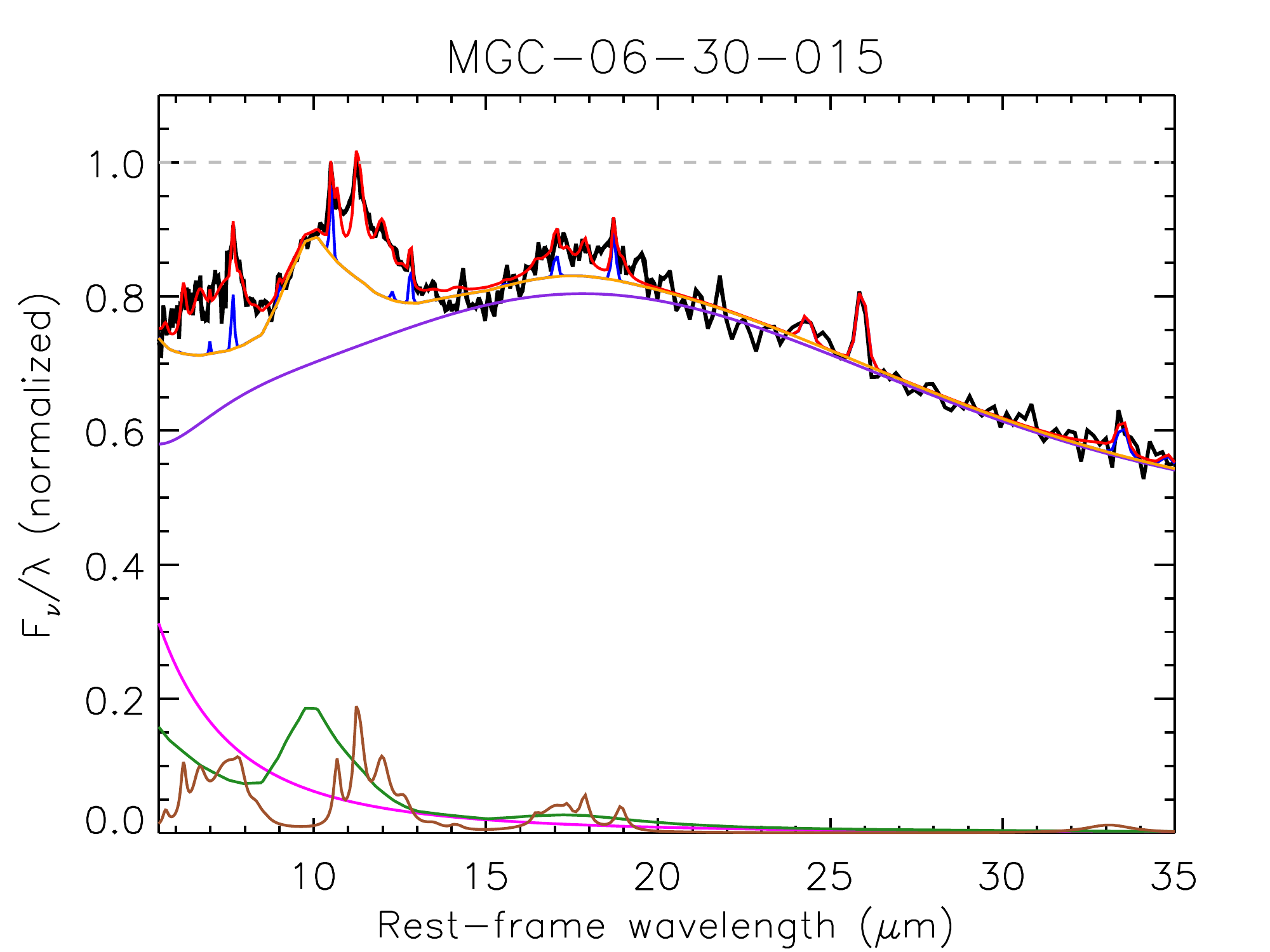}
\includegraphics[width=7.62cm]{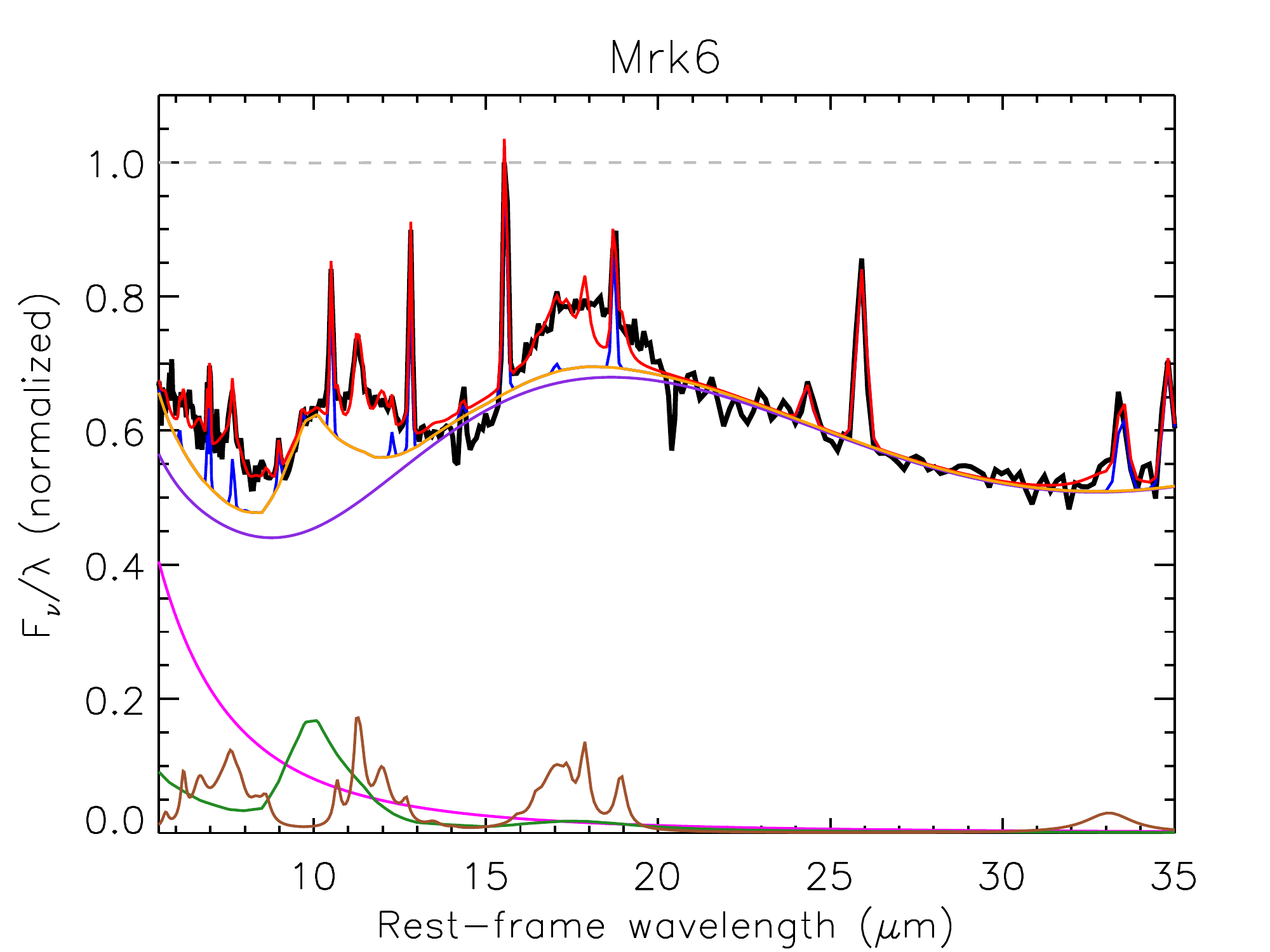}
\includegraphics[width=7.62cm]{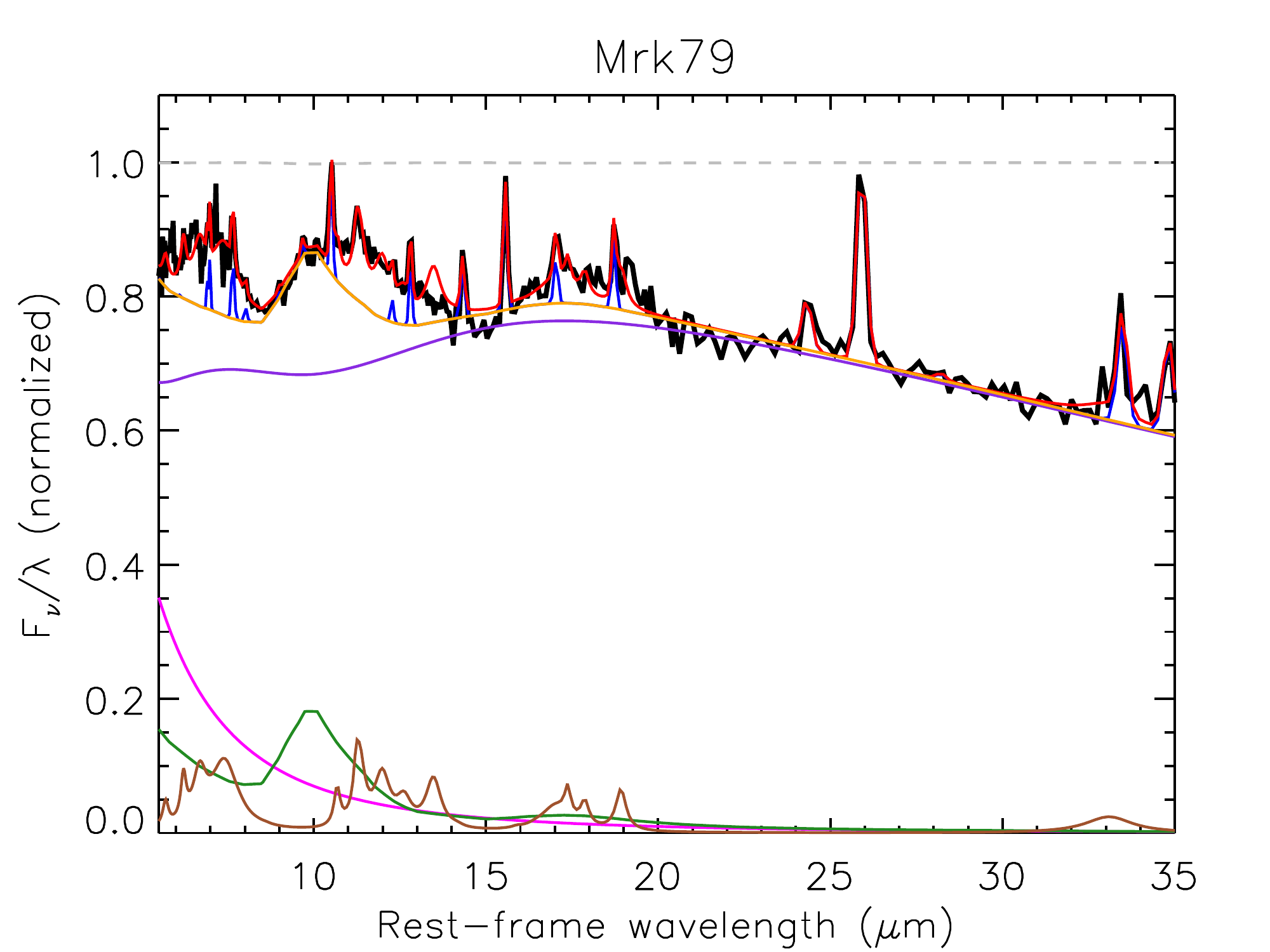}
\includegraphics[width=7.62cm]{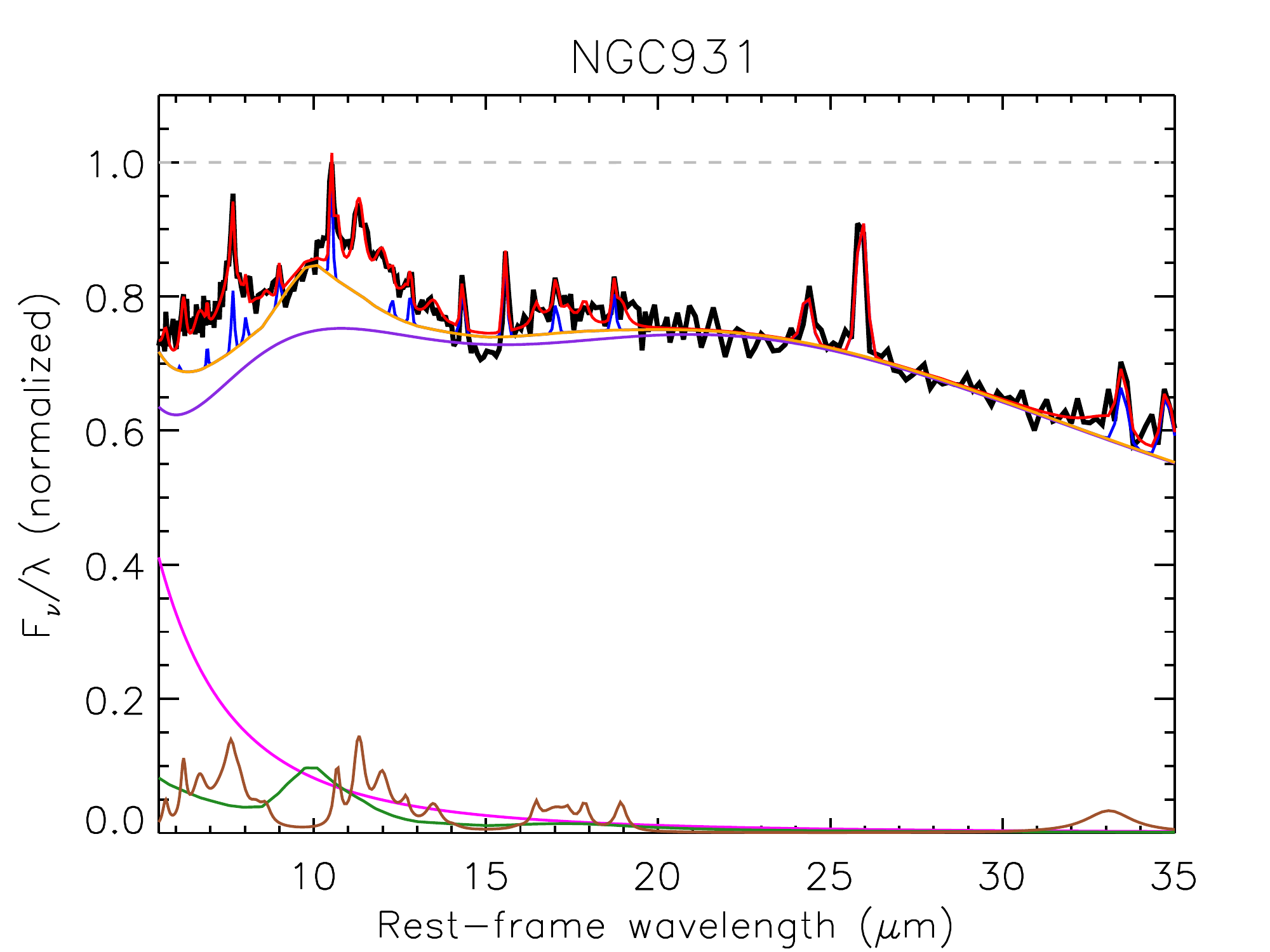}
\includegraphics[width=7.62cm]{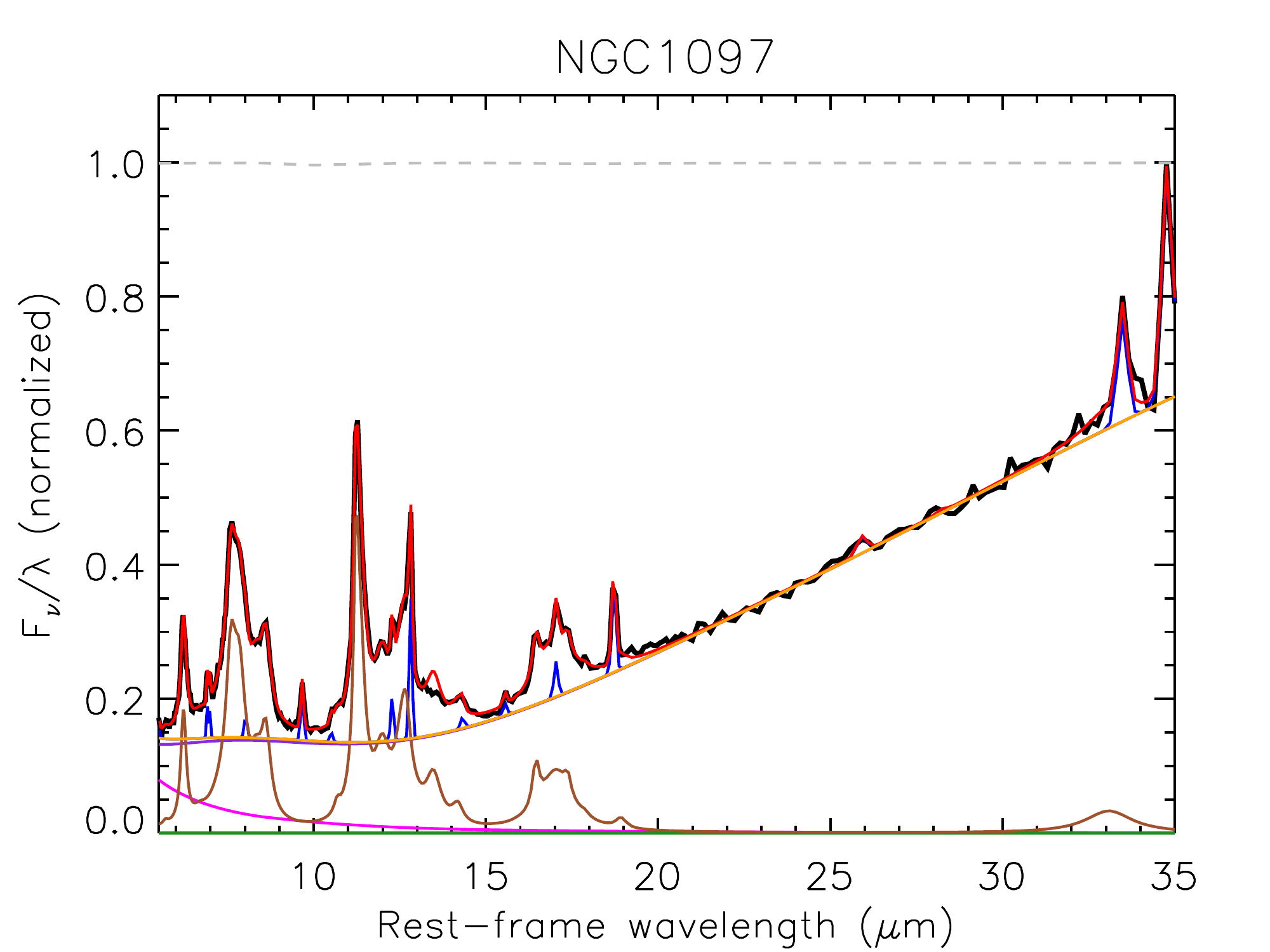}
\includegraphics[width=7.62cm]{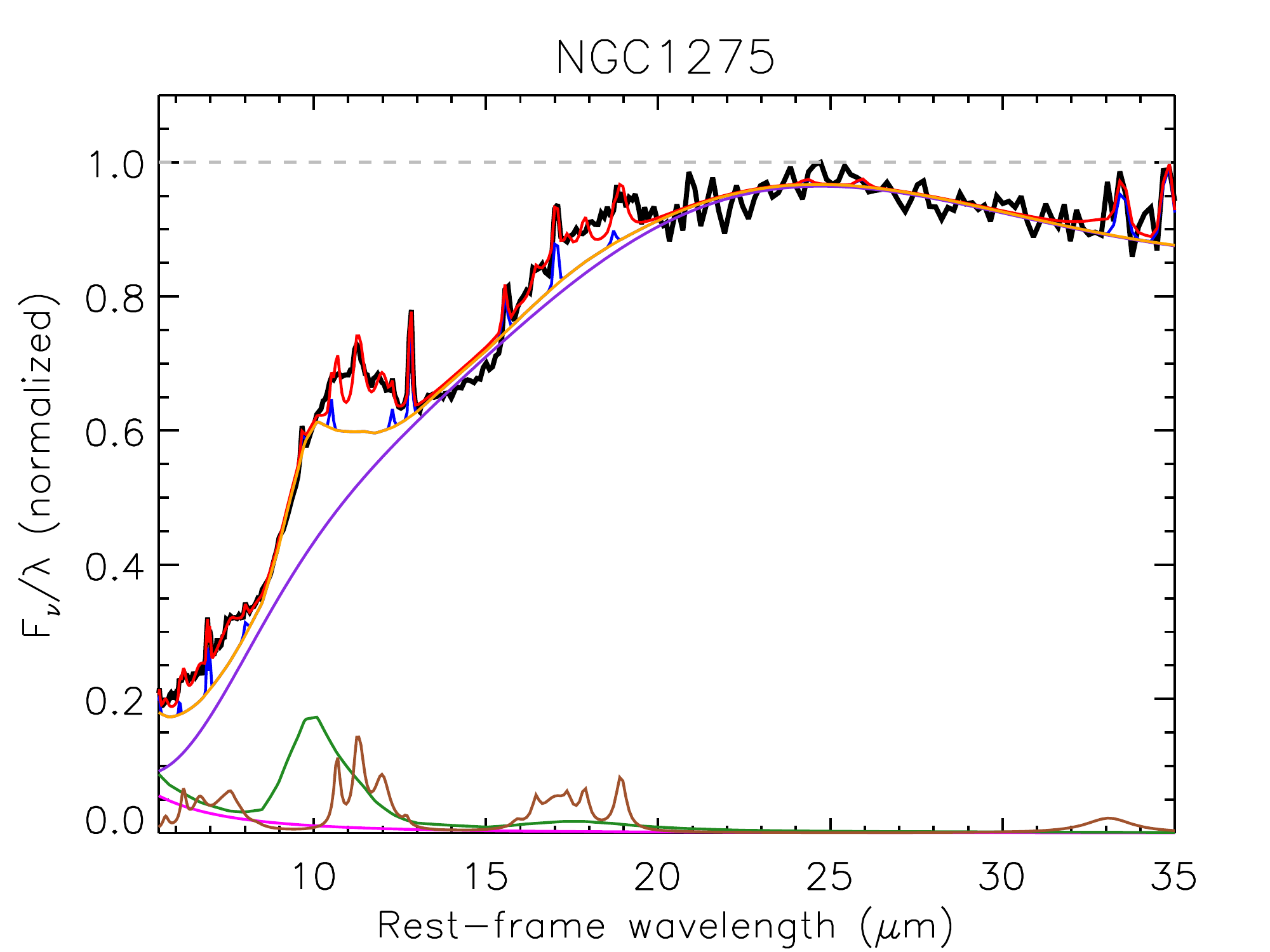}
\par}
\caption{MIR spectral modelling of our sample. The Spitzer/IRS rest-frame spectra and model fits correspond to the black and red solid lines. We show the dust continuum (purple solid lines), stellar continuum (pink solid lines), silicate feature in emission (green solid line), and their sum (orange solid line). The grey dashed line indicates the extinction profile. We also show the fitted
PAH features (brown solid lines) and emission lines (blue solid lines).}
\label{appendix_pahfit_fig}
\end{figure*}

\begin{figure*}
\contcaption
\centering
\par{
\includegraphics[width=7.62cm]{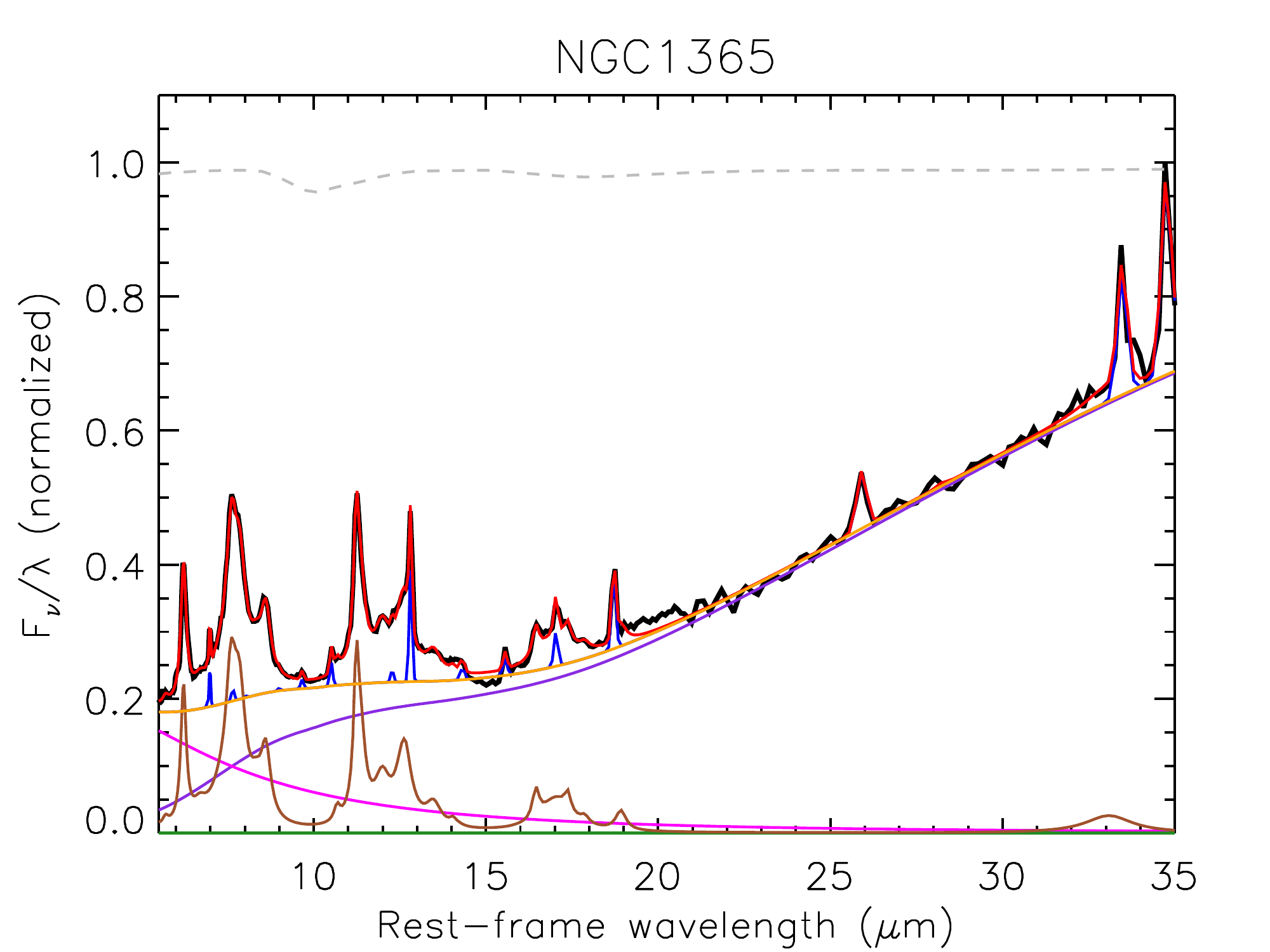}
\includegraphics[width=7.62cm]{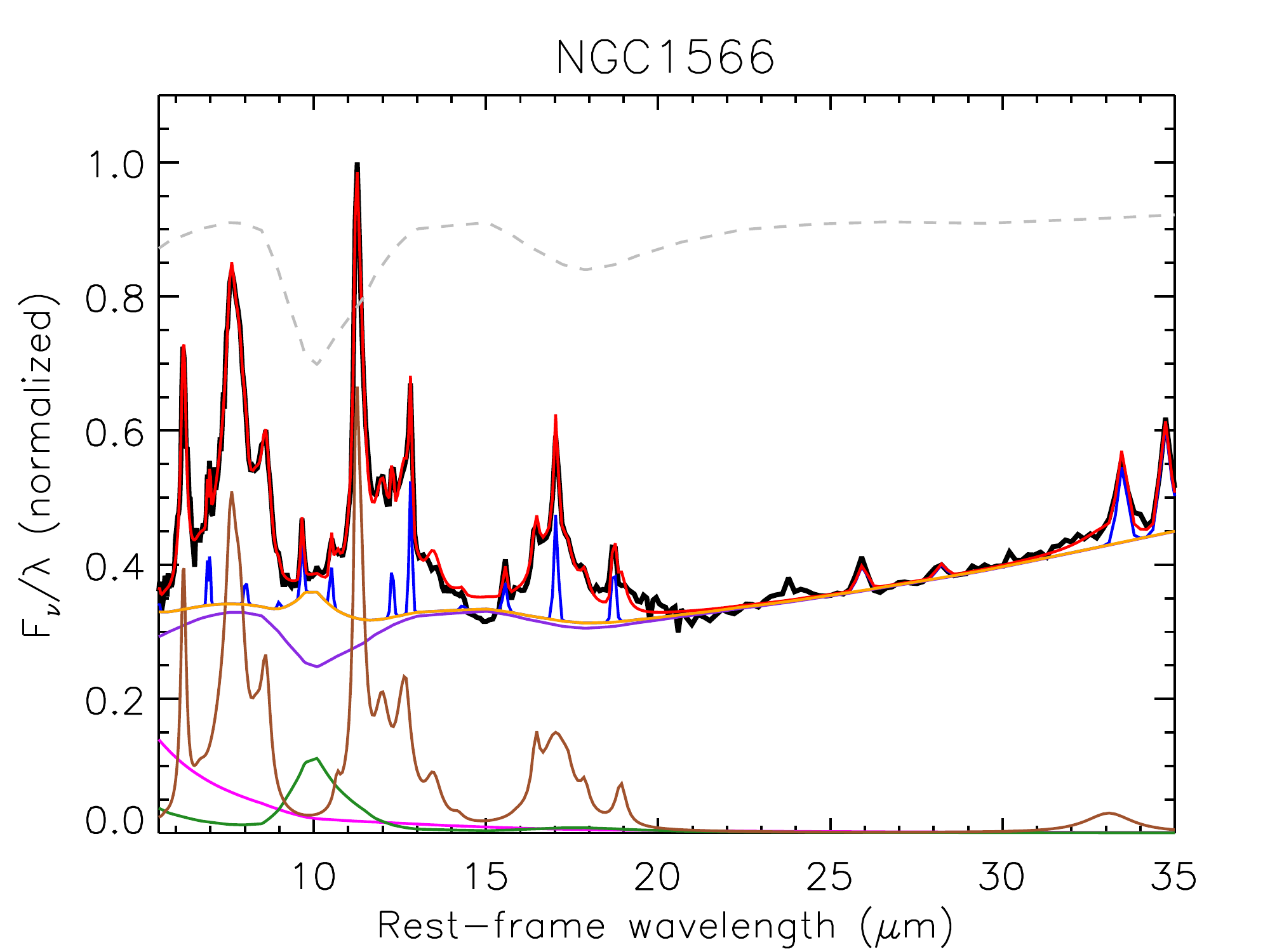}
\includegraphics[width=7.62cm]{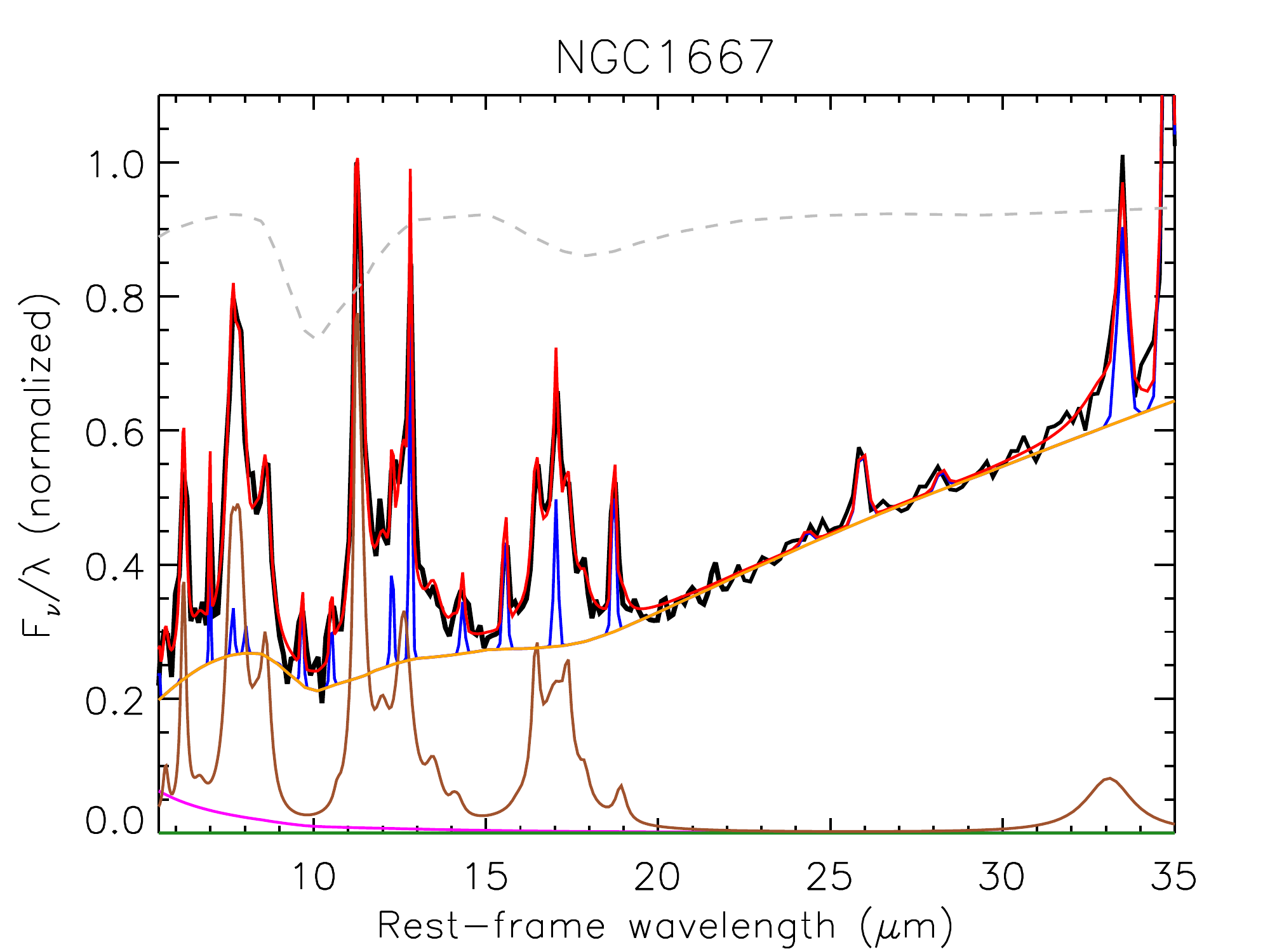}
\includegraphics[width=7.62cm]{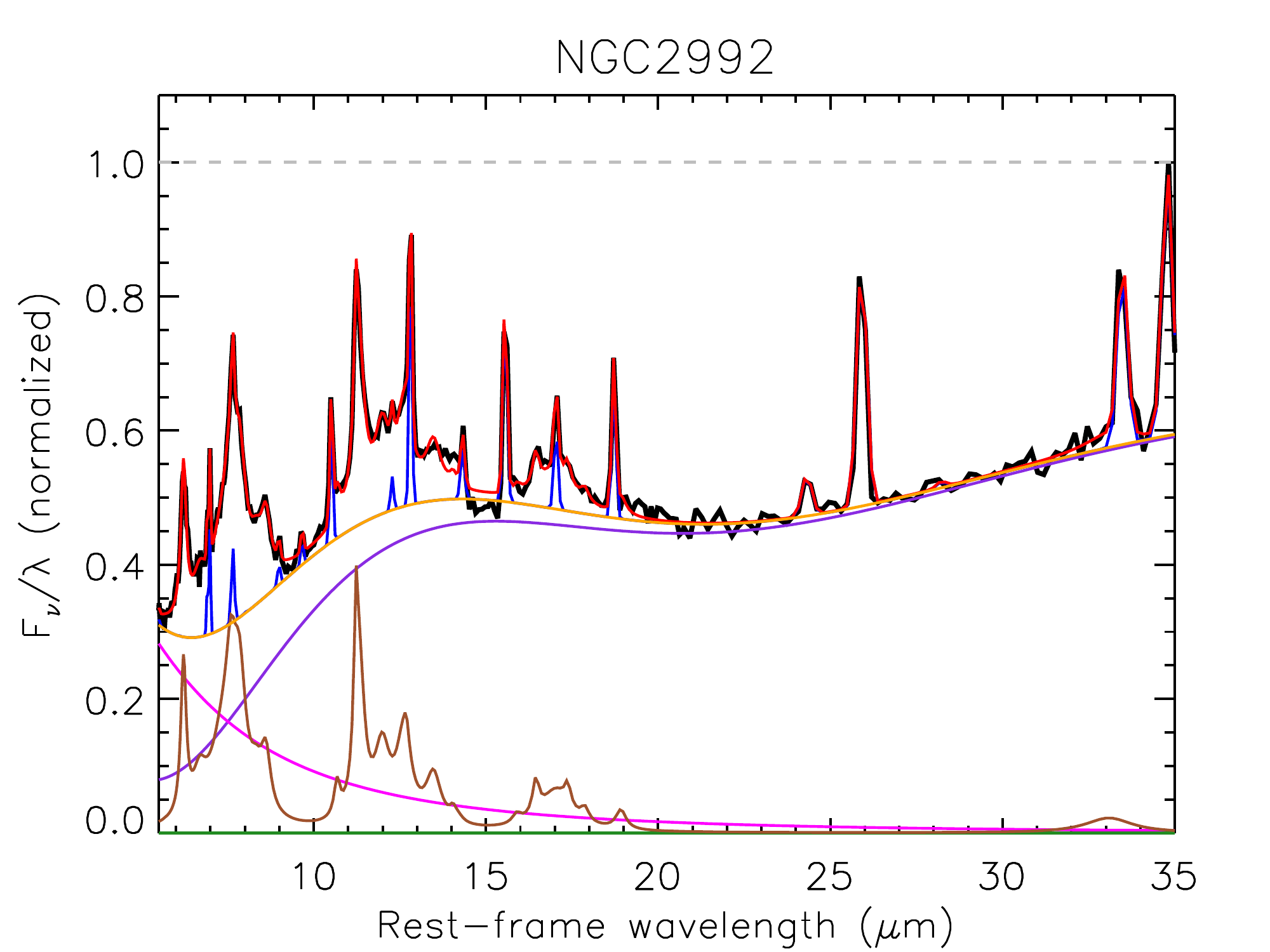}
\includegraphics[width=7.62cm]{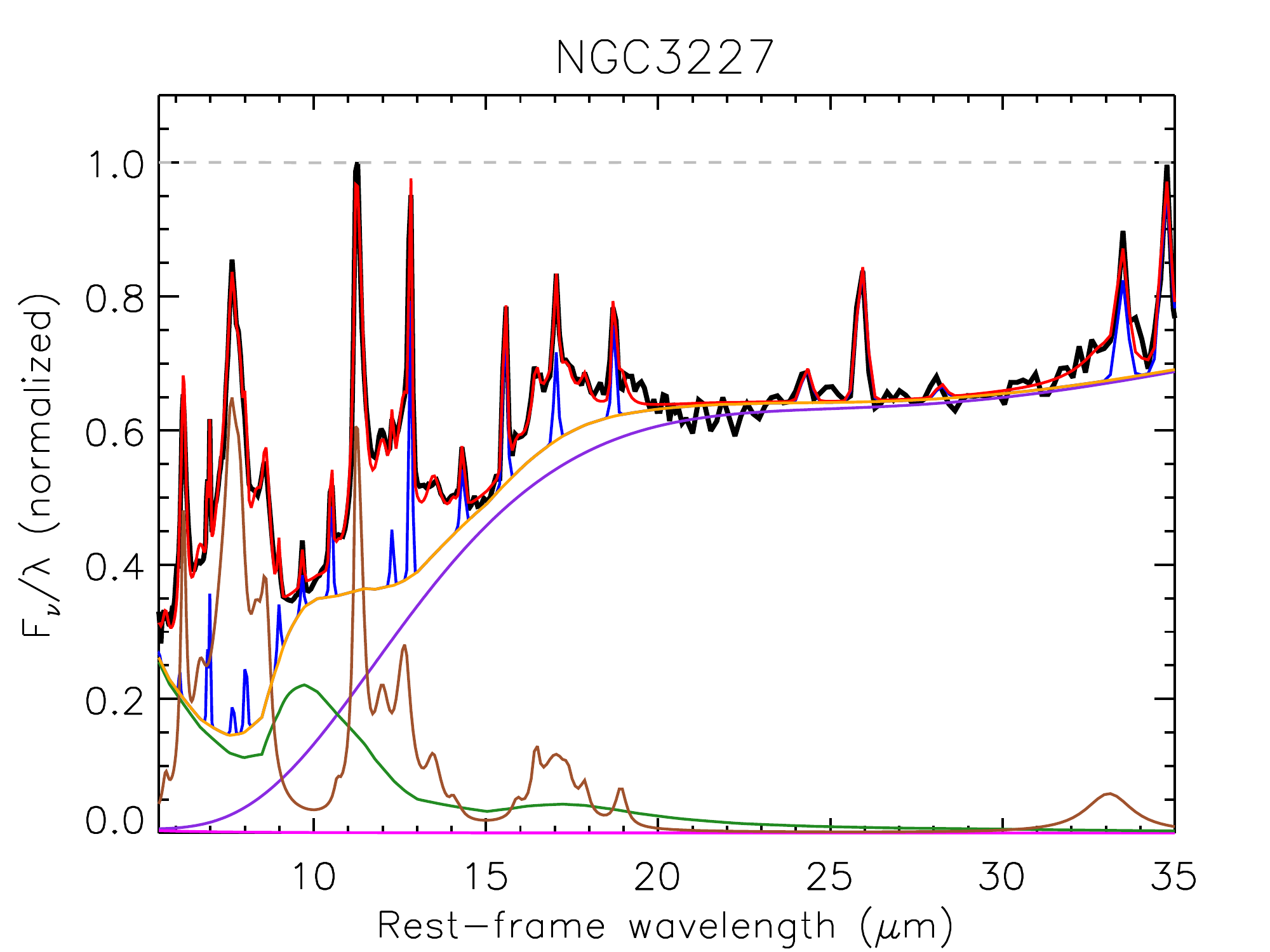}
\includegraphics[width=7.62cm]{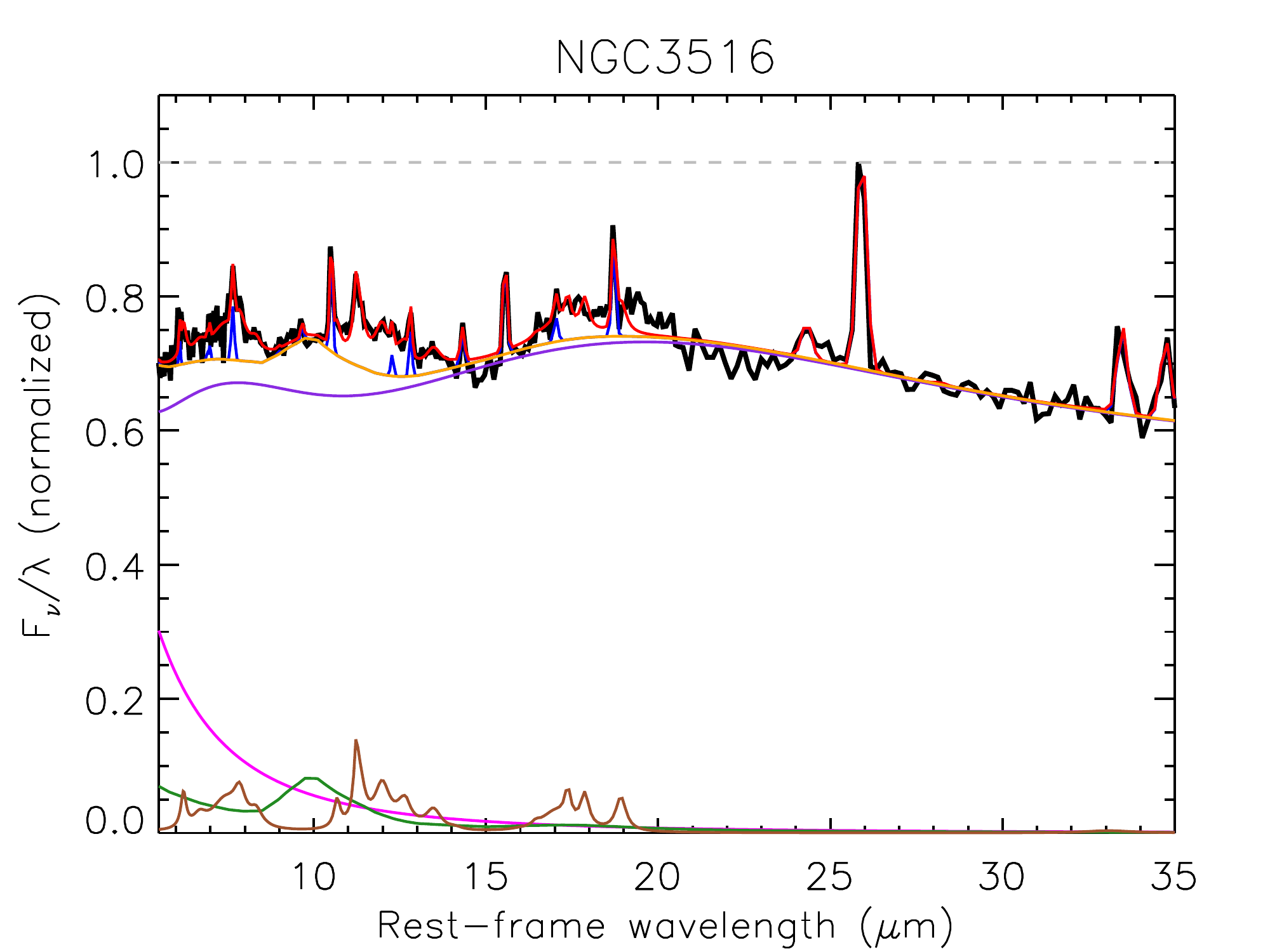}
\includegraphics[width=7.62cm]{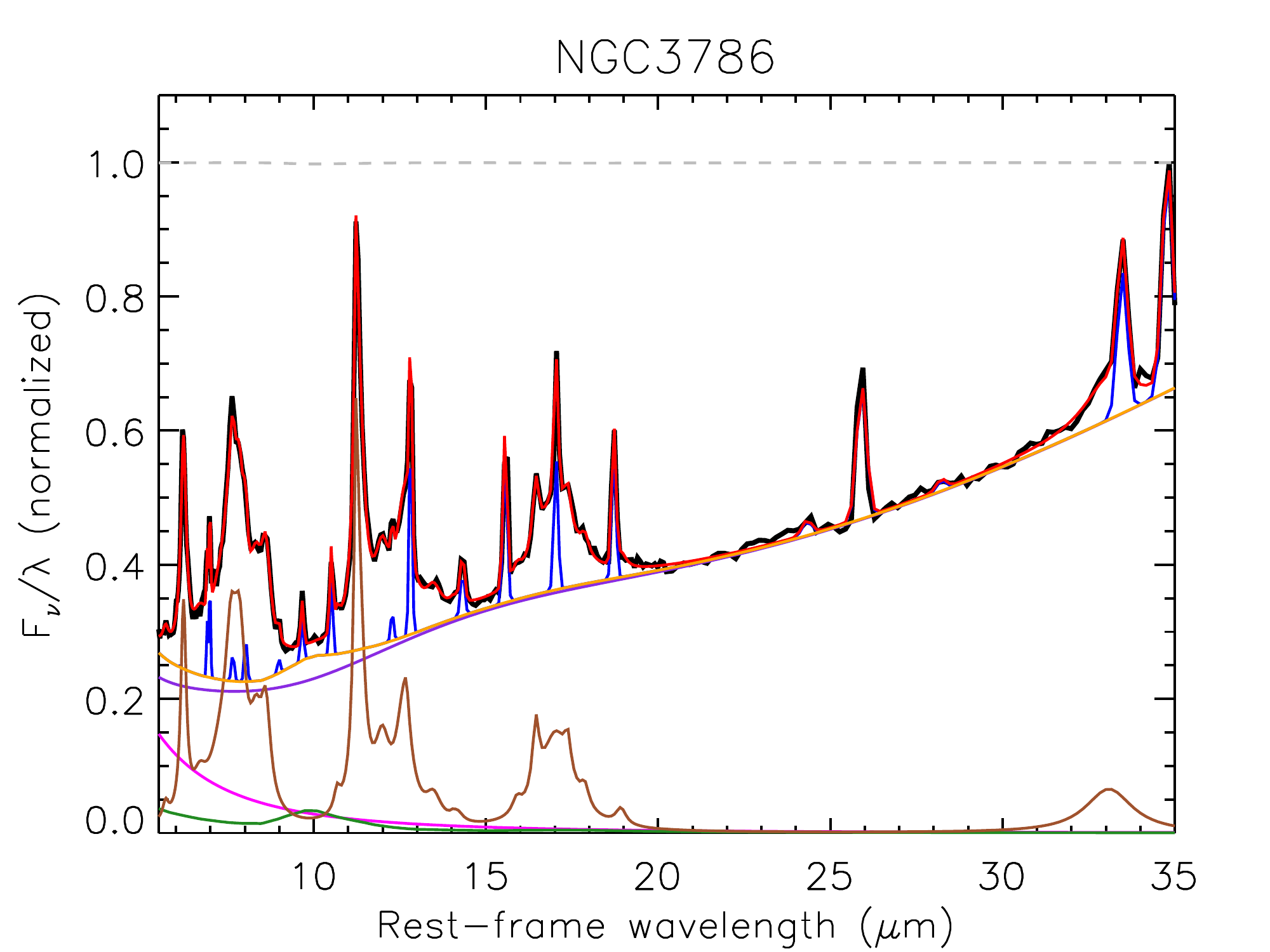}
\includegraphics[width=7.62cm]{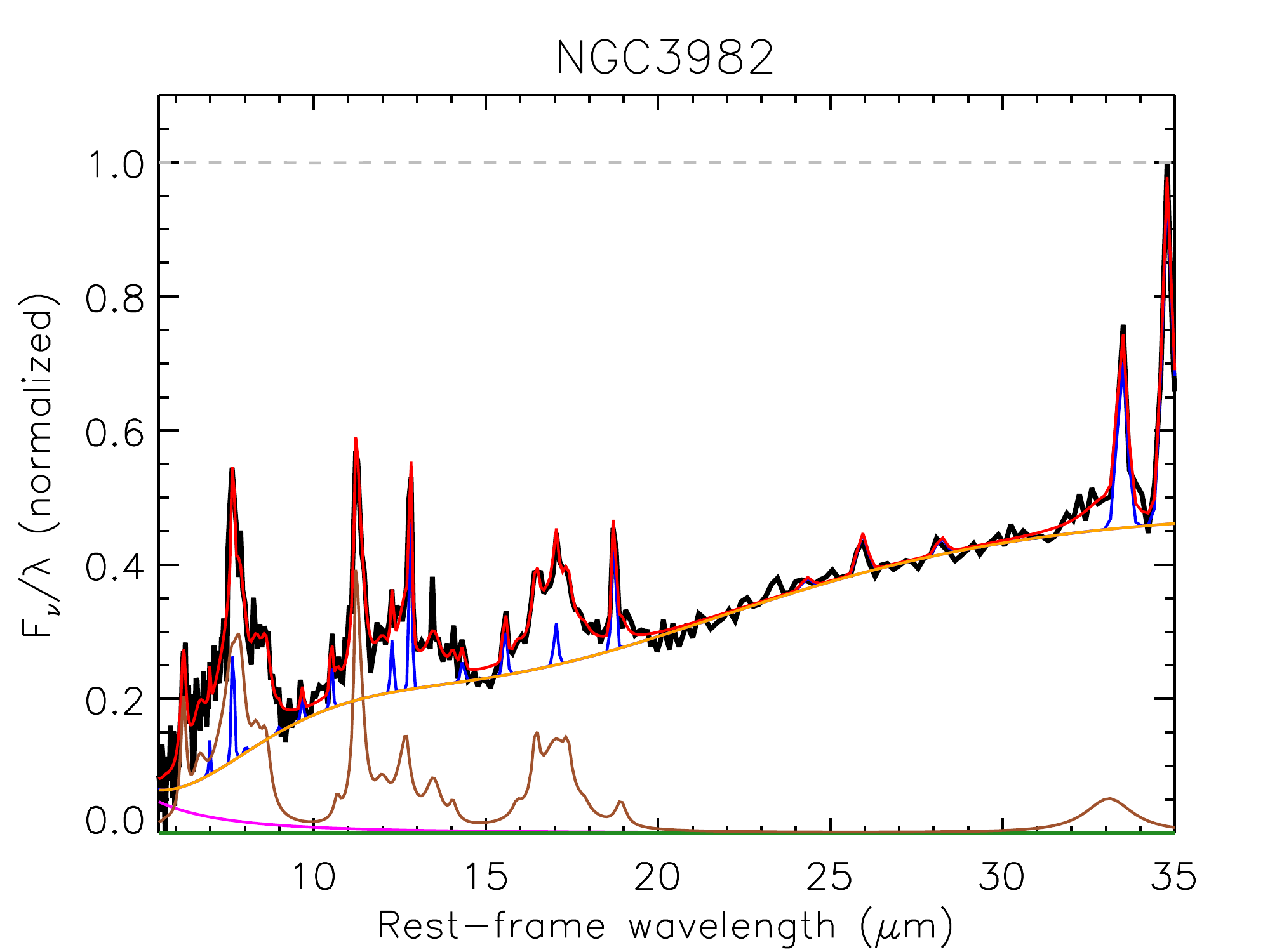}
\par} 
\end{figure*}

\begin{figure*}
\contcaption
\centering
\par{
\includegraphics[width=7.62cm]{figs/ngc4051_nuc_ap_cor_pahfit-eps-converted-to.pdf}
\includegraphics[width=7.62cm]{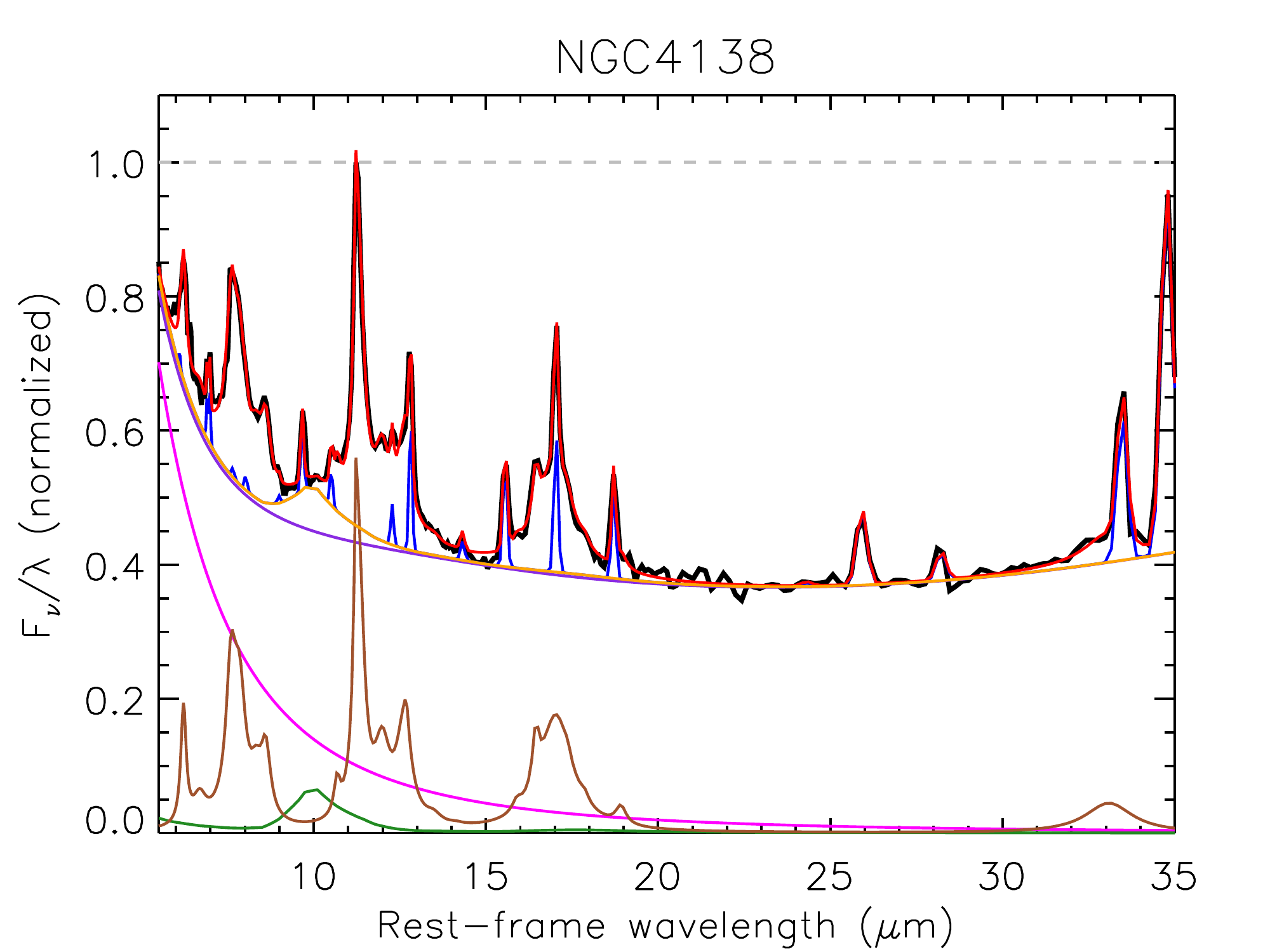}
\includegraphics[width=7.62cm]{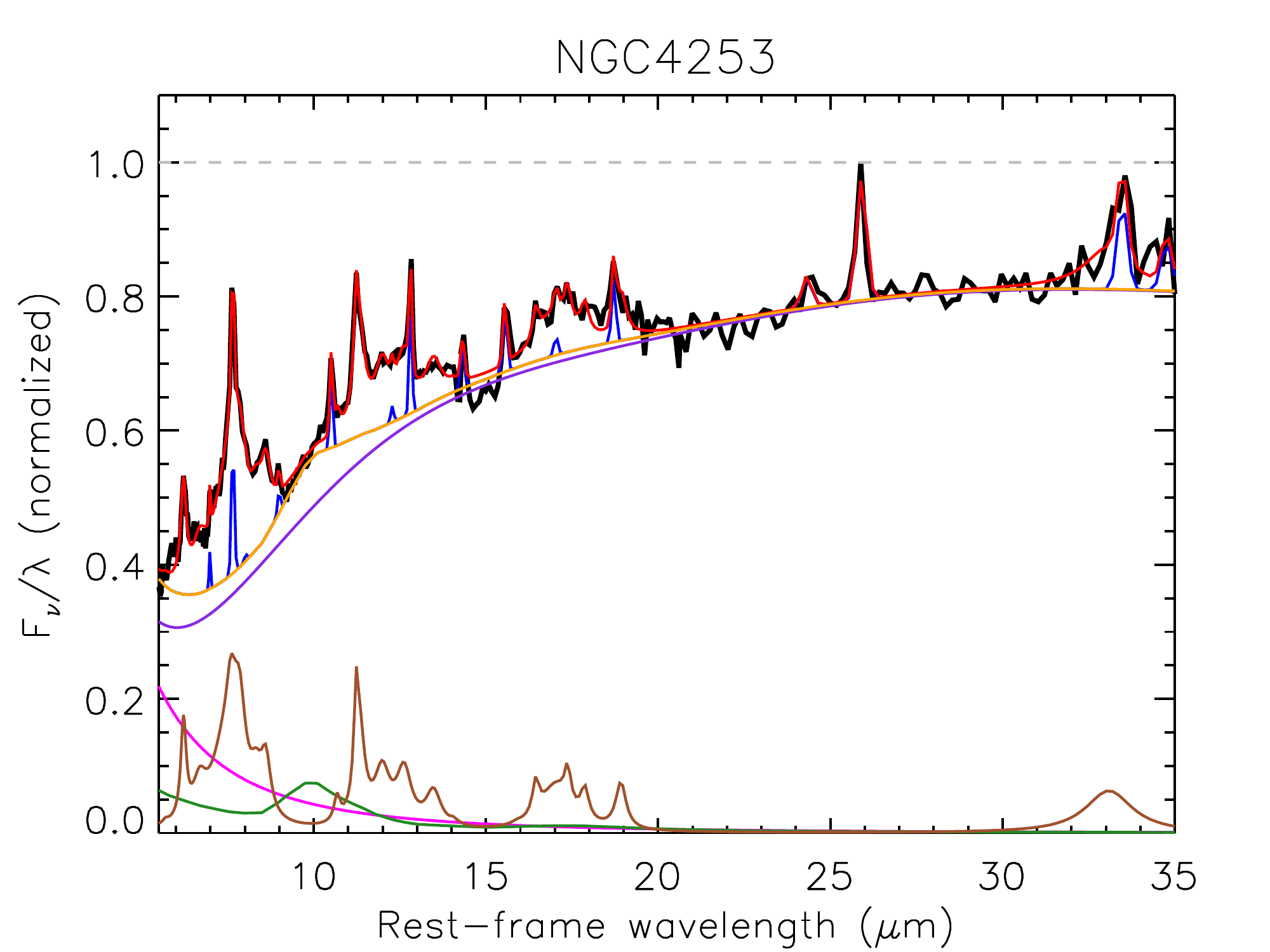}
\includegraphics[width=7.62cm]{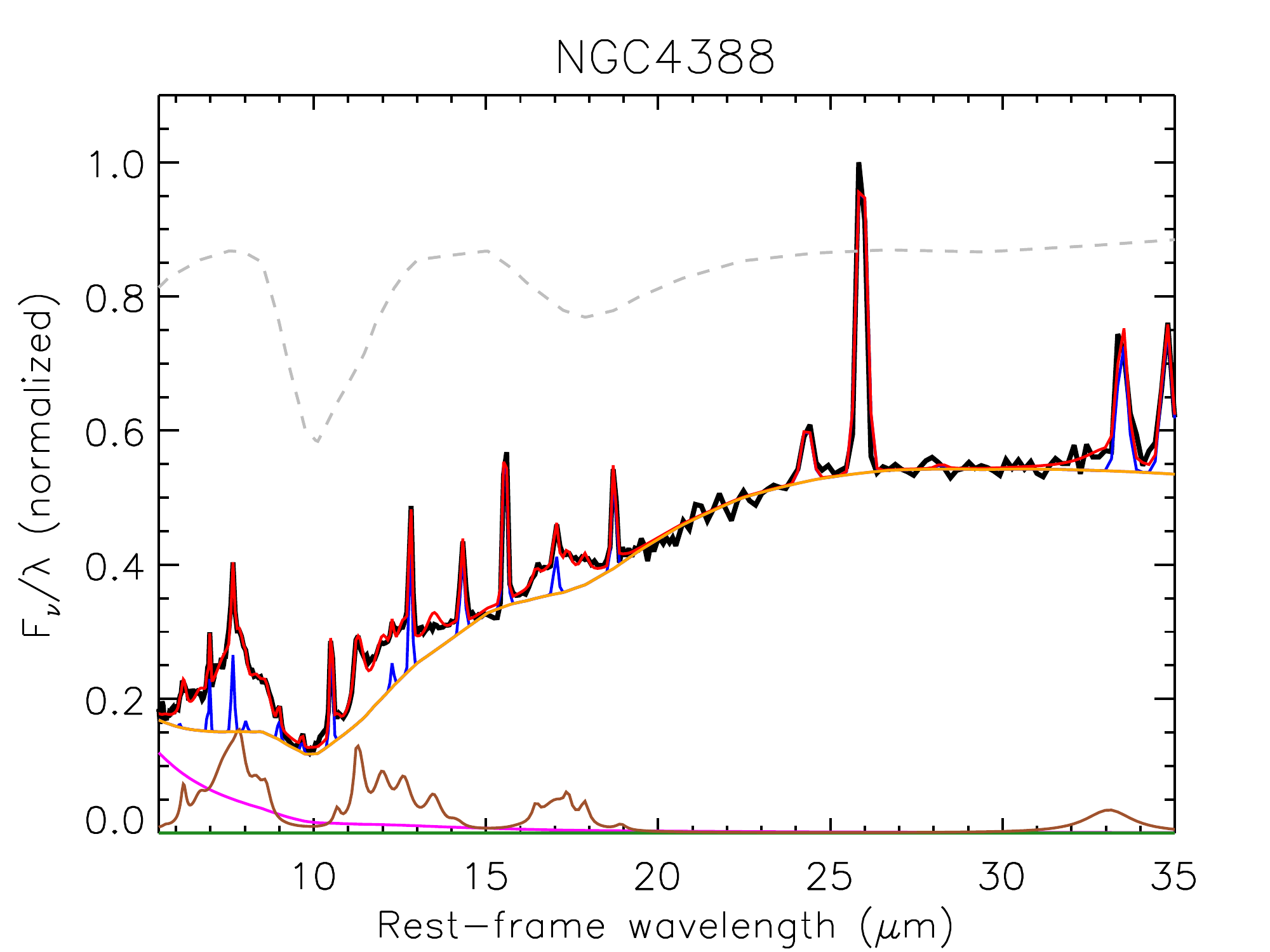}
\includegraphics[width=7.62cm]{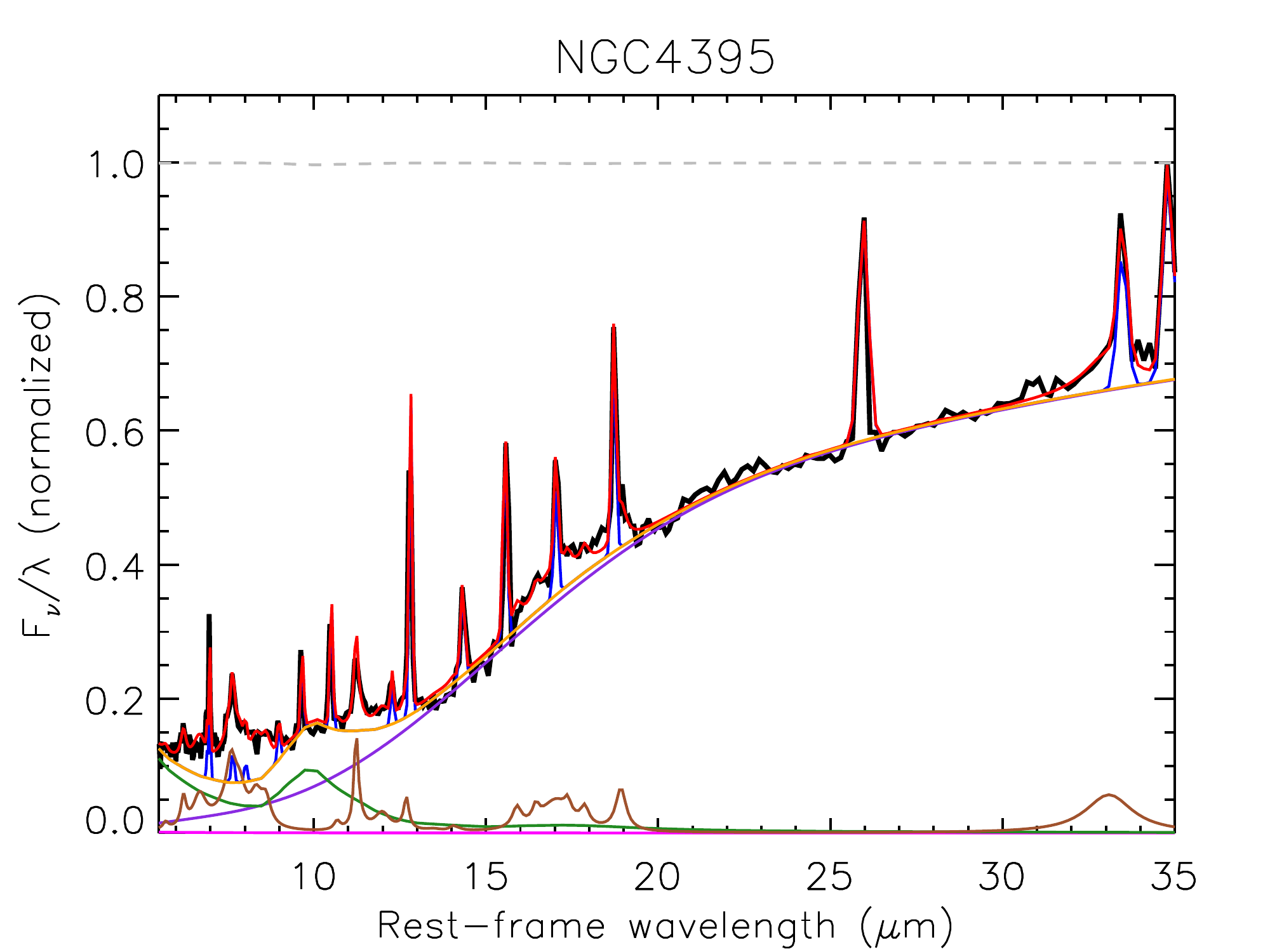}
\includegraphics[width=7.62cm]{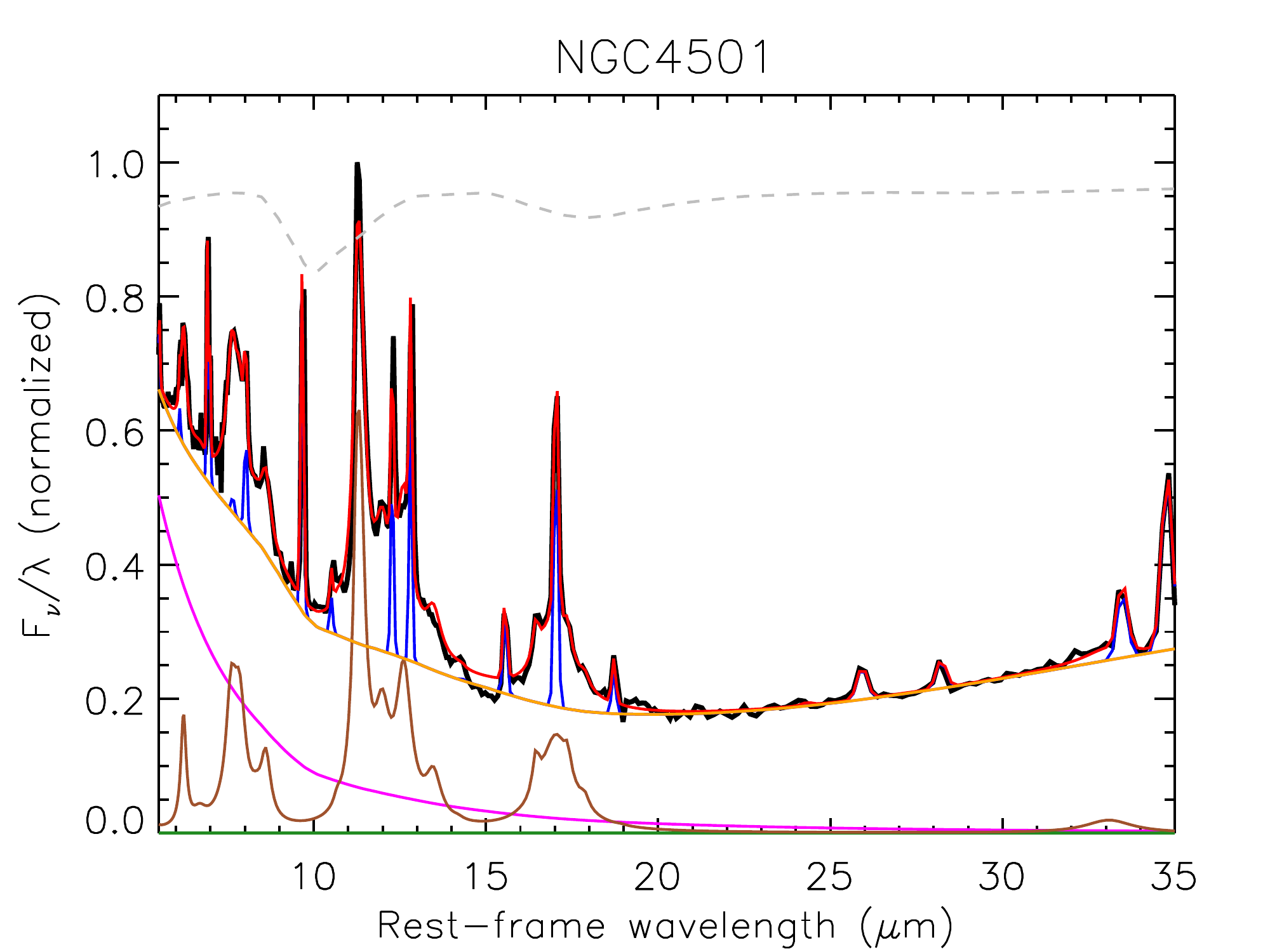}
\includegraphics[width=7.62cm]{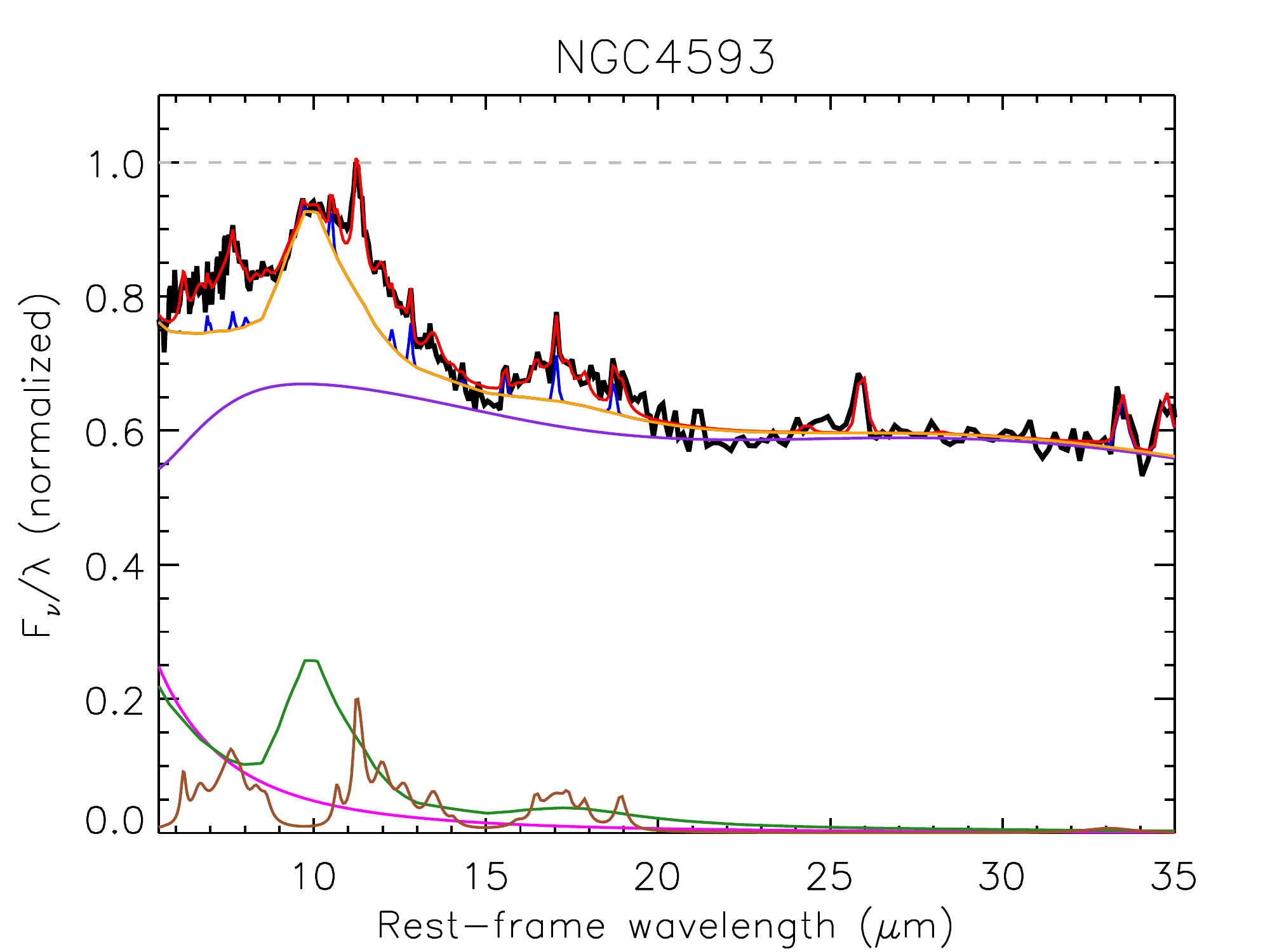}
\includegraphics[width=7.62cm]{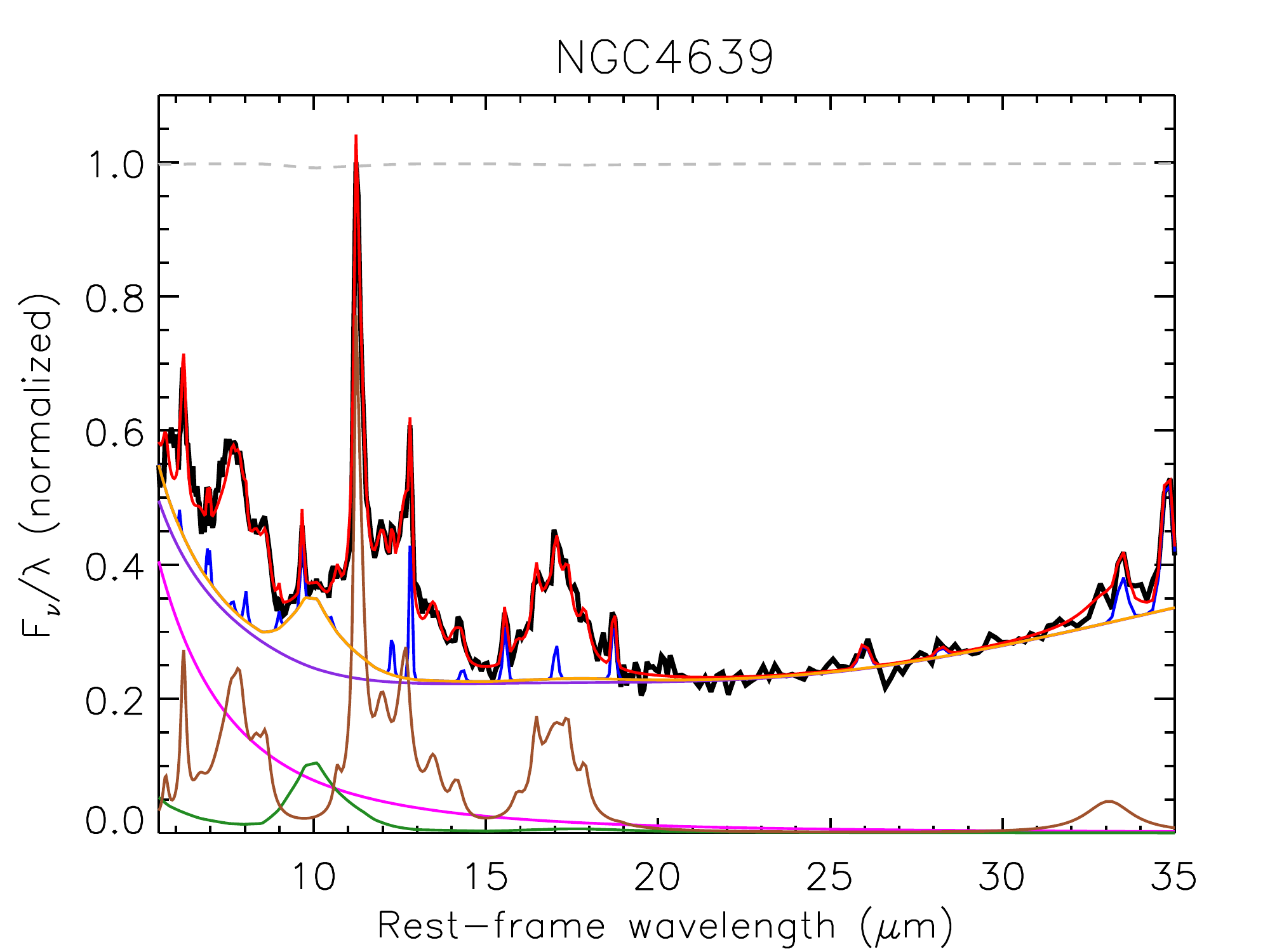}
\par} 
\end{figure*}
\begin{figure*}
\contcaption
\centering
\par{
\includegraphics[width=7.62cm]{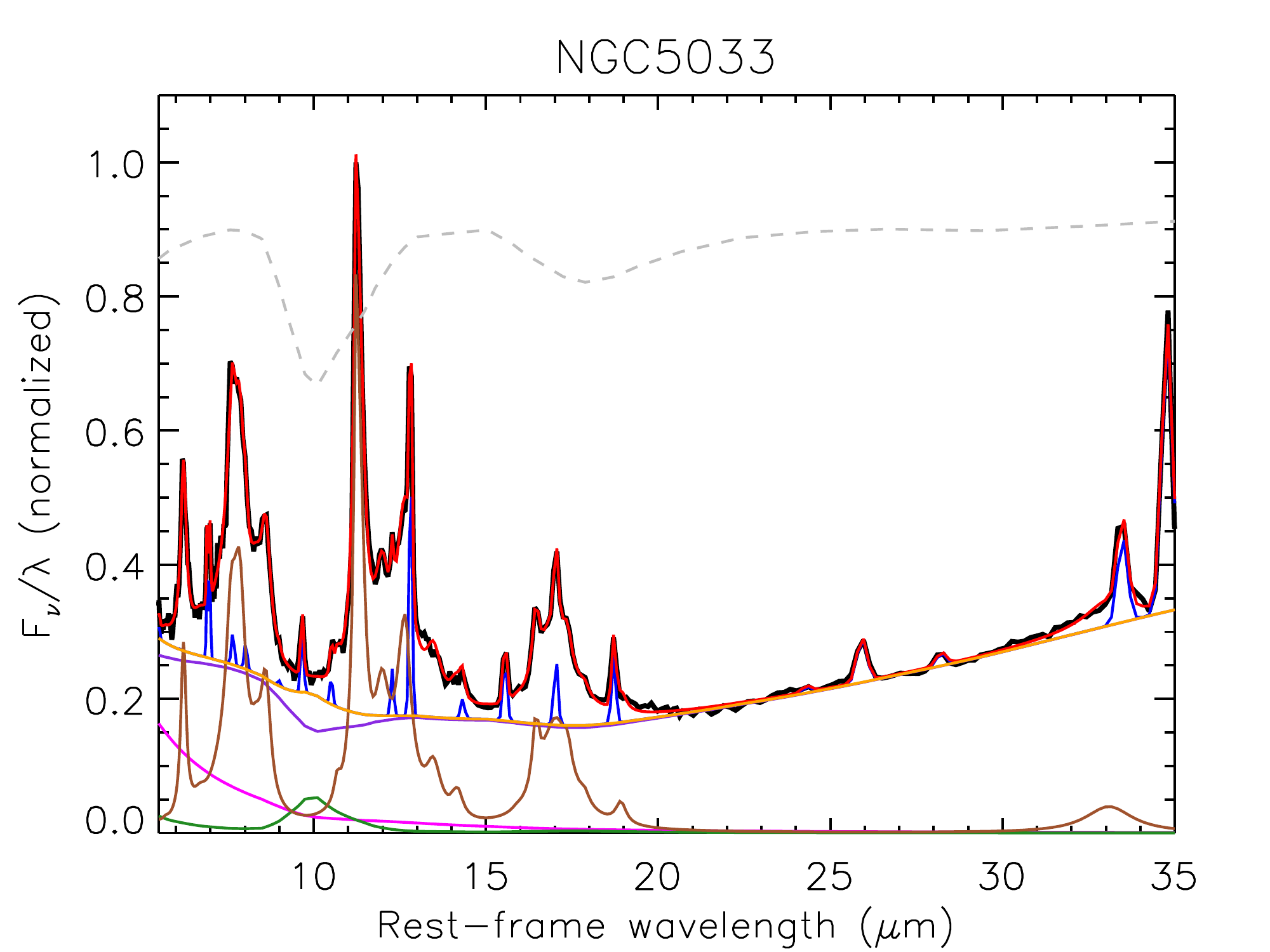}
\includegraphics[width=7.62cm]{figs/ngc5135_nuc_ap_cor_pahfit-eps-converted-to.pdf}
\includegraphics[width=7.62cm]{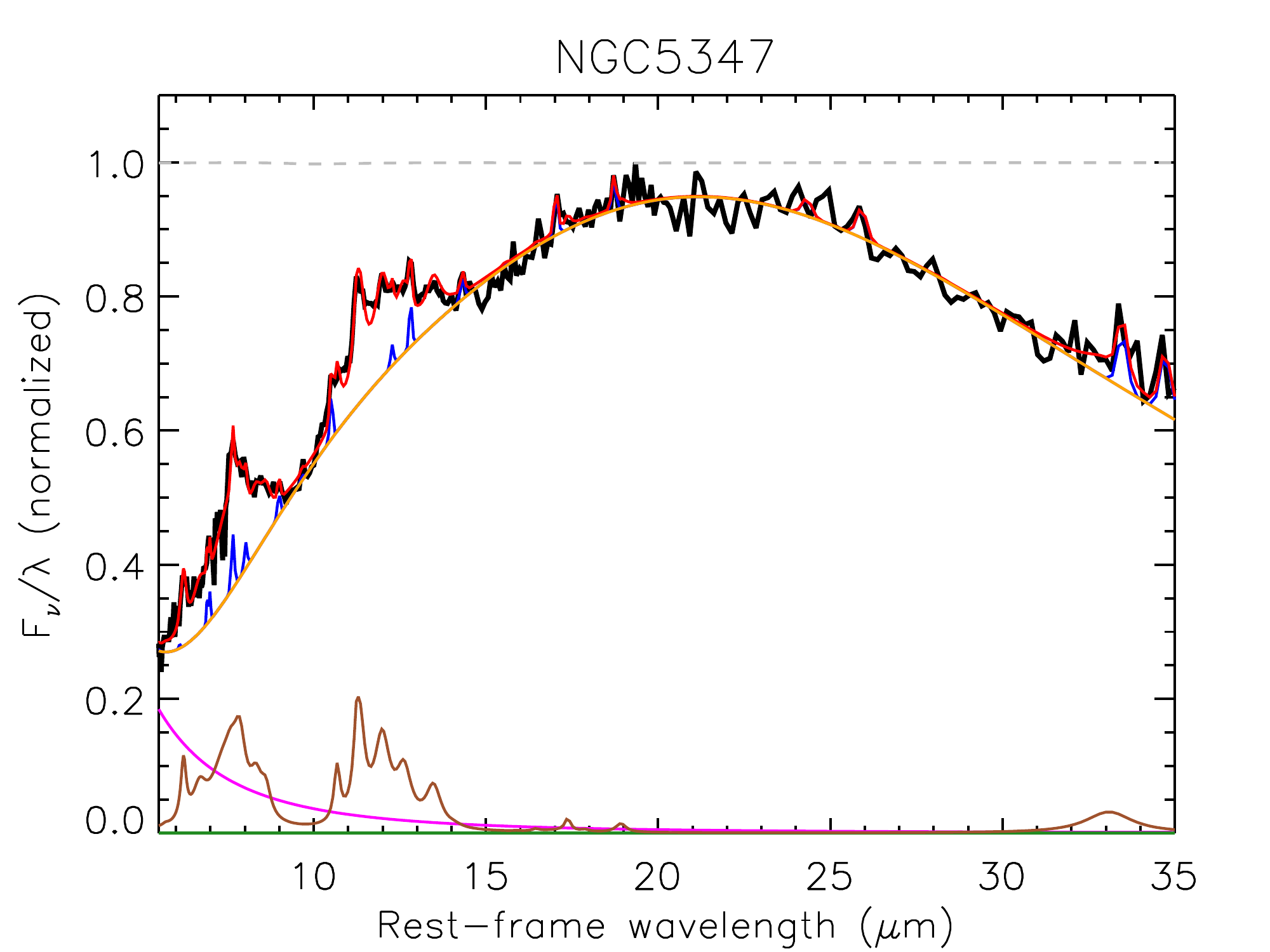}
\includegraphics[width=7.62cm]{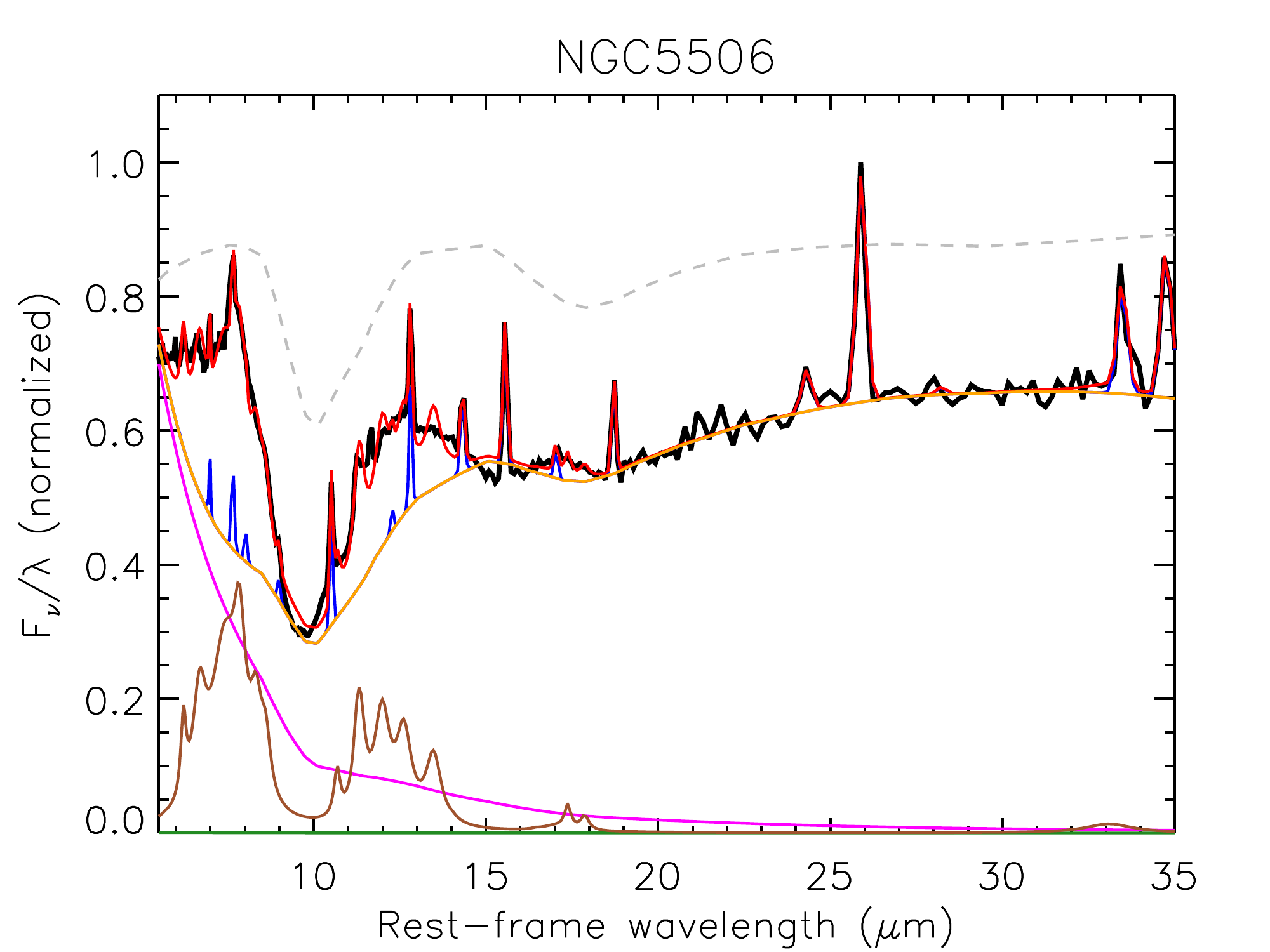}
\includegraphics[width=7.62cm]{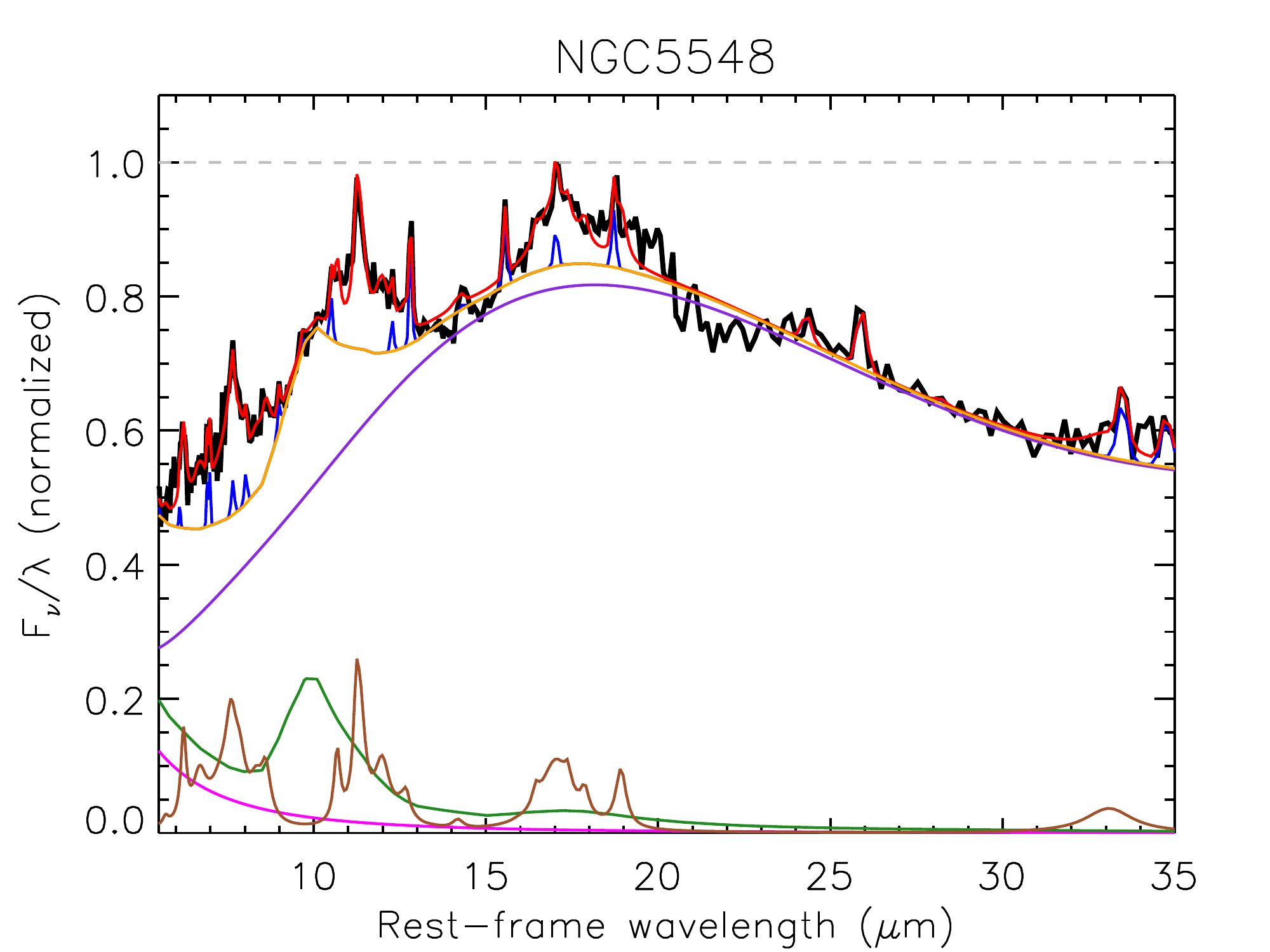}
\includegraphics[width=7.62cm]{figs/ngc5953_nuc_ap_cor_pahfit-eps-converted-to.pdf}
\includegraphics[width=7.62cm]{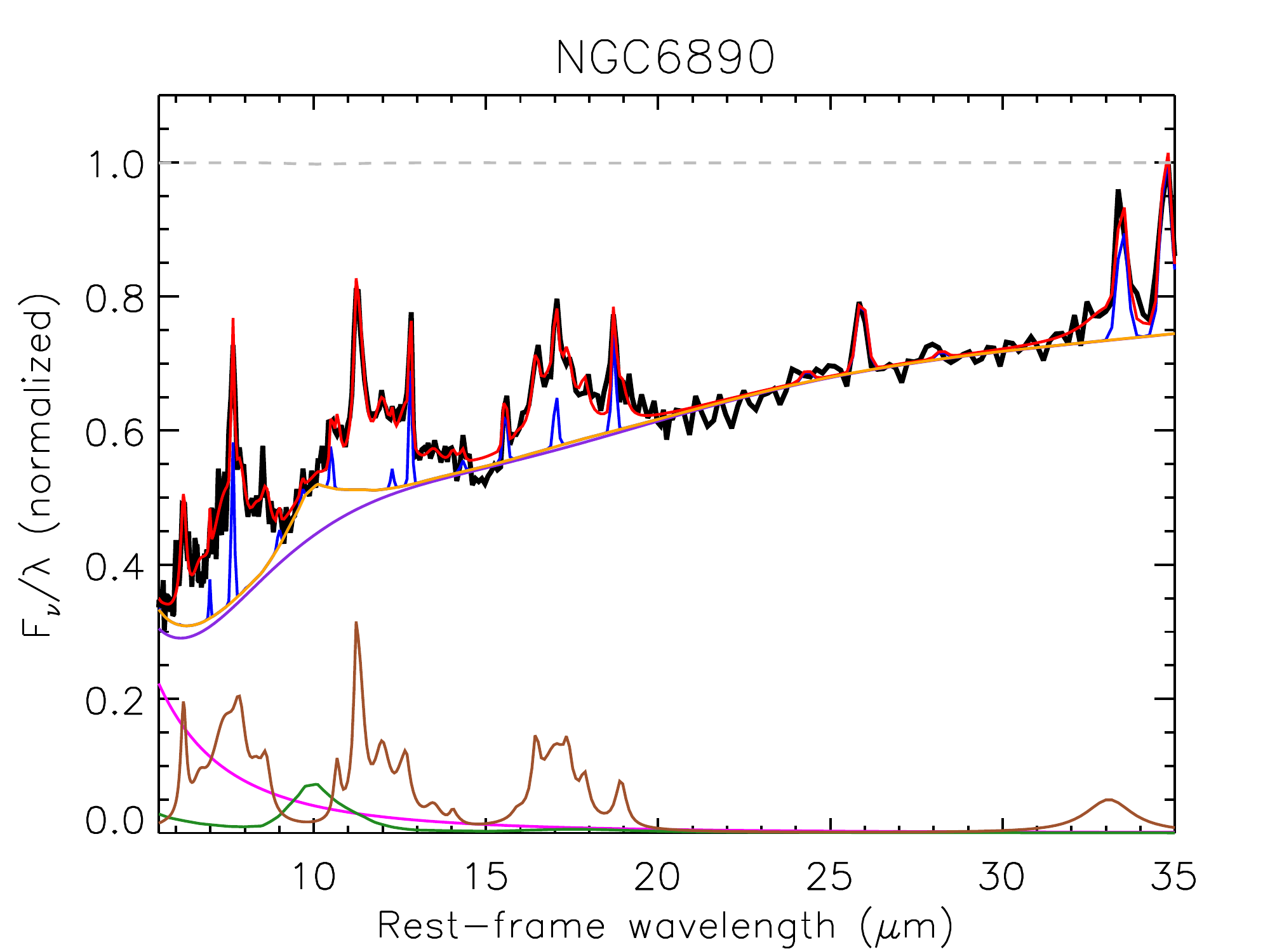}
\includegraphics[width=7.62cm]{figs/ngc7130_nuc_ap_cor_pahfit-eps-converted-to.pdf}
\par} 
\end{figure*}
\begin{figure*}
\contcaption
\centering
\par{
\includegraphics[width=7.62cm]{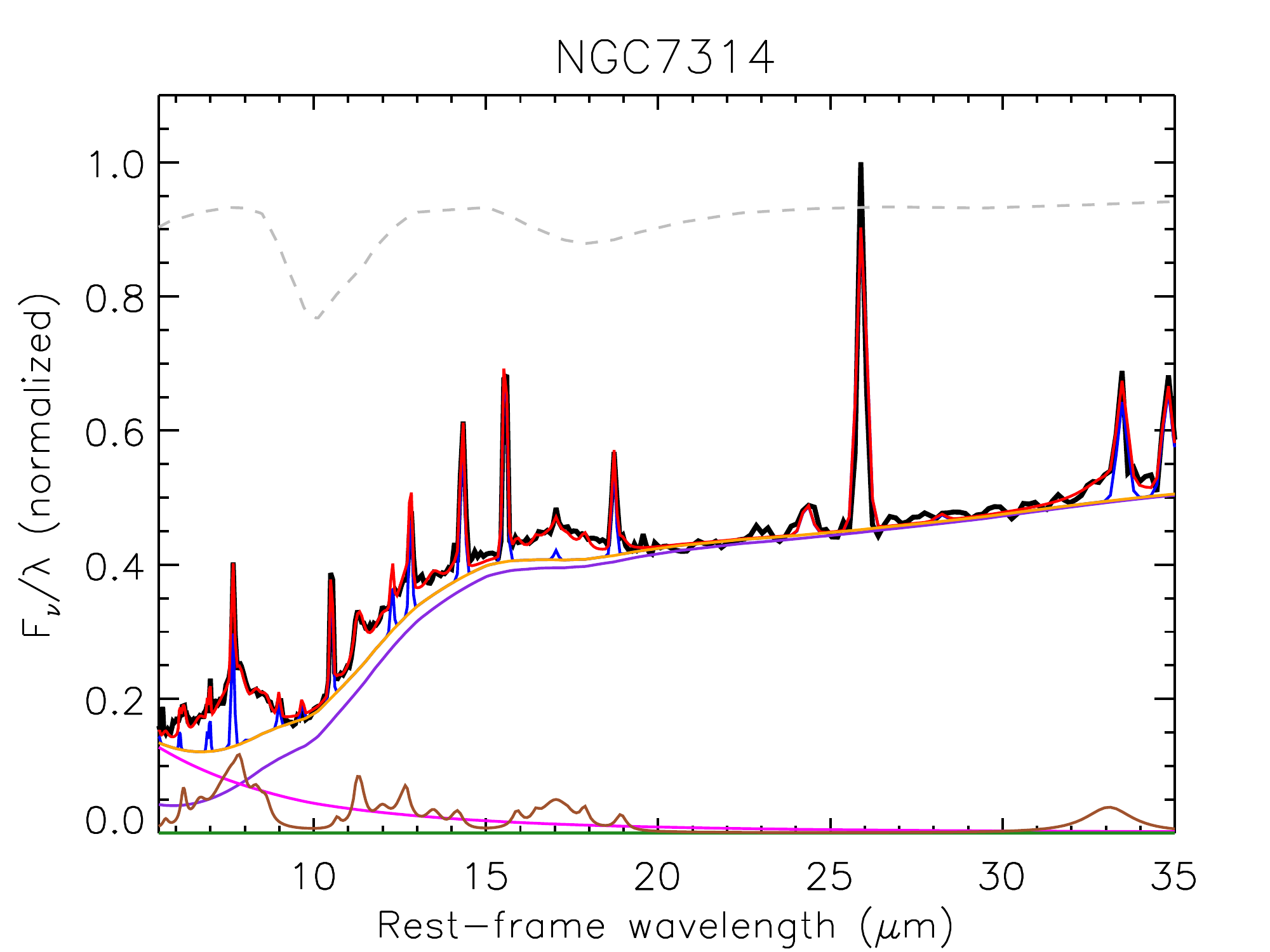}
\includegraphics[width=7.62cm]{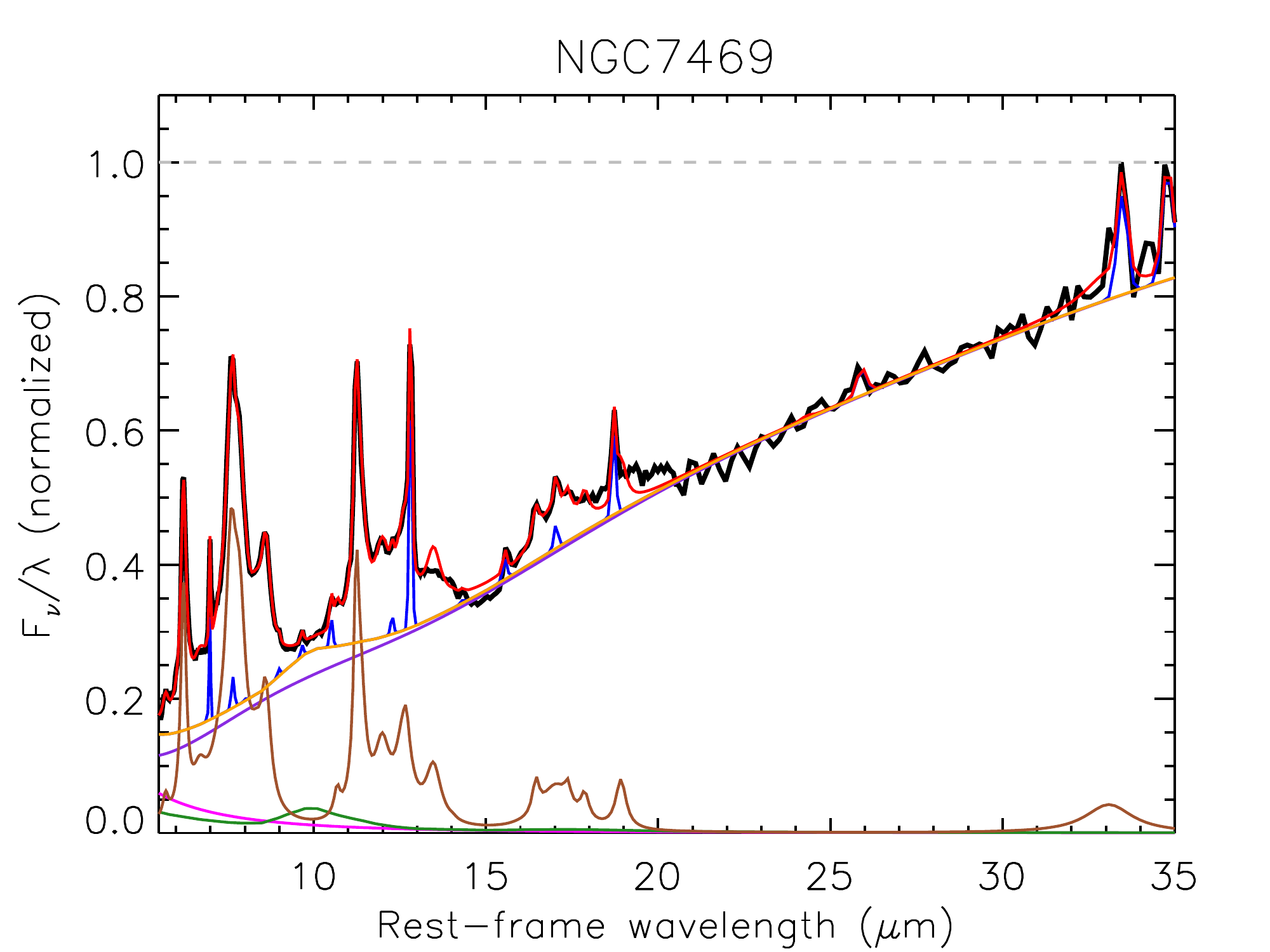}
\includegraphics[width=7.62cm]{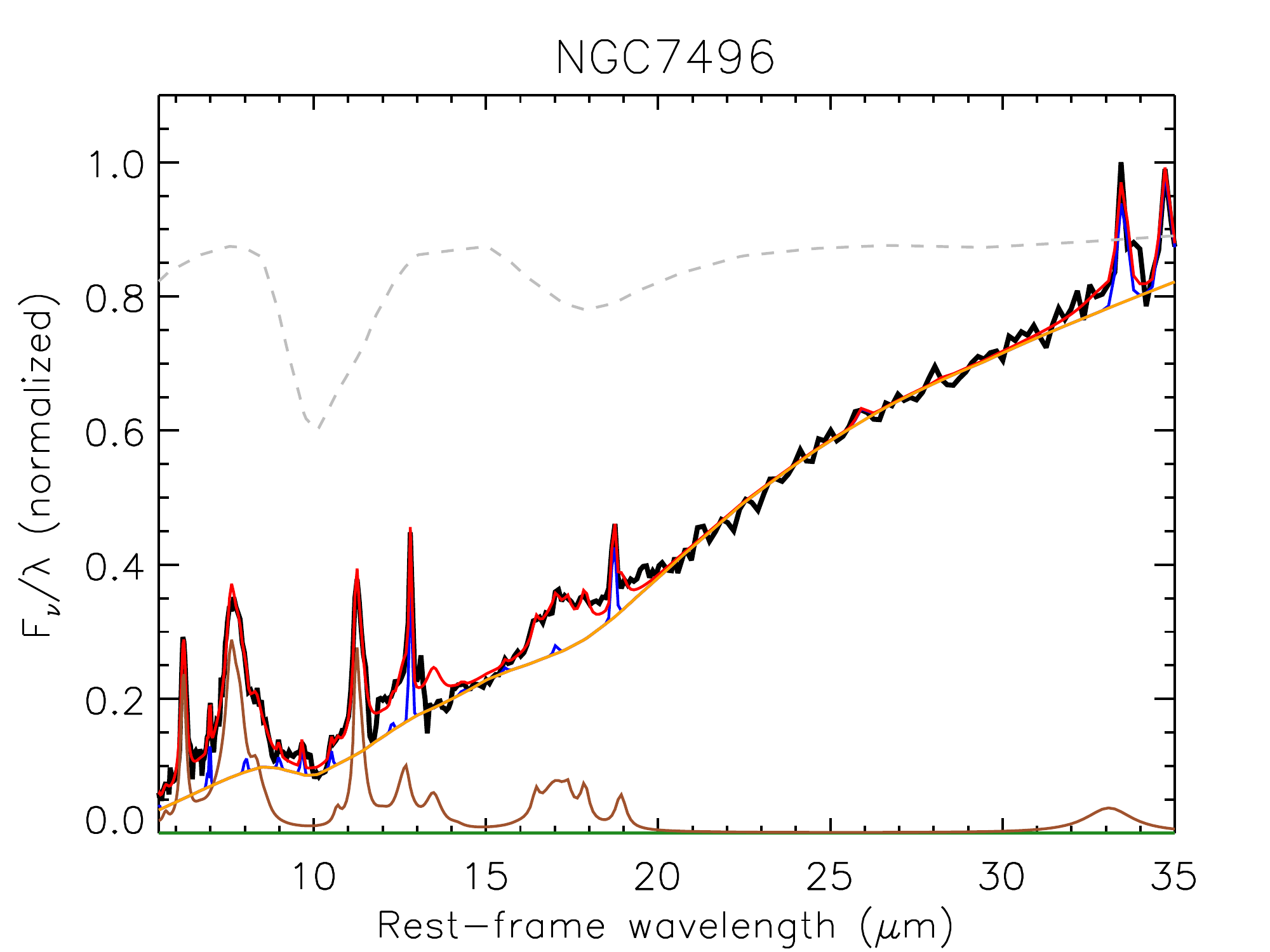}
\includegraphics[width=7.62cm]{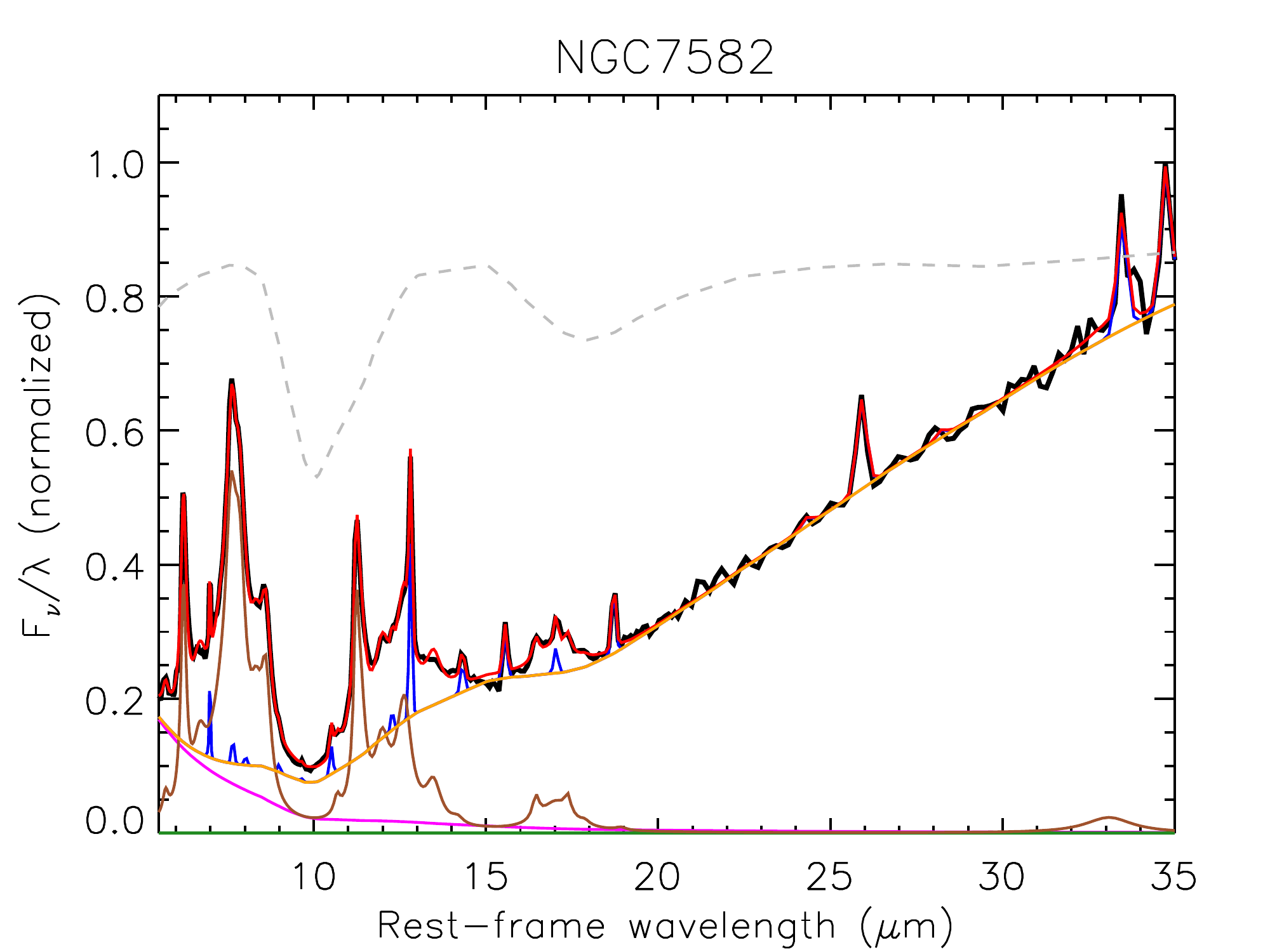}
\includegraphics[width=7.62cm]{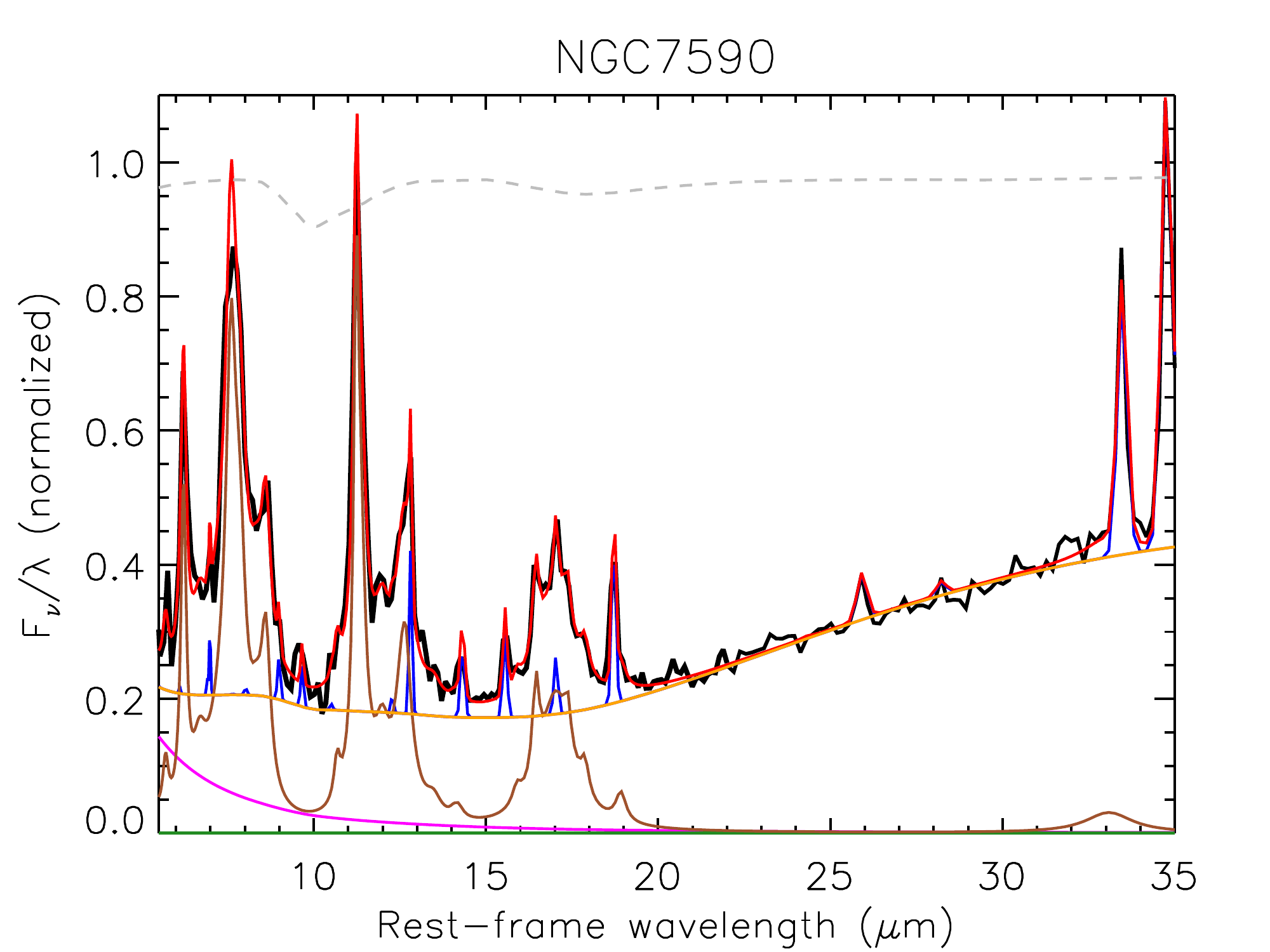}
\includegraphics[width=7.62cm]{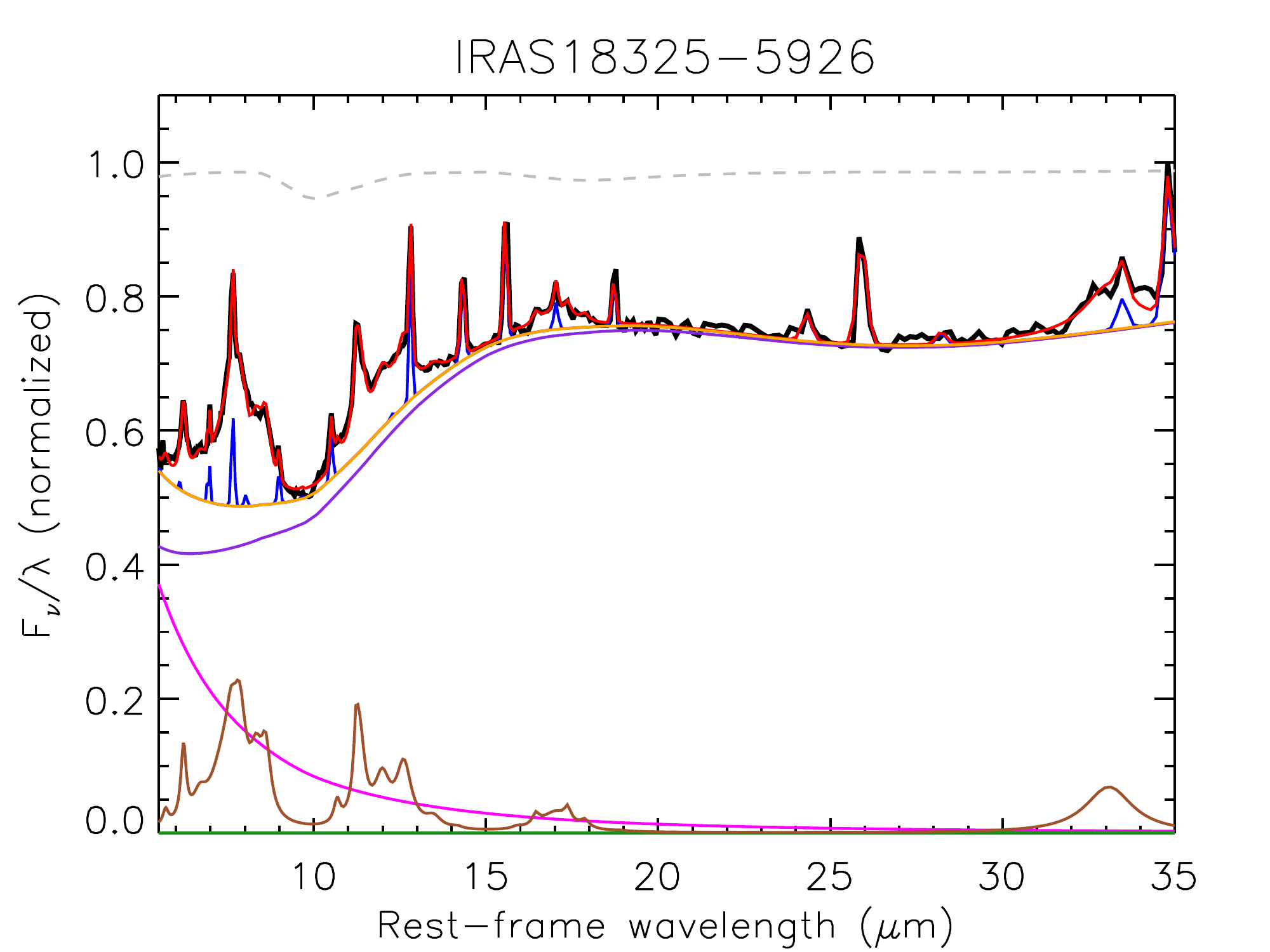}
\par} 
\end{figure*}
\begin{figure*}
\contcaption
\centering
\par{
\includegraphics[width=7.62cm]{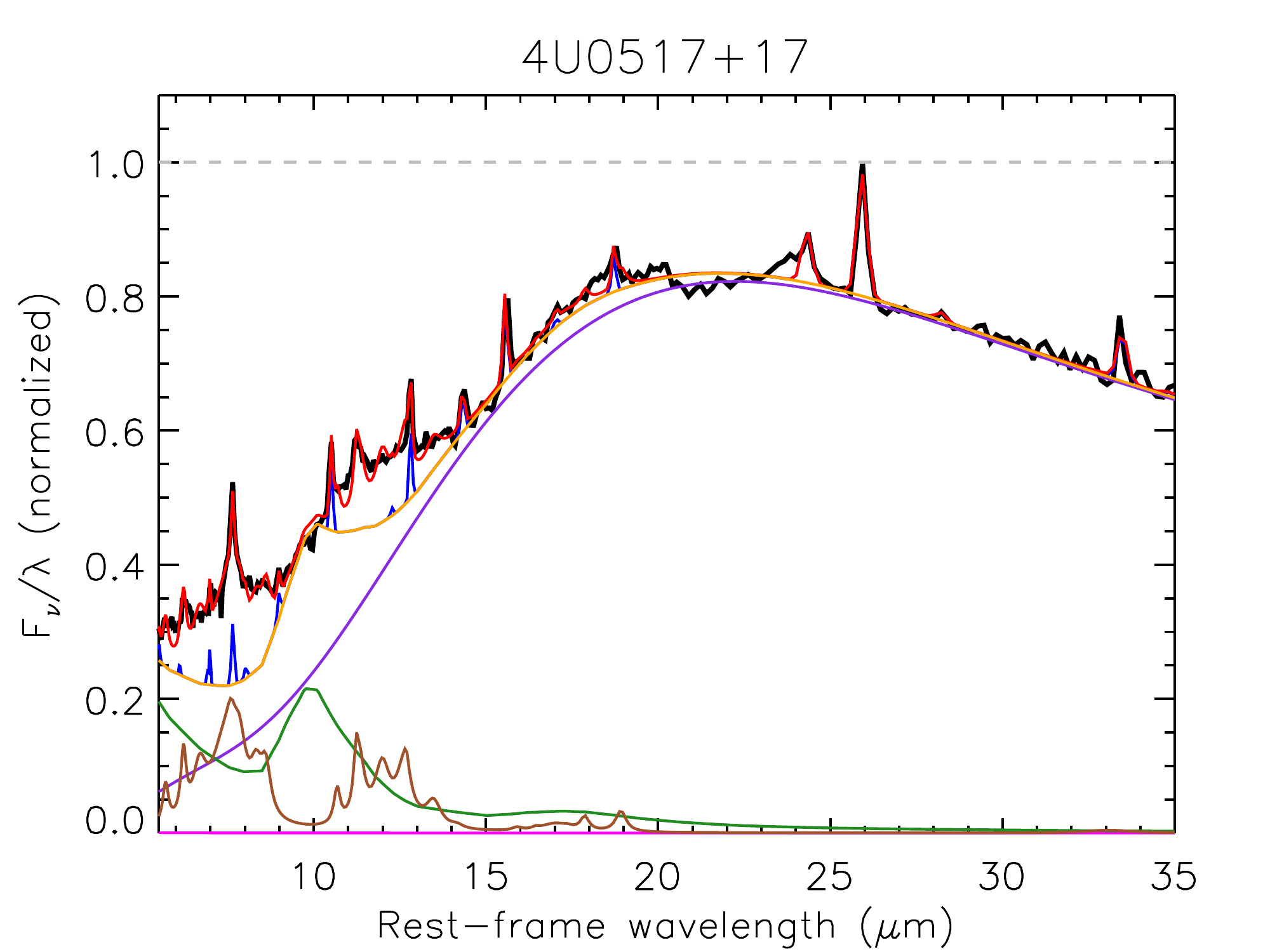}
\includegraphics[width=7.62cm]{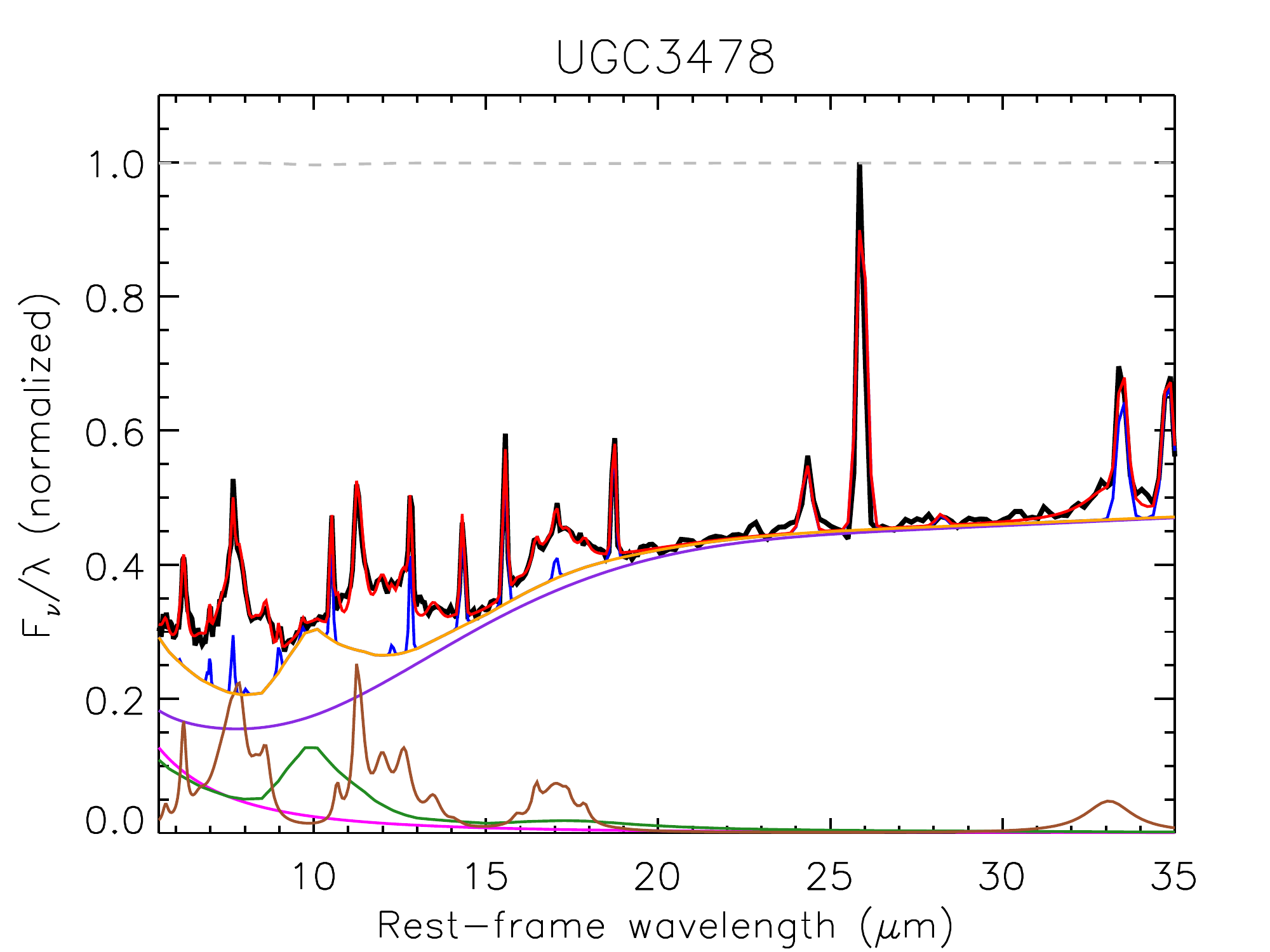}
\par} 
\end{figure*}

\section{Other hardness of the radiation field tracers}
\label{optical}
\begin{figure*}
\centering
\par{
\includegraphics[width=7.3cm]{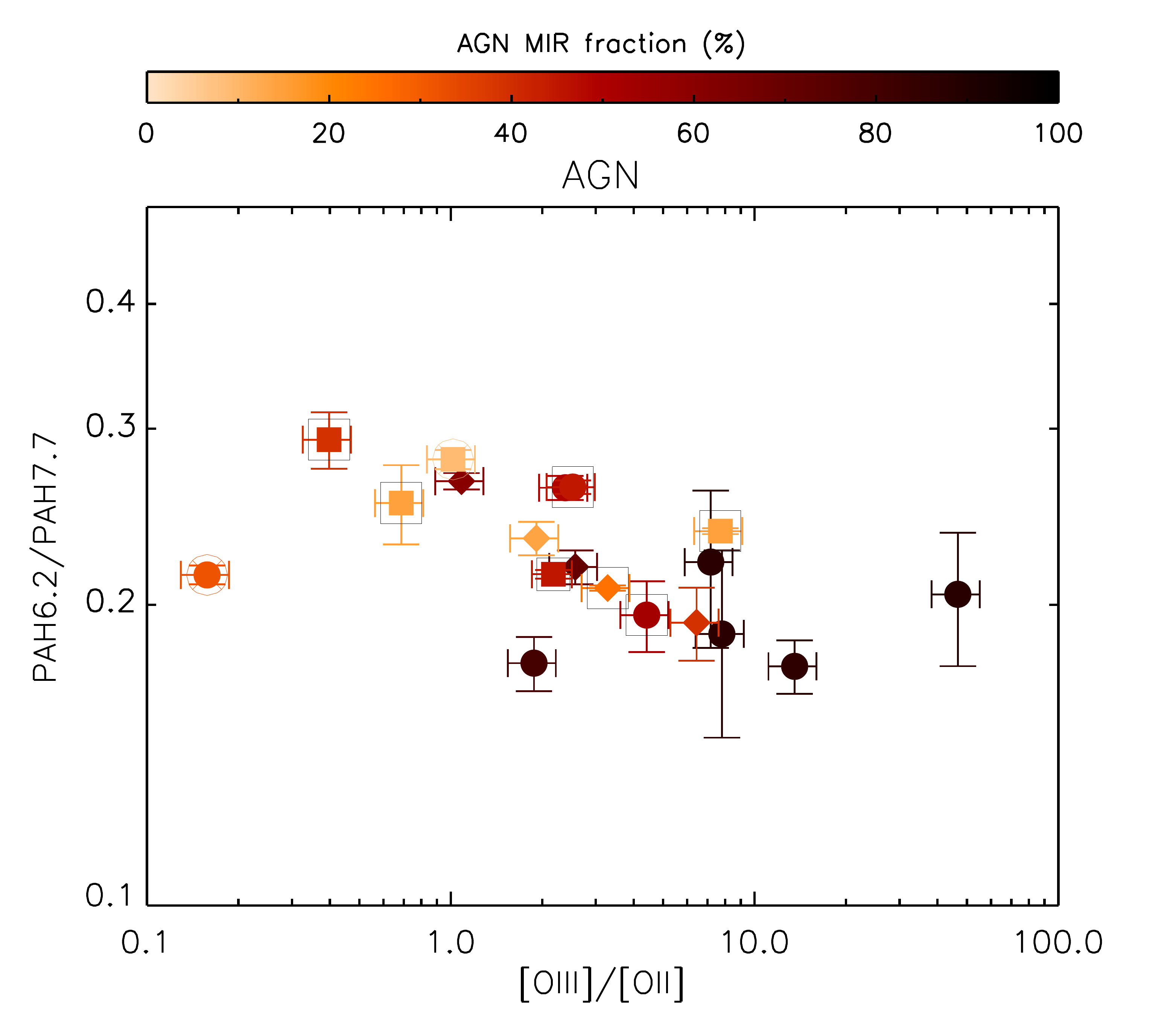}
\includegraphics[width=7.3cm]{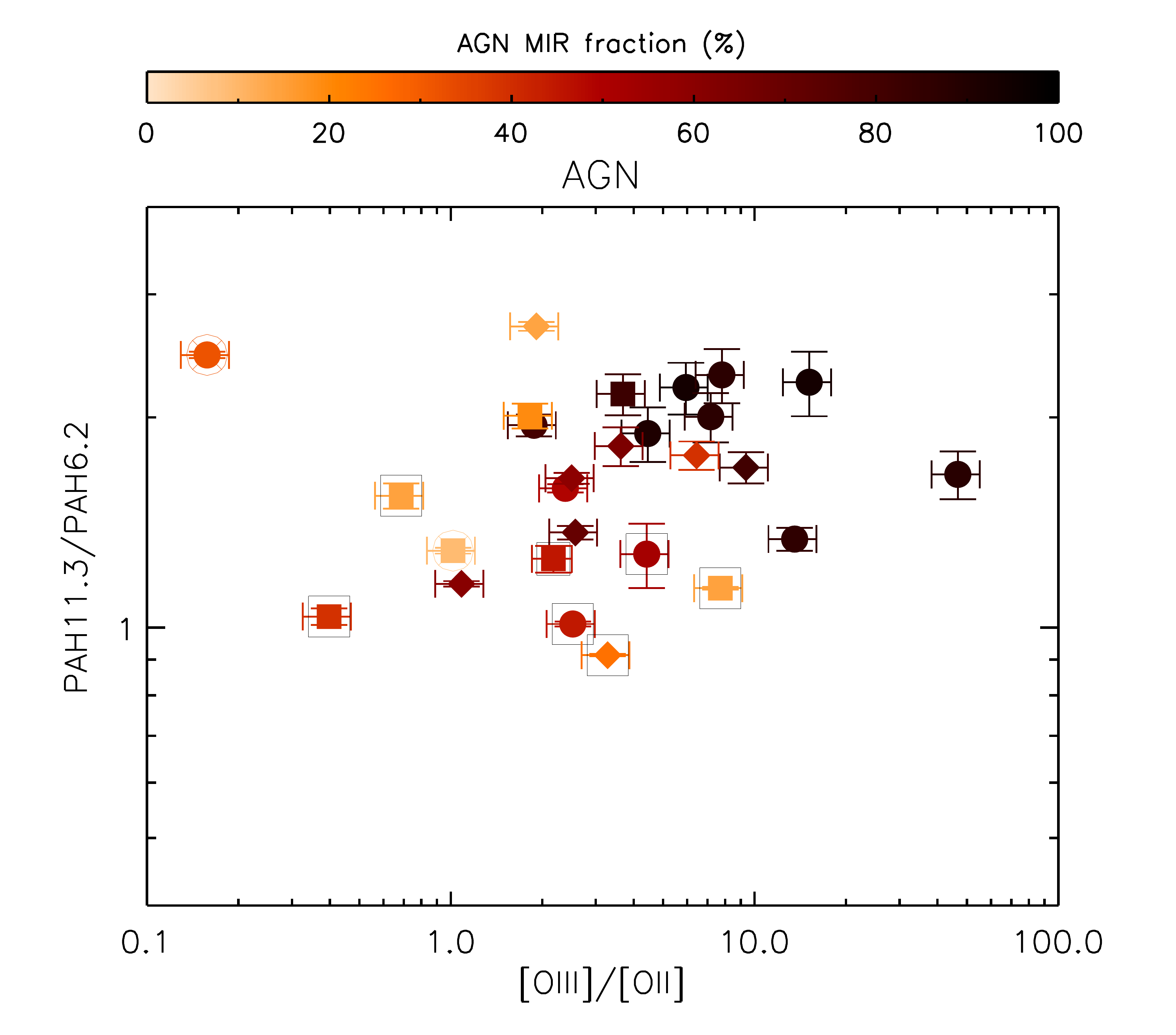}
\par}
\caption{Left panel: 6.2/7.7 PAH ratio vs. [O\,III]$\lambda$5007\AA/[O\,II]$\lambda$3727\AA ~ratio of AGN. Right panel: same as left panel but for the 11.3/6.2 PAH ratio. Filled circles, diamond and squares correspond to Sy1, Sy1.8/1.9 and Sy2, respectively. Note that open squares and circles represent galaxies that have been also classified as AGN/SF composite and LINER, respectively.}
\label{ratio_oiiioii}
\end{figure*}

\begin{figure*}
\centering
\par{
\includegraphics[width=7.3cm]{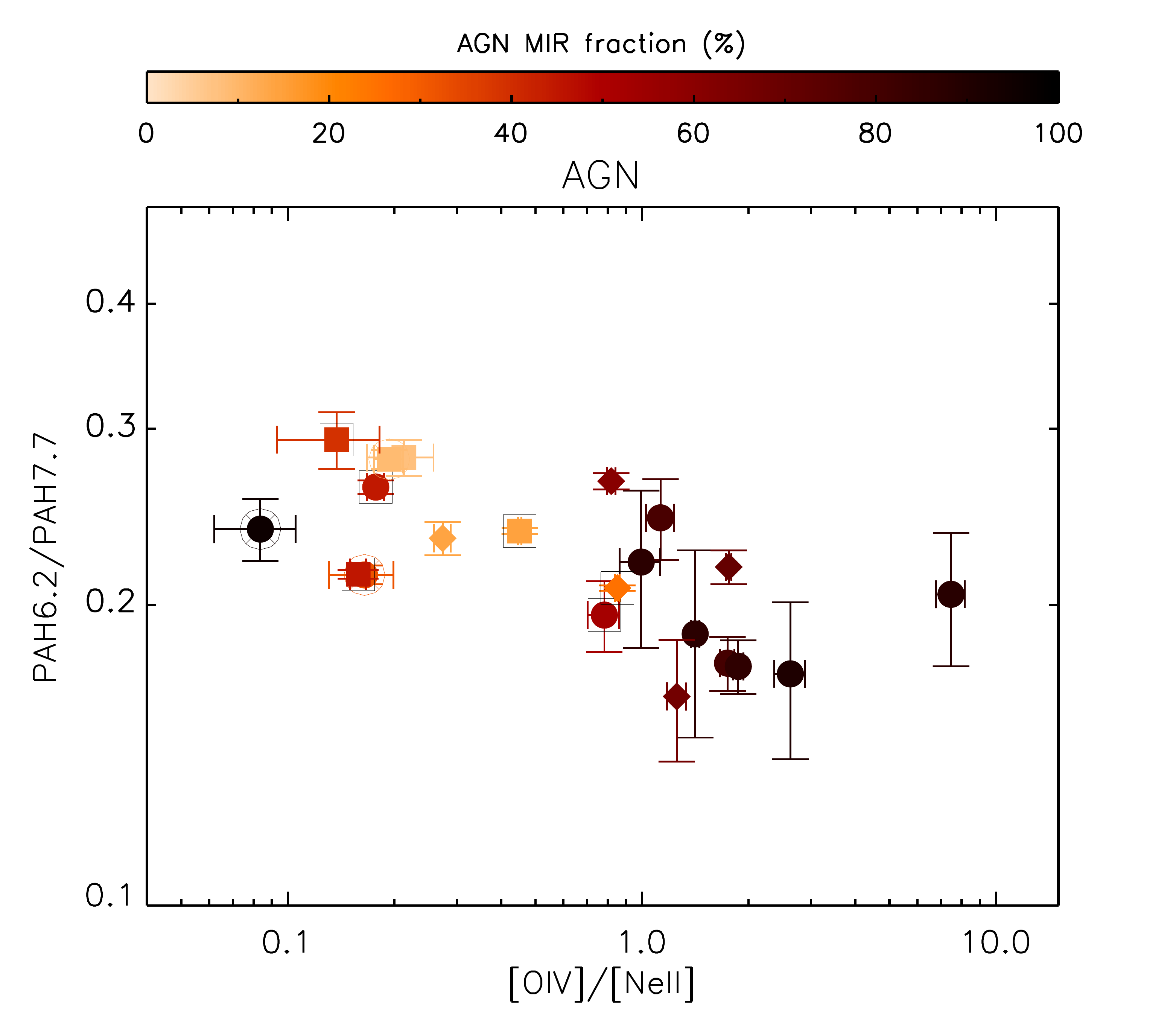}
\includegraphics[width=7.3cm]{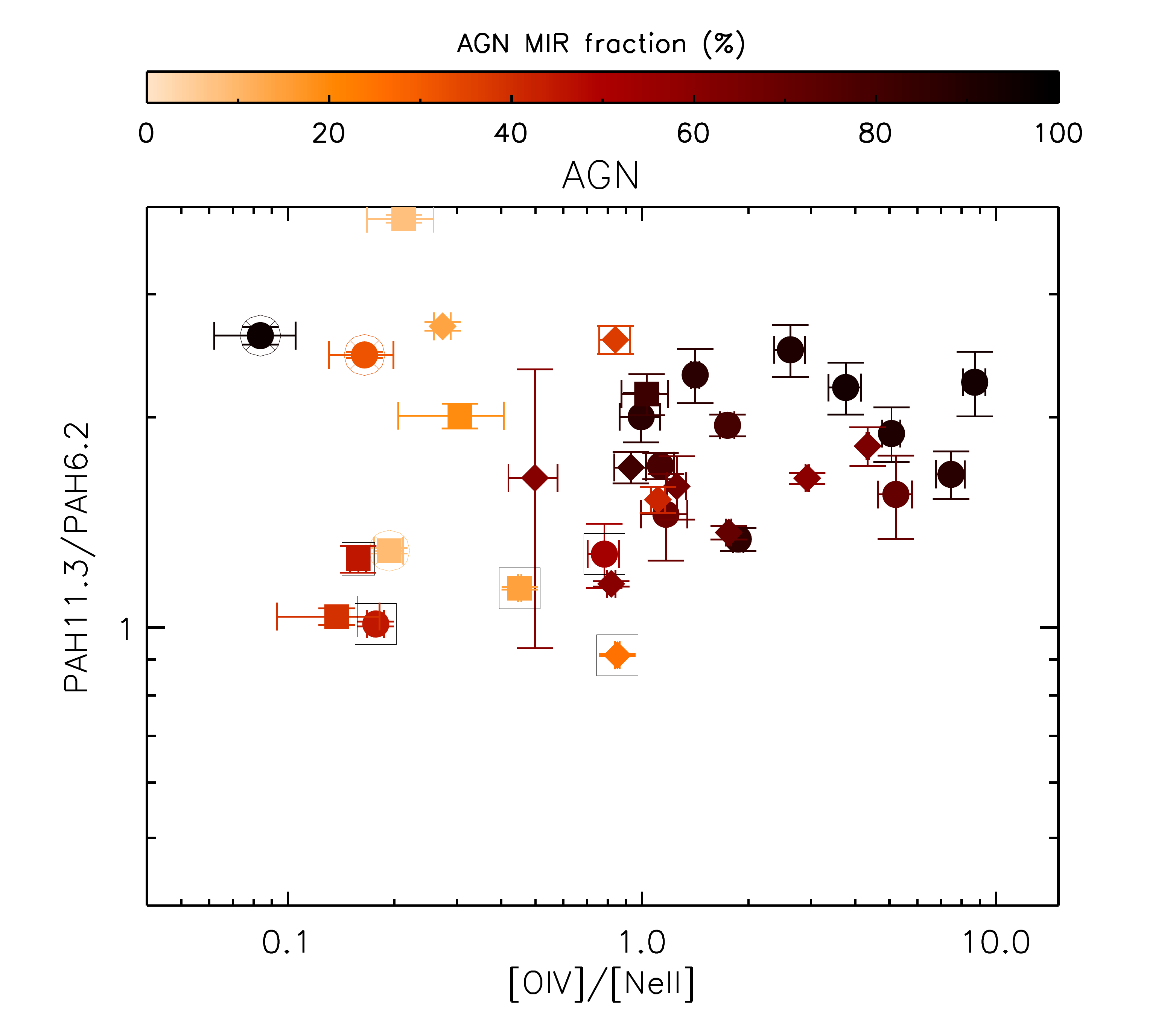}
\par}
\caption{Left panel: 6.2/7.7 PAH ratio vs. [O\,IV]/[Ne\,II] ratio of AGN. Right panel: same as left panel but for the 11.3/6.2 PAH ratio. Filled circles, diamond and squares correspond to Sy1, Sy1.8/1.9 and Sy2, respectively. Note that open squares and circles represent galaxies that have been also classified as AGN/SF composite and LINER, respectively.}
\label{ratio_o4neii}
\end{figure*}

\begin{figure*}
\centering
\par{
\includegraphics[width=7.3cm]{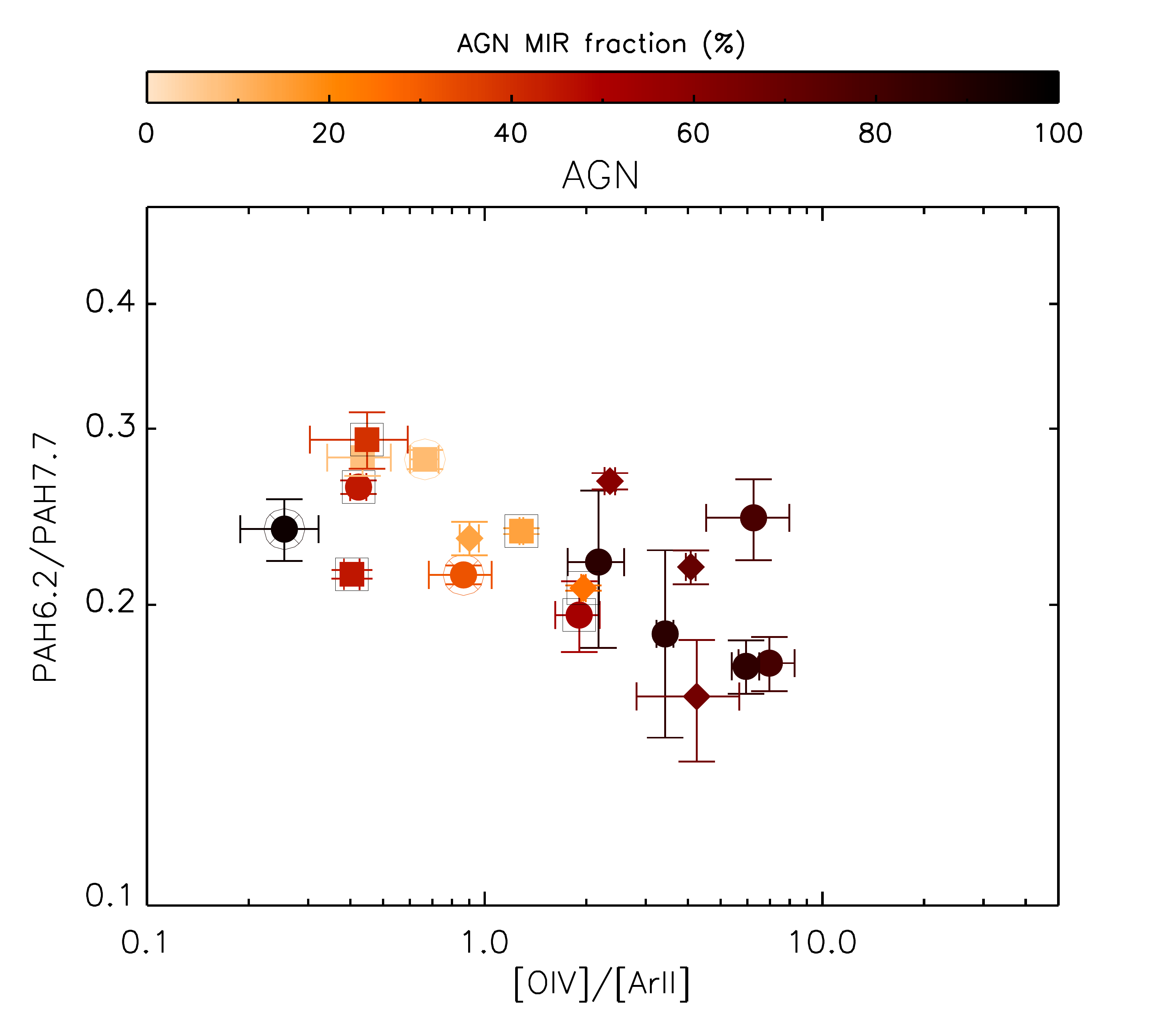}
\includegraphics[width=7.3cm]{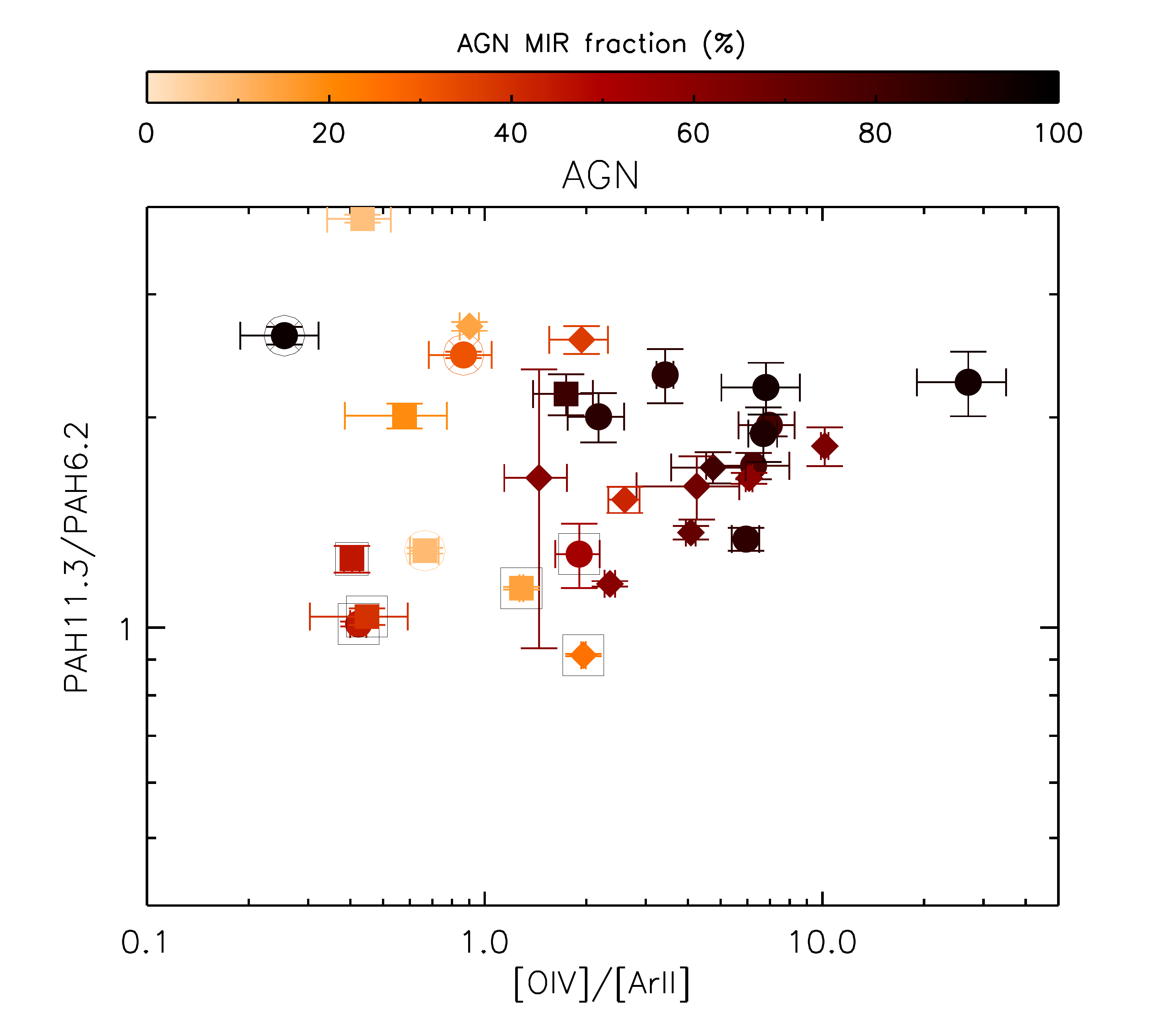}
\par}
\caption{Left panel: 6.2/7.7 PAH ratio vs. [O\,IV]/[Ar\,II] ratio of AGN. Right panel: same as left panel but for the 11.3/6.2 PAH ratio. Filled circles, diamond and squares correspond to Sy1, Sy1.8/1.9 and Sy2, respectively. Note that open squares and circles represent galaxies that have been also classified as AGN/SF composite and LINER, respectively.}
\label{ratio_o4arii}
\end{figure*}

Here we present the results of the comparison of the various PAH ratios used in Section \ref{results} but using the optical [O\,III]$\lambda$5007\AA/[O\,II]$\lambda$3727\AA ~ratio (see Fig. \ref{ratio_oiiioii}), and other MIR emission line ratios (e.g. [O\,IV]/[Ne\,II], [O\,IV]/[Ar\,II]; see Figs. \ref{ratio_o4neii} and \ref{ratio_o4arii}) as tracer of hardness of the radiation field instead of the MIR [Ne\,III]/[Ne\,II] ratio. We find essentially the same trends as when using both approaches (see also Section \ref{results}). 

\section{PAH ratio diagram for upper limits}
\label{PAH_ratio_upper_limits}
\begin{figure*}
\centering
\par{
\includegraphics[width=14.72cm]{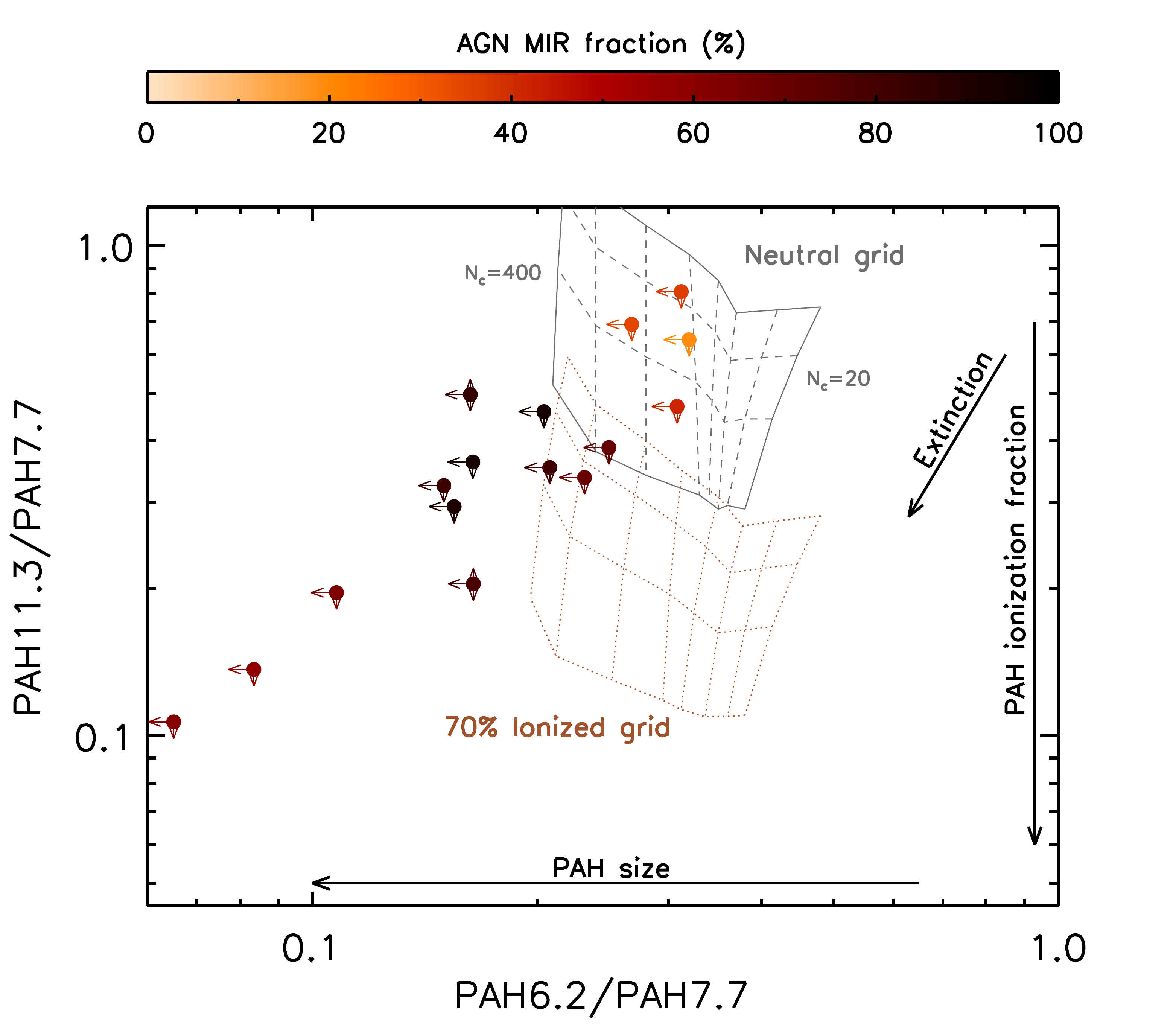}
\par}
\caption{Relative strengths of the 6.2, 7.7 and 11.3~$\mu$m PAH features for those AGN with upper limits in these PAH bands. Filled color code circles correspond with the AGN fractional contribution to the MIR spectra of Seyfert galaxies. The grey grid corresponds to neutral PAHs ranging from small PAHs (N$_{\rm c}$=20; right side of the grid) to large PAH molecules (N$_{\rm c}$=400; left side of the grid). The top and bottom tracks correspond to molecules exposed to a radiation field of ISRF and 10$^3\times$ISRF, respectively. Dashed grey lines correspond to intermediate number of carbons. The brown grid corresponds to a mixture of 70\% ionized - 30\% neutral PAH molecules exposed to the same radiation fields and for the same number of carbons as in the neutral grid. Dotted brown lines correspond to intermediate number of carbons.}
\label{ratio_diagram_upper}
\end{figure*}

In this Section we show the 11.3/7.7 vs. 6.2/7.7 PAH ratio diagram for the various type of astronomical sources used in this work. We find that a significant fraction of AGN has upper limits in these PAH band ratios. We find that AGN (filled color-coded circles in Fig. \ref{ratio_diagram_upper} are located within a diagonal line in this plot. 

\bsp	
\label{lastpage}
\end{document}